\documentclass[oneside,12pt]{book}	
\usepackage{uf_thesis}
\usepackage{amssymb}
\usepackage{amsfonts}
\usepackage{amsmath}
\usepackage{graphicx}
\usepackage{psfrag}
\usepackage{cite}
\usepackage{citesort}
\usepackage{portland}

\ifthenelse{\equal{\isfinal}{false}} {

}{

}

\Title{Radiation Reaction on Moving Particles in General Relativity}
\Author{Eirini Messaritaki}
\Degree{Doctor of Philosophy}
\WorkName{Dissertation}
\DegreeYear{2003}
\DegreeMonth{August}
\Department{Physics}
\College{Liberal Arts and Sciences}
\AddChairman{Professor of Physics}{Steven L. Detweiler}
\AddCommitteeMember{Associate Professor of Physics}{Bernard F. Whiting}
\AddCommitteeMember{Professor of Physics}{James N. Fry}
\AddCommitteeMember{Assistant Professor of Physics}{Andrew Rinzler}
\AddCommitteeMember{Assistant Professor of Astronomy}{Vicki L. Sarajedini}
\newcommand{\tx}{\text}

\begin{document}
   \frontmatter		
      \maketitle
      
\vskip 20pt 

First and foremost I would like to thank Dr.~Steve Detweiler, Professor of Physics at the University of Florida, who was my advisor during my graduate studies.
He has always been very supportive of my work, very willing to help and extremely patient with me.
I feel very lucky to have worked with him during the past 5 years.

I would also like to thank Dr.~Bernard Whiting for his help on writing this dissertation and for his contributions to the research.

Many thanks go to the other members of my committee:  Dr.~Jim Fry, Dr.~Andrew Rinzler and Dr.~Vicki Sarajedini, as well as to Dr.~Richard Woodard and Dr.~Luz Diaz, for their useful comments on the dissertation.

I would also like to thank my brothers, my family and my close friends.
Their help and support has always been very important and I could not have come this far without them.

My research was partly supported by the Institute of Fundamental Theory at the 
University of Florida and I am grateful for that.

      \tableofcontents
   \mainmatter		
      \chapter{Introduction}

The General Theory of Relativity, developed by Albert Einstein at the beginning of the twentieth century, has been successfully used by scientists for many years to explain physical phenomena and to make predictions about physical systems.
One of the most exciting predictions of General Relativity is the existence of gravitational radiation, which can be thought of as a wave-like distortion of spacetime. 
It is expected that gravitational radiation from remote astrophysical systems will yield very important information about those systems, thus answering many questions that scientists currently have. 
Even though the detection of gravitational waves has proved to be a challenging task,
recent developments in technology have made scientists confident that gravitational waves will indeed be detected in the near future. 
Both earth-based detectors (such as LIGO and VIRGO) and space-based detectors (such as LISA) are expected to start operating in the next decade or two.
For the data collected by these detectors to be useful, accurate information about the sources of gravitational radiation is required. 
Only if scientists know what the gravitational waves emitted by specific systems should look like, can they compare them with the patterns observed and draw conclusions about the physical characteristics of those systems.
That is why there has been a lot of interest lately in predicting the gravitational radiation emitted by different astrophysical systems.

One astrophysical system for which gravitational waves are expected to be detected, specifically by space-based detectors, is the binary system of a small neutron star or black hole (of mass equal to a few times the mass of the Sun) and a supermassive black hole (of mass equal to a few million times the mass of the Sun) \cite{schutz}. 
Supermassive black holes are believed to exist at the centers of many galaxies, including our own. 
It is also believed that their strong gravitational field can capture smaller stars and black holes, which then move toward the supermassive black hole, until they are absorbed by it.
The exact evolution of the system with time can give the pattern of the gravitational radiation that the system is expected to emit.
Knowing that pattern is crucial in distinguishing whether the gravitational waves measured by the detector come from such a system. 
It can also help extract information about many characteristics of the system, such as the masses and the angular momenta of the two components of the binary system.

To determine how the system evolves, the path of the smaller star or black hole needs to be calculated. 
That path is largely determined by the influence of the gravitational field of the supermassive black hole on the small star or black hole;
it is also affected by the interaction of the small star or black hole with its own gravitational field. 
This latter interaction, commonly referred to as the self-force, is responsible for the radiation reaction effects,
which cause the decay of the small star's or black hole's orbit toward the supermassive black hole.

Even though my motivation for studying the self-force effects stems from the need to know the evolution of this system, radiation reaction effects are present in other systems that are easier to deal with mathematically.
One can think, for example, of a particle of a certain scalar charge, which moves in a Schwarzschild background spacetime and creates its own scalar field. 
There is also the case of a particle that carries an electric charge and creates an electromagnetic field as it moves in spacetime.
The scalar field of the particle in the first case and the electromagnetic field of the electric charge in the second case will affect the motion of each particle, causing its worldline to differ from what it would be if radiation reaction effects were not present.
It is useful to predict the evolution of such systems, mainly because they can give an idea of how the more difficult system can be handled successfully and not because they are realistic systems, since they are not expected to be observed in nature.

In the past, various efforts have been made to develop a concrete scheme for including the radiation reaction effects for scalar, electromagnetic and gravitational fields in the equations of motion.
One approach that has been used extensively involves calculating the flux, both at infinity and at the event horizon of the central supermassive black hole, of quantities (such as the energy and momentum of the particle) that are constants of the motion when radiation reaction is not present \cite{davis,TeukPress,poiss1,shibata,IyerWill,poiss2,poiss3,poiss4}.
That flux is then associated with the rate of change of those quantities at the location of the particle and the evolution of the particle's motion can be predicted.
However, this approach is generally not applicable in the case in which the central black hole is a rotating black hole, except under some special circumstances \cite{hughes}.
One reason is that the evolution of the Carter constant \cite{BaPrTe}, which is a constant of the motion that is not derived by a Killing vector,  cannot be calculated using the flux at infinity and at the event horizon.
In addition, this approach does not take into account the very significant non-dissipative effects of the self-force that the particle's field exerts on the particle \cite{wiseman}.
That is why it is important for the self-force, and not just its radiation reaction effects, to be calculated.
The calculation of the self-force is the subject of this dissertation.

In Chapters \ref{RRF} and \ref{RRC} of this dissertation, I present the innovative work of Dirac \cite{dirac} for the radiation reaction effects on an electron moving in flat spacetime and the theoretical generalization of it to the case of curved spacetime by DeWitt and Brehme \cite{dewbre}. 
In Chapter \ref{UFM}, I present different methods for practical calculations of the self-force that have been proposed in the past, with emphasis on one in particular.
In Chapters \ref{SF}-\ref{CONC} of this dissertation, I present an implementation of that particular method for calculating the radiation reaction effects for a scalar, an electromagnetic and a gravitational field. 

      \chapter{Radiation Reaction in Flat Spacetime}
\label{RRF}

The first successful attempt to provide an expression for the radiation reaction effects on a particle in special relativity was made by Dirac in 1938 \cite{dirac}.
In his famous paper, he studied the motion of an electron moving in flat spacetime by using the concept of conservation of energy and momentum along the electron's worldline.
In this chapter I present the decomposition of the electromagnetic field and the derivation of the
equations of motion, which include the radiation damping effects on the electron.
The scheme for calculating the radiation reaction effects in curved spacetime presented in this dissertation extends Dirac's work.

For the equations to look as simple as possible, the speed of light is set equal to 1 in this chapter.

It is noted that the metric signature used by Dirac is $(+ - - -)$ and that is what is used in this chapter, in order to keep the results identical to those derived by Dirac.
In all subsequent chapters, however, the signature is changed to $(- + + +)$, in order to adhere to the currently widely used convention.
Thus, some of the equations given in this chapter change in subsequent chapters.
Since none of the results described in this chapter are used in the chapters that follow, that should not cause any confusion at all.

\section{Definitions of the Fields}
\label{DefFields}

Dirac's analysis starts by assuming that an electron of electric charge $q$ is moving in flat spacetime, so that the background metric is the Minkowski metric $\eta_{\mu \nu} $. 
The worldline of the electron is denoted by
\begin{equation}
z_{\mu} = z_{\mu} (s)
\label{zmu}
\end{equation}
where $s$ is the proper time along the worldline. 
The electromagnetic 4-vector potential that the electron creates is assumed to satisfy the Lorentz gauge condition
\begin{equation}
\nabla^{\mu} A_{\mu} = 0
\label{LorG}
\end{equation}
and obeys Maxwell's equations
\begin{equation}
\nabla ^{2} A_{\mu} = 4 \pi j_{\mu}.
\label{Maxwell}
\end{equation}
It is well-known that the Lorentz gauge condition leaves some arbitrariness for the electromagnetic potential.
In Equation (\ref{Maxwell}), $j_{\mu}$ is the charge-current density vector of the electron, namely
\begin{equation}
j_{\mu} = q \int \frac{dz_{\mu}}{ds} \: \delta(x_0 - z_0) \: \delta(x_1 - z_1) \: \delta(x_2 - z_2) \: \delta(x_3 - z_3) \: ds.
\end{equation}
The electromagnetic field associated with a 4-vector potential $A^{\mu}$ is, in general, given by the equation
\begin{equation}
F^{\mu \nu} = \nabla^{\nu} A^{\mu} - \nabla^{\mu} A^{\nu}.
\label{field}
\end{equation}

It is clear that, unless the boundary conditions of the problem are specified, Equations (\ref{LorG}) and (\ref{Maxwell}) do not have a unique solution. 
In fact, adding any solution of the homogeneous equation
\begin{equation}
\nabla ^{2} A_{\mu} = 0
\label{HomEq}
\end{equation}
(which represents a source-free radiation field and obeys the Lorentz gauge condition) to a solution of Equation (\ref{Maxwell}), gives a different solution of Equation (\ref{Maxwell}) that also obeys Equation (\ref{LorG}).

One solution of interest for the problem at hand is the retarded electromagnetic field $F_{\tx{ret}}^{\mu \nu}$ created by the electron, which is derived by the well-known Lienard-Wiechert potentials $A_
{\tx{ret}}^{\mu}$ that obey Equations (\ref{LorG}) and (\ref{Maxwell}).
Assuming that there is also a radiation field $F_{\tx{in}}^{\mu \nu}$ incident on the electron, the actual field in the neighborhood of the electron is the sum of those two fields
\begin{equation}
F_{\tx{act}}^{\mu \nu} = F_{\tx{ret}}^{\mu \nu} + F_{\tx{in}}^{\mu \nu}.
\label{Fact1}
\end{equation}
Another interesting solution is the one given by the advanced potentials $A_{\tx{adv}}^{\mu}$. 
It is reasonable to expect that the advanced potentials play a role symmetrical to the retarded potentials in this problem. 
If the outgoing radiation field leaving the neighborhood of the electron is $F_{\tx{out}}^{\mu \nu}$, then
\begin{equation}
F_{\tx{act}}^{\mu \nu} = F_{\tx{adv}}^{\mu \nu} + F_{\tx{out}}^{\mu \nu}.
\label{Fact2}
\end{equation}
The radiation emitted by the electron is given by the difference of the outgoing and the incoming radiation
\begin{equation}
F_{\tx{rad}}^{\mu \nu} = F_{\tx{out}}^{\mu \nu} - F_{\tx{in}}^{\mu \nu}.
\label{Frad1}
\end{equation}
Using Equations (\ref{Fact1}) and (\ref{Fact2}) for the actual field, an alternative expression for the radiation field can be obtained
\begin{equation}
F_{\tx{rad}}^{\mu \nu} = F_{\tx{ret}}^{\mu \nu} - F_{\tx{adv}}^{\mu \nu}.
\label{Frad2}
\end{equation}
The important implication of Equation (\ref{Frad2}) is that, because the retarded and the advanced fields are uniquely determined by the worldline of the electron, so is the radiation emitted by the electron.
It becomes clear from the analysis of Section (\ref{DEqM}) that the radiation field is responsible for the deviation of the electron's worldline from a background geodesic, because that field shows up in the equations of motion, in the radiation damping term.

It is significant to realize that Equation (\ref{Frad2}) for the radiation field is consistent with what one would expect for the radiation of an accelerating electron.
To understand that, one should recall that the radiation produced by an accelerating electron can be calculated using the retarded field $F_{\tx{ret}}^{\mu \nu}$ that the electron creates, at very large distances away from the electron and at much later times than the time when the acceleration takes place.
At those distances and times the advanced field vanishes and Equation (\ref{Frad2}) gives the correct result.
The advantage of Equation (\ref{Frad2}) is that it gives the radiation produced by the electron at any point in spacetime, so it can be used for the radiation in the neighborhood of the electron.

A field that is needed for the analysis of Section (\ref{DEqM}) is the difference of the average of the retarded and advanced fields from the actual field, denoted by $f^{\mu \nu}$
\begin{equation}
f^{\mu \nu} = F_{\tx{act}}^{\mu \nu} - \frac{1}{2} ( F_{\tx{ret}}^{\mu \nu} + F_{\tx{adv}}^{\mu \nu} ).
\label{f1}
\end{equation}
Using Equations (\ref{Fact1}) and (\ref{Fact2}) with Equation (\ref{f1})
\begin{equation}
f^{\mu \nu} = \frac{1}{2} ( F_{\tx{in}}^{\mu \nu} + F_{\tx{out}}^{\mu \nu} ).
\label{f2}
\end{equation}
Because the incoming and outgoing radiation fields are derived from potentials that satisfy the homogeneous Equation (\ref{HomEq}), the field $f^{\mu \nu}$ is sourceless. 
It is also free from singularities at the worldline of the electron.

\section{Derivation of the Equations of Motion}
\label{DEqM}

The main concept used by Dirac in deriving the equations of motion for the electron moving in flat spacetime is the conservation of energy and momentum.
The first step is to surround the worldline of the electron by a thin cylindrical tube of constant radius.
This radius is assumed to be very small, specifically smaller than any length of physical significance in the problem.
The objective is to calculate the flow of energy and momentum across the three-dimensional surface of the tube, using the stress-energy tensor $T_{\mu \rho}$ calculated from the actual electromagnetic field.
The stress-energy tensor is given by
\begin{equation}
4 \: \pi \: T_{\mu \rho} = F_{\tx{act} \: \mu \nu} \: F_{\tx{act} \; \; \rho}^{\; \; \; \; \; \nu} \: + \: \frac{1}{4} \: \eta_{\mu \rho}  \: F_{\tx{act} \: \alpha \beta} \: F_{\tx{act}}^{\alpha \beta}.
\label{SETensor}
\end{equation}
The flow of energy and momentum out of the surface of the tube is equal to the difference in the energy and momentum at the two ends of the tube.

In the following, dots over quantities denote differentiation with respect to the proper time $s$.
To simplify the notation, $(dz_{\mu}/ds)$ is set equal to $v_{\mu}$.
One then obtains the equations
\begin{equation}
v_{\mu} v^{\mu} = 1,
\label{v1}
\end{equation}
\begin{equation}
v_{\mu} \dot{v^{\mu}} = 0,
\label{v2}
\end{equation}
\begin{equation}
v_{\mu} \ddot{v^{\mu}} + \dot{v_{\mu}} \dot{v^{\mu}} = 0.
\label{v3}
\end{equation}

To calculate the stress-energy tensor, the electromagnetic 4-vector potential and the electromagnetic field need to be calculated first. 
The retarded potential at a point $x_{\mu}$ generated by an electron moving on the worldline $z_{\mu}(s)$ is given by
\begin{equation}
A_{\tx{ret} \: \mu} = \frac{q \: \dot{z_{\mu}} }{ \dot{z_{\nu}} \: (x^{\nu} - z^{\nu})}
\label{Aret1}
\end{equation}
calculated at the retarded proper time, which is the value of $s$ that solves the equation 
\begin{equation}
(x_{\nu} - z_{\nu}) \: (x^{\nu} - z^{\nu}) = 0.
\end{equation}
An equivalent expression for the retarded potential, obtained from Equation (\ref{Aret1}) 
by using elementary properties of the $\delta$ function, is
\begin{equation}
A_{\tx{ret} \: \mu} = 2 \: q \int_{-\infty}^{\tau_{\tx{int}}} \: \dot{z_{\mu}} \: \delta[(x_{\nu} - z_{\nu}) \: (x^{\nu} - z^{\nu})] \: ds
\label{Aret2}
\end{equation}
where $\tau_{\tx{int}}$ is a proper time between the retarded and advanced proper times.
By differentiating Equation (\ref{Aret2}) and using Equation (\ref{field}), the retarded electromagnetic field is obtained
\begin{equation}
F_{\tx{ret} \: \mu \nu} = - 2 \: q \int_{-\infty}^{\tau_{\tx{int}}} \frac{d}{ds} \bigg [ \frac{\dot{z_{\mu}} \: (x_{\nu} - z_{\nu}) - \dot{z_{\nu}} \: (x_{\mu} - z_{\mu})}{\dot{z_{\kappa}} \: (x^{\kappa} - z^{\kappa})} \bigg ] \: \delta[(x_{\lambda} - z_{\lambda}) \: (x^{\lambda} - z^{\lambda})] \: ds.
\label{FretInt}
\end{equation}
By using again some of the elementary properties of the $\delta$ function, an equivalent expression for the retarded field is derived
\begin{equation}
F_{\tx{ret} \: \mu \nu} = - \frac{q}{ \dot{z_{\lambda}} \: (x^{\lambda} - z^{\lambda})} \:  \frac{d}{ds} \bigg [  \frac{\dot{z_{\mu}} \:
 (x_{\nu} - z_{\nu}) - \dot{z_{\nu}} \: (x_{\mu} - z_{\mu})}{\dot{z_{\kappa}} \:
 (x^{\kappa} - z^{\kappa})} \bigg ]. 
\label{Fret}
\end{equation}
Again, all quantities are calculated at the retarded proper time.

Since the goal is to calculate the stress-energy tensor at the two ends of the world-tube surrounding the worldline of the electron, it can be assumed that the point $x_{\mu}$ is very close to the worldline. 
Specifically
\begin{equation}
x_{\mu} = z_{\mu} (s_0) + \gamma _{\mu}
\label{defgamma}
\end{equation}
where the $\gamma_{\mu}$'s are very small.
Then, the fields can be Taylor-expanded in the $\gamma_{\mu}$'s.
In the expansions that follow, all the coefficients are taken at the proper time $s_0$.
It can be assumed that the retarded proper time is $s_0 - \sigma$,
where it is reasonable that $\sigma$ is a small positive quantity of the same order as $\gamma_{\mu}$.

The detailed calculations needed to obtain the Taylor expansions of the electromagnetic fields are tedious but straightforward.
They can be found in Dirac's paper; here I present only the results of that calculation.

Since $\gamma^{\mu}$ is a space-like vector, it can be assumed that $\gamma_{\mu} \gamma^{\mu} = - \varepsilon^{2}$, where $\varepsilon$ is a positive number.
Taylor-expanding the retarded field and keeping  only the terms that do not vanish in the limit $\varepsilon \rightarrow 0$ gives
\begin{equation}
\begin{split}
F_{\tx{ret} \: \mu \nu} = q \: (1 - \gamma_{\kappa} \dot{v^{\kappa}})^{-\frac{1}{2}}
 \bigg [ &- \: \varepsilon^{-3} \: (v_{\mu} \gamma_{\nu} - v_{\nu} \gamma_{\mu})  
          \: - \: \frac{1}{2} \: \varepsilon^{-1} \: ( \dot{v_{\mu}} v_{\nu} - \dot{v_{\nu}} v_{\mu} )
\: (1 - \gamma_{\lambda} \dot{v^{\lambda}}) \\
         &+ \: \frac{1}{8} \: \varepsilon^{-1} \: \dot{v_{\lambda}} \dot{v^{\lambda}}
\: (v_{\mu} \gamma_{\nu} - v_{\nu} \gamma_{\mu}) 
         \: + \: \frac{1}{2} \: \varepsilon ^{-1} ( \ddot{v_{\mu}} \gamma_{\nu} - \ddot{v_{\nu}} \gamma_{\mu} ) \\ 
         &+ \: \frac{2}{3} \: ( \ddot{v_{\mu}} v_{\nu} - \ddot{v_{\nu}} v_{\mu} ) \bigg ].
\end{split}
\end{equation}
The advanced field can be obtained by changing $\varepsilon$ to $-\varepsilon$ and changing the sign of the whole expression.
By using the retarded and advanced fields in Equation (\ref{Frad2}) the radiation field on the worldline is obtained
\begin{equation}
F_{\tx{rad} \: \mu \nu} = \frac{4 \: q}{3} \: (\ddot{v_{\mu}} v_{\nu} - \ddot{v_{\nu}} v_{\mu}).
\label{Fradv}
\end{equation}
The actual field then becomes
\begin{equation}
\begin{split}
F_{\tx{act} \: \mu \nu} = f_{\mu \nu} + q [1-\gamma_{\lambda} \dot{v^{\lambda}}]^{-\frac{1}{2}}
 \bigg [ & ( \varepsilon^{-3} - \frac{1}{8} \varepsilon^{-1} \dot{v_{\kappa}} \dot{v^{\kappa}}) (\gamma_{\mu} v_{\nu} - \gamma_{\nu} v_{\mu}) \\ 
  	 & + \frac{1}{2} \varepsilon^{-1} (1+\gamma_{\kappa} \dot{v^{\kappa}}) (v_{\mu} \dot{v_{\nu}} - v_{\nu} \dot{v_{\mu}} \\
	 & + \frac{1}{2} \varepsilon^{-1} (\ddot{v_{\mu}} \gamma_{\nu} - \ddot{v_{\nu}} \gamma_{\mu}) \bigg ].
\end{split}
\label{FactTaylor}
\end{equation}

To calculate the energy and momentum flow out of the tube, the stress-energy tensor needs to be calculated. 
In fact, only its component along the direction of $\gamma_{\mu}$ is necessary.
Substituting Equation (\ref{FactTaylor}) into Equation (\ref{SETensor}):
\begin{equation}
\begin{split}
4 \: \pi \: T_{\mu \rho} \: \gamma^{\rho} &= F_{\tx{act} \: \mu \nu} \: F_{\tx{act} \; \; \rho}^{\; \; \; \; \;
 \nu} \: \gamma^{\rho} \: - \: \frac{1}{4} \: F_{\tx{act} \: \alpha \beta} \: F_{\tx{act}}^{\beta \alpha} \gamma_{\mu}\\
                                          &= q^2 \: (1 - \gamma_{\kappa} \dot{v^{\kappa}})^{-1} \: \bigg [ \: (\frac{1}{2} \: \varepsilon^{-4} \: + \: \frac{1}{2} \: \varepsilon^{-2} \dot{v_{\lambda}} \dot{v^{\lambda}}) \: \gamma_{\mu} \: - \: \frac{1}{2} \: \varepsilon^{-2} (1 + \frac{3}{2} \: \gamma_{\lambda} \dot{v^{\lambda}}) \: \dot{v_{\mu}} \: \bigg ] \\ 
                                          & \; \; \; \; + \: q \: \varepsilon^{-1} v^{\nu} f_{\mu \nu}.   \\
\end{split}
\end{equation}

The flow of energy and momentum out of the surface of the tube is given by integrating this component of the stress-energy tensor over the surface of the tube.
The result is
\begin{equation}
\int \: ( \: \frac{1}{2} \: q^2 \: \varepsilon^{-1} \: \dot{v_{\mu}} \: - \: q \: v^{\nu} \: f_{\mu \nu} \: ) \: ds
\label{flow}
\end{equation}
where the integration is over the length of the tube.

As mentioned earlier, this flow of energy-momentum depends only on the conditions at the two ends of the tube.
That means that the integrand must be a perfect differential, namely
\begin{equation}
\frac{1}{2} \: q^2 \: \varepsilon^{-1} \: \dot{v_{\mu}} \: - \: q \: v^{\nu} \: f_{\mu \nu} = \dot{B_{\mu}}
\label{B}
\end{equation}
for some $B_{\mu}$.
Equation (\ref{v2}) and the fact that $f^{\mu \nu}$ is antisymmetric in its indices put a restriction on $B_{\mu}$, specifically
\begin{equation}
v_{\mu} \dot{B^{\mu}} = \frac{1}{2} \: q^2 \: \varepsilon^{-1} \: v_{\mu} \: \dot{v^{\mu}} \: - \: q \: v_{\mu} \: v_{\nu} \: f^{\mu \nu} = 0.
\label{vmBm}
\end{equation}
Thus, the simplest acceptable expression for $B_{\mu}$ is
\begin{equation}
B_{\mu} = k v_{\mu}
\end{equation}
for some $k$ independent of $s$.
From Equation (\ref{B}) one obtains
\begin{equation}
k = \frac{1}{2} \: q^2 \: \varepsilon^{-1} - m
\end{equation}
where $m$ must be a constant independent of $\varepsilon$ in order for Equations (\ref{B}) and (\ref{vmBm}) to have a well-defined behavior in the limit $\varepsilon \to 0$.
Finally, the equations of motion for the electron are
\begin{equation}
m \: \dot{v_{\mu}} = q \: v_{\nu} \: f_{\mu}^{\; \nu}
\label{motion}
\end{equation}
and $m$ plays the part of the rest mass of the electron.
This result is used in Section (\ref{RadReacDirac}) to derive an expression for the radiation damping effects on the moving electron.

\section{Radiation Reaction}
\label{RadReacDirac}

The field $f^{\mu \nu}$ defined in Section (\ref{DefFields}) allows the equations of motion (\ref{motion}) to be expressed in a simple form. 
However, in practical applications one would prefer to have the incident radiation field $F_{\tx{in}}^{\mu \nu}$ in the equations of motion, as $F_{\tx{in}}^{\mu \nu}$ is usually given. 
By substituting Equations (\ref{Fact1}) and (\ref{Frad2}) into Equation (\ref{f1}), an expression for $f^{\mu \nu}$ is obtained, that involves the incident field and the radiation field:
\begin{equation}
f^{\mu \nu} = F_{\tx{in}}^{\mu \nu} + \frac{1}{2} \: F_{\tx{rad}}^{\mu \nu}.
\end{equation}
Then, the equations of motion become
\begin{equation}
m \: \dot{v_{\mu}} = q \: v_{\nu} \: F_{\tx{in} \: \mu}^{\;\;\;\;\;\; \nu} + \frac{1}{2} \: q \: v_{\nu} \: F_{\tx{rad} \: \mu}^{\;\;\;\;\;\;\;\; \nu}.
\label{resultD1}
\end{equation}

The first term of the right-hand side of Equation (\ref{resultD1}) involves the incident radiation field and gives the work done by that field on the electron.
The second term of the right-hand side involves the radiation field emitted by the electron and is the term of particular interest when one considers radiation reaction.
This term gives the effect of the electron radiation on itself and is present even if there is no external radiation field, namely even if $F_{\tx{in}}^{\mu \nu} = 0$.
The fact that the radiation reaction effects on the electron can be described by using the electromagnetic field $F_{\tx{rad}}^{\mu \nu}$ is one of the important results of Dirac's work and is the idea used when motion of particles in curved spacetime is considered, to calculate radiation reaction effects on those particles.

It is worth writing the equations of motion in terms of the characteristics of the worldline of the electron and the external radiation field, which can be achieved by using Equation (\ref{Fradv}) for the radiation field of the electron
\begin{equation}
m \: \dot{v_{\mu}} = q \: v_{\nu} \: F_{\tx{in} \: \mu}^{\;\;\;\;\;\; \nu} + \frac{2}{3} \: q^2 \: \ddot{v_{\mu}} + \frac{2}{3} \: q^2 \: \dot{v_{\nu}} \dot{v^{\nu}} \: v_{\mu}.
\label{resultD2}
\end{equation}
It is an important and a very interesting feature of Equation (\ref{resultD2}) that, in addition to the first derivative of $v_{\mu}$, its second derivative shows up as well.
A discussion of that fact and of some of its implications is presented in \cite{dirac}, but that is beyond the scope of this dissertation.

      \chapter{Radiation Reaction in Curved Spacetime}
\label{RRC}

Dirac's study of the radiation reaction effects on an electron gave a relatively simple result, because the electron was assumed to be moving in flat spacetime.
The analysis becomes significantly more complicated for particles moving in curved spacetime.
DeWitt and Brehme \cite{dewbre} were the first to study the motion of an electrically charged particle in a curved spacetime using Green's functions.
Later, Mino, Sasaki and Tanaka \cite{mst} generalized that analysis for the case of a particle creating a gravitational field while moving in curved spacetime.
Since Green's functions are used extensively in subsequent chapters to determine the self-force effects on different particles, it is appropriate to present the analyses of DeWitt and Brehme and of Mino, Sasaki and Tanaka before proceeding.

In this and all subsequent chapters, geometrized units are used, meaning that Newton's gravitational constant and the speed of light are both set equal to 1.
Note also that the signature of the metric is changed to $(- + + +)$ from now on, so minor changes in some of the equations previously mentioned should not be surprising.

\section{Nonlocal Quantities}
\label{NonlocQ}

One of the main characteristics of any Green's function is that it connects two points in spacetime. 
It does that by propagating the effect of the source, from the point where that source is located (source point) to the point where the field needs to be calculated (field point).
Since Green's functions are inherently nonlocal quantities, the discussion about them can be facilitated if some more elementary nonlocal quantities are introduced first.

The most general class of nonlocal quantities is the class of bitensors, which are simply tensors whose indices refer to two points in spacetime.
In the following, $x$ denotes the field point and $z$ denotes the source point.
In order for the indices for each point to be easily identifiable, all unprimed indices refer to the field point $x$ and all primed indices refer to the source point $z$.

A very important quantity for the study of nonlocal properties of spacetime is the biscalar of geodesic interval $s(x,z)$.
It is the magnitude of the invariant distance between the points $x$ and $z$ as measured along a geodesic that joins them and is a non-negative quantity.
It is defined by the equations
\begin{equation}
\nabla^{\alpha} s \: \nabla_{\alpha} s \: = \: \nabla^{\alpha'} s \: \nabla_{\alpha'} s = \pm 1,
\label{defs1}
\end{equation}
\begin{equation}
\lim_{x \to z} s = 0,
\end{equation} 
where $\nabla^{\alpha}$ denotes covariant differentiation with respect to the background metric $g^{\mu \nu}$.
It is obvious that $s$ must be symmetric under interchange of its two arguments, namely
\begin{equation}
s(x,z) = s(z,x).
\end{equation}
With the signature of the metric being $(- + + +)$, the interval $s$ is spacelike when the $+$ sign holds in Equation (\ref{defs1}) and timelike when the $-$ sign holds.
The points $x$ for which $s=0$ define the null cone of $z$.

It is more convenient to use a different quantity to measure the invariant distance between the source point and the field point, a quantity that is, however, related to $s(x,z)$.
That quantity is Synge's \cite{Synge} world function and is defined by
\begin{equation}
\sigma(x,z) \equiv \: \pm \: \frac{1}{2} \: s^2(x,z).
\end{equation}
By using the defining equations for $s$ one can deduce that $\sigma$ has the properties
\begin{equation}
\frac{1}{2} \: \nabla^{\alpha} \sigma \: \nabla_{\alpha} \sigma \: = \frac{1}{2} \: \nabla^{\alpha'} \sigma \: \nabla_{\alpha'} \sigma \: = \sigma,
\end{equation}
\begin{equation}
\lim_{x \to z} \sigma = 0.
\end{equation}
Also, $\sigma$ is positive for spacelike intervals and negative for timelike intervals.

Another very significant nonlocal quantity is the bivector of geodesic parallel displacement, denoted by $\bar{g}_{\alpha \alpha'}(x,z)$.
The defining equations for it are
\begin{equation}
\nabla^{\beta} \bar{g}_{\alpha \alpha'} \: \nabla_{\beta} \sigma = 0 , \: \: \nabla^{\beta'} \bar{g}_{\alpha \alpha'} \: \nabla_{\beta'} \sigma = 0 
\label{defgbar1}
\end{equation}
\begin{equation}
\lim_{x \to z} \bar{g}_{\alpha}^{\: \: \alpha'} = \delta_{\alpha}^{\: \: \alpha'}.
\label{defgbar2}
\end{equation}
Equation (\ref{defgbar1}) signifies that the covariant derivatives of $\bar{g}_{\alpha \alpha'}$ are equal to zero in the directions tangent to the geodesic joining $x$ and $z$.
Equation (\ref{defgbar2}) expresses the fact that $\bar{g}_{\alpha \alpha'}$ is equal to the Kronecker-$\delta$ when $x = z$. 
The bivector of geodesic parallel displacement also has the property that
\begin{equation}
\bar{g}_{\alpha \alpha'}(x,z) = \bar{g}_{\alpha' \alpha}(z,x).
\end{equation}
In the following, the determinant of $\bar{g}_{\alpha \alpha'}$ is denoted by
\begin{equation}
\bar{g} = - \mid \bar{g}_{\alpha \alpha'} \mid,
\end{equation}
and
\begin{equation}
\bar{\delta} = \mid \bar{g}_{\: \: \alpha'}^{\alpha} \mid.
\end{equation}
The effect of applying the bivector of geodesic parallel displacement to a local vector $A^{\alpha'}$ at $z$ is a parallel transport of that vector from point $z$ to point $x$, along the geodesic that connects the two points.
The result is a local vector $A^{\alpha}$ at point $x$.
In general, $\bar{g}_{\alpha \alpha'}$ can be used to transform any bitensor that has indices refering to the two points $x$ and $z$ to a tensor whose indices refer to one point only, for example
\begin{equation}
\bar{g}^{\alpha}_{\: \: \alpha'} \: \bar{g}^{\beta}_{\: \: \beta'} \: T^{\alpha' \: \: \beta'}_{\: \: \: \: \gamma \: \: \: \delta \epsilon} = T^{\alpha \: \: \beta}_{\: \: \: \: \gamma \: \: \: \delta \epsilon}.
\end{equation} 
How useful the bivector of geodesic parallel displacement is in studying nonlocal prorerties of spacetime can be understood if expansions of a bitensor about one point are considered.
In order for a bitensor to be expanded about a certain point in spacetime, all its indices must refer to that specific point.
For bitensors for which that is not the case, their indices must first be homogenized by applying $\bar{g}_{\alpha \alpha'}$ and then the expansion about the specific point can be taken.

A bivector that is very useful for the Hadamard \cite{hadamard} expansion of the Green's functions is defined by
\begin{equation}
D_{\alpha \alpha'}(x,z) = - \nabla_{\alpha} \nabla_{\alpha'} \sigma(x,z)
\end{equation}
and a biscalar relating to it is its determinant
\begin{equation}
D = - \mid D_{\alpha \alpha'} \mid.
\end{equation}
DeWitt and Brehme proved that
\begin{equation}
\lim_{x \to z} D_{\alpha \alpha'}(x,z) = g_{\alpha \alpha'}(z)
\end{equation}
which shows that the biscalar $D$ is nonvanishing, at least when $x$ and $z$ are close to each other. 
In fact, $D$ is the Jacobian of the transformation from the set of variables $\{ z^{\alpha'},x^{\alpha} \}$, which specify the geodesic between $x$ and $z$ in terms of its two end points, to the set of variables $\{ z^{\alpha'},\nabla^{\alpha'} \sigma \}$, which specify the geodesic in terms of one of its end points and the tangent to the geodesic at that end point.
DeWitt and Brehme also showed that $D$ obeys the differential equation
\begin{equation}
D^{-1} \: \nabla_{\alpha} (D \: \nabla^{\alpha} \sigma) = 4.
\end{equation}

Instead of the biscalar $D$, a different biscalar is used in the Hadamard expansions of the Green's functions.
That biscalar is denoted by $\Delta$ and is defined as
\begin{equation}
\Delta(x,z) = [\bar{g}(x,z)]^{-1} D(x,z).
\end{equation}
DeWitt and Brehme proved that $\Delta$ can be expanded in terms of derivatives of $\sigma$ and that the expansion is
\begin{equation}
\Delta = 1 + \frac{1}{6} \: R^{\alpha \beta} \: \nabla_{\alpha} \sigma \: \nabla_{\beta} \sigma + O(s^{3}), \:  \tx{for} \: x \to z
\label{Delta}
\end{equation}
where $R^{\alpha \beta}$ is the background Ricci tensor.

\section{Green's Functions}
\label{GreensFuncs}

The goal of the work of DeWitt and Brehme was to study the radiation damping effects on a particle of a given electric charge that moves in curved spacetime. 
To do that it is necessary to study the vector field that represents the electromagnetic potential created by the particle.
However, the equations for the scalar field are less complicated and thus easier to deal with. 
Also, the lack of indices makes the results more transparent.
For those reasons the scalar Green's functions and their corresponding fields are discussed first.

\subsection{Green's Functions for a Scalar Field}
\label{GFS}

A point particle of scalar charge $q$ which is moving on a worldline $\Gamma: z^{\alpha'}(\tau)$, where $\tau$ is the proper time along the geodesic, creates a scalar field $\psi$ which obeys Poisson's equation:
\begin{equation}
\nabla^{2} \: \psi = - 4 \: \pi \: \varrho. 
\label{scinh}
\end{equation}
In Equation (\ref{scinh}), $\varrho$ is the source function for the point particle.
Specifically
\begin{equation}
\varrho(y) = q \: \int \: (-g)^{-\frac{1}{2}} \: \delta^{4}(y - z(\tau)) \: d\tau
\label{sourcesc}
\end{equation}
where $y$ is some point in spacetime.
It is desired to express the different solutions of Equation (\ref{scinh}) as integrals containing Green's functions, namely
\begin{equation}
\begin{split}
\psi(x) & = 4 \: \pi \: \int_{spacetime} (-g)^{\frac{1}{2}} \: G(x,y) \: \varrho(y) \: dy^{4} \\
        & = 4 \: \pi \: q \: \int_{\Gamma} G[x,z(\tau)] \: d\tau, \\
\end{split}
\end{equation}
so the properties of the various Green's functions are emphasized in this section.

One function of importance is the symmetric Green's function, $G^{\tx{sym}} (x,z)$, which satisfies the inhomogeneous differential equation
\begin{equation}
\nabla^{2} G^{\tx{sym}} (x,z) = -(-g)^{-\frac{1}{2}} \: \delta^{4}(x-z).
\label{dGsym}
\end{equation}
The Hadamard form \cite{hadamard} of this function is
\begin{equation}
G^{\tx{sym}} (x,z) = \frac{1}{8 \pi} \: [U(x,z) \: \delta(\sigma) - V(x,z) \: \Theta(-\sigma)]
\label{Gsym}
\end{equation}
where $\Theta$ is the step function that equals 1 if the argument is greater than zero and equals 0 otherwise. 
$U(x,z)$ and $V(x,z)$ are biscalars that are free of singularities and symmetric under the interchange of $x$ and $z$, namely
\begin{equation}
U(x,z) = U(z,x)
\end{equation}
and
\begin{equation}
V(x,z) = V(z,x).
\end{equation}
They can be determined by expanding the solution of Equation (\ref{dGsym}) in powers of $\sigma$, in the vicinity of the geodesic $\Gamma$.
The biscalar $U(x,z)$ satisfies the differential equation
\begin{equation}
U^{-1} (x,z) \: \nabla^{\alpha}U(x,z) = \frac{1}{2} \Delta^{-1}(x,z) \: \nabla^{\alpha} \Delta(x,z)
\end{equation}
and the boundary condition
\begin{equation}
\lim_{x \to z} U(x,z) = 1
\end{equation}
and is given by
\begin{equation}
\begin{split}
U(x,z) &= [\Delta(x,z)]^{\frac{1}{2}} \\ 
       &= 1 \: + \: \frac{1}{12} \: R_{\alpha \beta} \: \nabla^{\alpha} \sigma \: \nabla^{\beta} \sigma \: + \: O(s^{3}), \: x \rightarrow \Gamma\\
\end{split}
\end{equation}
where $s$ is the proper distance from the point $x$ to $\Gamma$ measured along the spatial geodesic which is orthogonal to $\Gamma$.
The biscalar $V(x,z)$ satisfies the homogeneous differential equation
\begin{equation}
\nabla^{2} V(x,z) = 0
\label{DelV}
\end{equation}
and is given by
\begin{equation}
V(x,z) = - \frac{1}{12} \: R(z) \: + \: O(s), \: x \rightarrow \Gamma.
\label{ExpV}
\end{equation}
It is noteworthy that the symmetric Green's function vanishes for $\sigma > 0$, that is for spacelike separation of the points $x$ and $z$. 
Also, because both $U(x,z)$ and $V(x,z)$ are symmetric under the interchange of $x$ and $z$, so is $G^{\tx{sym}}$:
\begin{equation}
G^{\tx{sym}}(x,z) = G^{\tx{sym}}(z,x).
\end{equation}

If $\dot{\sigma} = d\sigma(x,z(\tau))/d\tau$, the scalar field $\psi^{\tx{sym}}(x)$ associated with the symmetric Green's function is
\begin{equation}
\begin{split}
\psi^{\tx{sym}}(x) &= \frac{q}{2} \: \int \: [U(x,z) \: \delta(\sigma) - V(x,z) \: \Theta(-\sigma)] \: d\tau \\
                   &= \Big [ \frac{q \: U(x,z)}{2 \: \dot{\sigma}} \Big ]_{\tau_{\tx{ret}}} + \Big [ \frac{q \: U(x,z)}{2 \: \dot{\sigma}} \Big ]_{\tau_{\tx{adv}}} - \frac{q}{2} \Big ( \int_{-\infty}^{\tau_{\tx{ret}}} + \int_{\tau_{\tx{adv}}}^{+\infty} \Big ) V(x,z) \: d\tau \\
\end{split}
\label{PsiSym} 
\end{equation}
and consists of two parts. 
The first part is the one that contains the biscalar $U(x,z)$ and the $\delta$-function $\delta(\sigma)$ and is referred to as the \textit{direct} part. 
This is the term that corresponds to $\sigma = 0$, which is the part of the field that comes from the retarded and advanced proper times ($\tau_{\tx{ret}}$ and $\tau_{\tx{adv}}$ respectively), namely the proper times that correspond to the intersection of the geodesic $\Gamma$ with the past and future null cone of the point $x$. 
In other words, the direct part has the same singularity on the null cone that the symmetric scalar field has in flat spacetime.
The second part is the one that contains the biscalar $V(x,z)$ and the step-function $\Theta(-\sigma)$ and is referred to as the \textit{tail} part.
This term gives the part of the field that corresponds to $\sigma < 0$, which is the part that comes from the interior of the past and future null cone of $x$.
It is the part of the field that is due to the curvature of spacetime and vanishes in flat spacetime.

The symmetric Green's function can be separated into the retarded and advanced parts, which constitute two very important Green's functions themselves.
Specifically it can be written as
\begin{equation}
G^{\tx{sym}}(x,z) = \frac{1}{2} \: [ G^{\tx{ret}}(x,z) + G^{\tx{adv}}(x,z) ]
\end{equation}
where the retarded and advanced Green's functions are given by
\begin{equation}
G^{\tx{ret}} (x,z) = 2 \: \Theta[\Sigma(x),z] \: G^{\tx{sym}}(x,z),
\label{GSigma1}
\end{equation}
\begin{equation}
G^{\tx{adv}} (x,z) = 2 \: \Theta[z,\Sigma(x)] \: G^{\tx{sym}}(x,z).
\label{GSigma2}
\end{equation}
In Equations (\ref{GSigma1}) and (\ref{GSigma2}), $\Sigma(x)$ is any spacelike hypersurface that contains $x$. 
The step-function $\Theta[\Sigma(x),z] = 1 - \Theta[z,\Sigma(x)]$ is equal to 1 when $z$ lies to the past of $\Sigma(x)$ and vanishes when $z$ lies to the future of $\Sigma(x)$.
Both the retarded and the advanced Green's functions satisfy the inhomogeneous differential equation
\begin{equation}
\nabla^{2} G^{\tx{ret}}(x,z) \: = \: \nabla^{2} G^{\tx{adv}}(x,z) \: = \: - (-g)^{- \frac{1}{2}} \: \delta^{4}(x,z).
\end{equation}
They also have the property that
\begin{equation}
\label{symretadv}
G^{\tx{ret}}(x,z) = G^{\tx{adv}}(z,x).
\end{equation}

The retarded and advanced scalar fields, both solutions of the inhomogeneous differential Equation (\ref{scinh}), can also be expressed as integrals of the respective Green's functions
\begin{equation}
\psi^{\tx{ret}}(x) \: = \: 4 \: \pi \: q \: \int_{\Gamma} G^{\tx{ret}}[x,z(\tau)] \: d\tau \: = \: 2 \Big [ \frac{q \: U(x,z)}{2 \: \dot{\sigma}} \Big ]_{\tau_{\tx{ret}}} - q \int_{-\infty}^{\tau_{\tx{ret}}} V(x,z) \: d\tau,
\label{PsiRet}
\end{equation}
\begin{equation}
\psi^{\tx{adv}}(x) \: = \: 4 \: \pi \: q \: \int_{\Gamma} G^{\tx{adv}}[x,z(\tau)] \: d\tau \: = \: 2 \Big [ \frac{q \: U(x,z)}{2 \: \dot{\sigma}} \Big ]_{\tau_{\tx{adv}}} - q \int_{\tau_{\tx{adv}}}^{+\infty} V(x,z) \: d\tau.
\label{PsiAdv} 
\end{equation}
The retarded field $\psi^{\tx{ret}}(x)$ is the actual scalar field that results from the scalar particle and is singular at the location of the particle.

The last scalar Green's function that is significant for the analysis that follows is the radiative Green's function $G^{\tx{rad}}$, which is defined in terms of the retarded and advanced Green's functions in a manner analogous to the radiation field defined by Dirac in flat spacetime, namely
\begin{equation}
G^{\tx{rad}}(x,z) = G^{\tx{ret}}(x,z) - G^{\tx{adv}}(x,z).
\end{equation}
It is obvious by this definition that $G^{\tx{rad}}(x,z)$ satisfies the homogeneous differential equation
\begin{equation}
\nabla^{2} G^{\tx{rad}}(x,z) = 0.
\end{equation}
By Equation (\ref{symretadv}) one can infer that $G^{\tx{rad}}(x,z)$ is antisymmetric under the interchange of its arguments
\begin{equation}
G^{\tx{rad}}(x,z) = - G^{\tx{rad}}(z,x).
\end{equation}

The corresponding radiation field equals the difference between the retarded and advanced fields
\begin{equation}
\psi^{\tx{rad}}(x) = \psi^{\tx{ret}}(x) \: - \: \psi^{\tx{adv}}(x) = 4 \: \pi \: q \: \int_{\Gamma} G^{\tx{rad}}[x,z(\tau)] \: d\tau
\end{equation}
and is a sourceless field because it satisfies the homogeneous differential equation
\begin{equation}
\nabla^{2} \psi^{\tx{rad}}(x) = 0.
\end{equation}
Consequently, $\psi^{\tx{rad}}$ is smooth and differentiable everywhere in space.

The Green's functions and the corresponding scalar fields that were mentioned in this section are used extensively in Chapter \ref{UFM}, where the self-force on scalar particles is discussed.

\subsection{Green's Functions for a Vector Field}
\label{GFV}

Now that the scalar Green's functions and their properties have been discussed, it is straightforward to define analogous Green's functions for the electromagnetic vector potential $A^{\alpha}$ generated by an electrically charged particle moving on a geodesic $\Gamma: z^{\alpha'}(\tau)$.
Assuming that $A^{\alpha}$ satisfies the Lorentz gauge condition
\begin{equation}
\nabla_{\alpha} A^{\alpha} = 0,
\end{equation}
Maxwell's equations for it become
\begin{equation}
\nabla^{2} A^{\alpha} - R^{\alpha}_{\: \beta} A^{\beta} = - 4 \: \pi \: J^{\alpha},
\label{maxcurved}
\end{equation}
and, again, the goal is to express the vector potential as an integral that contains a Green's function and the source $J^{\alpha}$, specifically
\begin{equation}
A_{\alpha}(x) = 4 \: \pi \: \int_{spacetime} (-g)^{\frac{1}{2}} \: G_{\alpha \beta}(x,y) \: J^{\beta}(y) \: d^{4}y.
\end{equation}

The symmetric Green's function for a vector field obeys the equation
\begin{equation}
\nabla^{2} G^{\tx{sym}}_{\alpha \alpha'}(x,z) \: - \: R_{\alpha}^{\: \: \beta} G^{\tx{sym}}_{\beta \alpha'}(x,z) =  - \: \bar{g}^{\; -\frac{1}{2}} \: \bar{g}_{\alpha \alpha'} \: \delta^{4}(x,z)
\end{equation}
and the Hadamard form of it is
\begin{equation}
G^{\tx{sym}}_{\alpha \alpha'}(x,z) = \frac{1}{8 \pi} [U_{\alpha \alpha'}(x,z) \: \delta(\sigma) - V_{\alpha \alpha'}(x,z) \: \Theta(-\sigma)].
\end{equation}
The biscalar $U_{\alpha \alpha'}$ is given by the differential equation
\begin{equation}
(2 \: \nabla^{\beta} U_{\alpha \alpha'} + U_{\alpha \alpha'} \: \Delta^{-1} \: \nabla^{\beta} \Delta) \: \nabla_{\beta} \sigma = 0
\end{equation}
with the boundary condition
\begin{equation}
\lim_{x \to z} U_{\alpha \alpha'}(x,z) = g_{\alpha \alpha'}(z).
\end{equation}
As shown in \cite{dewbre}, the solution is unique
\begin{equation}
\begin{split}
U_{\alpha \alpha'}(x,z) &= \sqrt{\Delta(x,z)} \: \bar{g}_{\alpha \alpha'}(x,z) \\
                   &= [1 + \frac{1}{12} \: R^{\beta' \gamma'} \: \nabla_{\beta'} \sigma \: \nabla_{\gamma'} \sigma + O(s^{3})] \: \bar{g}_{\alpha \alpha'}, \: \tx{for} \: x \to z \\
\end{split}
\end{equation}
where Equation (\ref{Delta}) was used to derive the final expression.
For the bivector $V_{\alpha \alpha'}$ DeWitt and Brehme prove that
\begin{equation}
\lim_{x \to z} V_{\alpha \alpha'}(x,z) = \frac{1}{2} \: \bar{g}_{\alpha}^{\: \: \beta'} \: (R_{\alpha' \beta'} - \frac{1}{6} \: g_{\alpha' \beta'} \: R) + O(s), \: \tx{for} \: x \to \Gamma.
\label{VR}
\end{equation}

The symmetric electromagnetic vector potential is calculated by
\begin{equation}
\begin{split}
A^{\tx{sym}}_{\alpha}(x) &= 4 \pi \int (-g)^{\frac{1}{2}} \: G^{\tx{sym}}_{\alpha \beta}(x,y) \: J^{\beta}(y) \: d^{4}y \\
                      &= \frac{1}{2} \int (-g)^{\frac{1}{2}} \Big  \{[\Delta(x,y)]^{\frac{1}{2}} \: \bar{g}_{\alpha \beta}(x,y) \: \delta(\sigma) - V_{\alpha \beta}(x,y) \: \Theta(-\sigma) \Big \} \: J^{\beta}(y) \: d^{4}y.\\
\end{split}
\end{equation}
Just as in the case of the symmetric scalar field, the symmetric electromagnetic potential also consists of the direct part, that contains the $\delta$-function and comes from the retarded and advanced proper times, and the tail part, that contains the $\Theta$-function and is the contribution from within the past and future null cone.
An interesting feature is the appearance of the bivector of geodesic parallel displacement in the direct part.
It signifies that the electromagnetic radiation is parallel propagated along the null geodesic that connects the points $x$ and $z$.

The symmetric Green's function can again be separated into the retarded and advanced parts
\begin{equation}
G^{\tx{sym}}_{\alpha \alpha'}(x,z) = \frac{1}{2} \: [G^{\tx{ret}}_{\alpha \alpha'}(x,z) + G^{\tx{adv}}_{\alpha \alpha'}(x,z)]
\end{equation}
which are given by the equations
\begin{equation}
G^{\tx{ret}}_{\alpha \alpha'}(x,z) = 2 \: \Theta[\Sigma(x),z] \: G^{\tx{sym}}_{\alpha \alpha'}(x,z)
\end{equation}
\begin{equation}
G^{\tx{adv}}_{\alpha \alpha'}(x,z) = 2 \: \Theta[z,\Sigma(x)] \: G^{\tx{sym}}_{\alpha \alpha'}(x,z).
\end{equation}
They are solutions of the inhomogeneous differential equation
\begin{equation}
\nabla^{2} G^{\tx{ret}}_{\alpha \alpha'}(x,z) \: - \: R_{\alpha}^{\: \: \beta} G^{\tx{ret}}_{\beta \alpha'}(x,z) \: = \: \nabla^{2} G^{\tx{adv}}_{\alpha \alpha'}(x,z) \: - \: R_{\alpha}^{\: \: \beta} G^{\tx{adv}}_{\beta \alpha'}(x,z) \:  = \: - \: \bar{g}^{\; -\frac{1}{2}} \: \bar{g}_{\alpha \alpha'}(x,z) \: \delta^{4}(x,z)
\end{equation}
and have the property that
\begin{equation}
G^{\tx{ret}}_{\alpha \alpha'}(x,z) = G^{\tx{adv}}_{\alpha' \alpha}(z,x).
\label{symretadvvector}
\end{equation}

The retarded and advanced electromagnetic potentials are then given by
\begin{equation}
A^{\tx{ret}}_{\alpha}(x) = 4 \: \pi \: \int_{spacetime} (-g)^{\frac{1}{2}} \: G^{\tx{ret}}_{\alpha \beta}(x,y) \: J^{\beta}(y) \: d^{4}y
\label{Aret}
\end{equation}
\begin{equation}
A^{\tx{adv}}_{\alpha}(x) = 4 \: \pi \: \int_{spacetime} (-g)^{\frac{1}{2}} \: G^{\tx{adv}}_{\alpha \beta}(x,y) \: J^{\beta}(y) \: d^{4}y
\label{Aadv}
\end{equation}
and $A^{\tx{ret}}_{\alpha}(x)$ is the actual potential generated by the moving charged particle.

The radiative Green's function is defined in terms of the retarded and advanced Green's functions by the equation
\begin{equation}
G^{\tx{rad}}_{\alpha \alpha'}(x,z) = G^{\tx{ret}}_{\alpha \alpha'}(x,z) - G^{\tx{adv}}_{\alpha \alpha'}(x,z)
\end{equation}
and satisfies the homogeneous differential equation
\begin{equation}
\nabla^{2} G^{\tx{rad}}_{\alpha \alpha'}(x,z) \: - \: R_{\alpha}^{\: \: \beta} G^{\tx{rad}}_{\beta \alpha'}(x,z) \: = 0.
\end{equation}
Using the symmetry expressed in Equation (\ref{symretadvvector}) for the retarded and advanced Green's functions it can be inferred that
\begin{equation}
G^{\tx{rad}}_{\alpha \alpha'}(x,z) = - G^{\tx{rad}}_{\alpha' \alpha}(z,x).
\end{equation}

The corresponding radiation vector potential is equal to the difference between the retarded and advanced vector potentials
\begin{equation}
\begin{split}
A^{\tx{rad}}_{\alpha}(x) &= A^{\tx{ret}}_{\alpha}(x) - A^{\tx{adv}}_{\alpha}(x) \\ 
&= 4 \: \pi \: \int_{spacetime} (-g)^{\frac{1}{2}} \: G^{\tx{rad}}_{\alpha \beta}(x,y) \: J^{\beta}(y) \: d^{4}y \\
\end{split}
\end{equation}
and is a source-free field, because it satisfies the homogeneous differential equation
\begin{equation}
\nabla^{2} A^{\tx{rad}}_{\alpha}(x) \: - \: R_{\alpha}^{\: \: \beta} A^{\tx{rad}}_{\beta}(x) \: = \: 0.
\end{equation}

The important quantities for the motion of a charged particle are not the vector potentials but rather the electromagnetic fields.
The electromagnetic field that corresponds to a vector potential $A_{\alpha}$ is calculated by:
\begin{equation}
F_{\alpha \beta}(x) = \nabla_{\beta} A_{\alpha}(x) - \nabla_{\alpha} A_{\beta}(x).
\label{EMfield}
\end{equation} 
For completeness, the definitions of the symmetric and radiation fields are given, in terms of the retarded and advanced fields:
\begin{equation}
F^{\tx{sym}}_{\alpha \beta} = \frac{1}{2} ( F^{\tx{ret}}_{\alpha \beta} + F^{\tx{adv}}_{\alpha \beta} )
\label{Fsym}
\end{equation}
\begin{equation}
F^{\tx{rad}}_{\alpha \beta} = F^{\tx{ret}}_{\alpha \beta} - F^{\tx{adv}}_{\alpha \beta}.
\label{Frad}
\end{equation}

\subsection{Green's Functions for a Gravitational Field}
\label{GFG}

Even though the work of DeWitt and Brehme did not cover the motion of a massive particle creating its own gravitational field, the Green's functions for a gravitational field are presented here, for completeness.
A more detailed description is given by Mino, Sasaki and Tanaka in \cite{mst} and it is that analysis that is followed here.

Assuming that a massive pointlike particle is moving on a geodesic $\Gamma: z^{\alpha'}(\tau)$ and is inducing a perturbation $h_{\mu \nu}(x)$ on the background, the trace-reversed metric perturbation is
\begin{equation}
\bar{h}_{\mu \nu}(x) = h_{\mu \nu}(x) - \frac{1}{2} g_{\mu \nu}(x) h(x)
\end{equation}
where $\bar{h}(x)$ and $h(x)$ are the traces of $\bar{h}^{\mu \nu}(x)$ and of $h^{\mu \nu}(x)$ respectively.
It is assumed that the trace-reversed metric perturbation obeys the harmonic gauge condition
\begin{equation}
\nabla_{\nu} \bar{h}^{\mu \nu}(x) = 0.
\end{equation}
In this gauge, the linearized Einstein equations \cite{mtw} become:
\begin{equation}
\nabla^{2} \bar{h}_{\alpha \beta} + 2 R_{\alpha \: \; \beta}^{\: \: \gamma \: \: \; \delta} \: \bar{h}_{\gamma \delta} = - 16 \pi T_{\alpha \beta}
\end{equation}
to first order in the metric perturbation.
The Green's functions of interest in this case are bitensors and are used to express the trace-reversed metric perturbation in integral form
\begin{equation}
\bar{h}^{\alpha \beta}(x) = 16 \pi \int_{spacetime} (-g)^{\frac{1}{2}} G^{\alpha \beta \gamma \delta}(x,y) \: T_{\gamma \delta}(y) \: d^{4}y.
\label{GravPert}
\end{equation}

The symmetric Green's function $G^{\alpha \beta \gamma' \delta'}(x,z)$ satisfies the differential equation
\begin{equation}
\nabla^{2} G_{\tx{sym}}^{\alpha \beta \gamma' \delta'}(x,z) + 2 \: R^{\alpha \: \: \beta}_{\: \: \mu \: \: \nu}(x) \: G_{\tx{sym}}^{\mu \nu \gamma' \delta'}(x,z) = -2 \: \bar{g}^{\gamma' ( \alpha}(x,z) \: \bar{g}^{\beta ) \delta'}(x,z) \: \frac{\delta^{4}(x-z)}{(-g)^\frac{1}{2}}
\end{equation}
and the Hadamard form for it is
\begin{equation}
G_{\tx{sym}}^{\alpha \beta \gamma' \delta'}(x,z) = \frac{1}{8 \pi} [ U^{\alpha \beta \gamma' \delta'}(x,z) \delta(\sigma) - V^{\alpha \beta \gamma' \delta'}(x,z) \Theta(- \sigma) ].
\end{equation}
The bitensor $U^{\alpha \beta \gamma' \delta'}(x,z)$ is the solution of the homogeneous differential equation
\begin{equation}
[ 2 \: \nabla^{\xi} U^{\alpha \beta \gamma' \delta'}(x,z) + \frac{\nabla^{\xi} \Delta}{\Delta} \: U^{\alpha \beta \gamma' \delta'}(x,z)] \: \nabla_{\xi}\sigma = 0
\end{equation}
with the boundary condition
\begin{equation}
\lim_{x \to z} U^{\alpha \beta \gamma' \delta'}(x,z) = \lim_{x \to z} 2 \bar{g}^{\gamma' ( \alpha}(x,z) \: \bar{g}^{\beta ) \delta'}(x,z).
\end{equation}
Mino, Sasaki and Tanaka \cite{mst} prove that the solution is
\begin{equation}
U^{\alpha \beta \gamma' \delta'}(x,z) = 2 \: \bar{g}^{\gamma' ( \alpha}(x,z) \: \bar{g}^{\beta ) \delta'}(x,z) \: \sqrt{\Delta(x,z)}.
\end{equation}
The bitensor $V^{\alpha \beta \gamma' \delta'}(x,z)$ is divergence-free
\begin{equation}
\nabla_{\beta} V^{\alpha \beta \gamma' \delta'}(x,z) = 0
\end{equation}
and satisfies the homogeneous differential equation
\begin{equation}
\nabla^{2} V^{\alpha \beta \gamma' \delta'}(x,z) + 2 \: R^{\alpha \: \: \beta}_{\: \mu \: \: \nu} \: V^{\mu \nu \gamma' \delta'}(x,z) = 0
\end{equation}
with the boundary condition
\begin{equation}
\lim_{\sigma \to 0} V^{\alpha \beta \gamma' \delta'}(x,z) = 0.
\end{equation}
It is proven in \cite{mst} that the solution is
\begin{equation}
V_{\alpha \beta \gamma' \delta'}(x,z) = - \bar{g}_{\alpha}^{\: \alpha'} \: \bar{g}_{\beta}^{\beta'} \: R_{\gamma' \alpha' \delta' \beta'}(z) + O(r), \: \tx{for} \: x \to \Gamma.
\label{VgR}
\end{equation}

The symmetric gravitational field is obtained by
\begin{equation}
\begin{split}
\bar{h}_{\tx{sym}}^{\alpha \beta}(x) &= 16 \pi \int_{spacetime} (-g)^{\frac{1}{2}} \: G_{\tx{sym}}^{\alpha \beta \gamma \delta}(x,y) \: T_{\gamma \delta}(y) \: d^{4}y \\
				     &= 2 \int_{spacetime} (-g)^{\frac{1}{2}} \: [ U^{\alpha \beta \gamma \delta}(x,y) \delta(\sigma) - V^{\alpha \beta \gamma \delta}(x,y) \Theta(-\sigma) ] \: T_{\gamma \delta}(y) \: d^{4}y.\\
\end{split}
\end{equation}
Its direct part is the part that contains the $\delta$-function and gives the contribution from the null cone of point $x$.
Its tail part is the part that contains the $\Theta$-function and gives the contribution that comes from within the null cone of point $x$.

The symmetric Green's function can be separated to the retarded and advanced Green's functions
\begin{equation}
G_{\tx{sym}}^{\alpha \beta \gamma' \delta'}(x,z) = \frac{1}{2} \: [G_{\tx{ret}}^{\alpha \beta \gamma' \delta'}(x,z) + G_{\tx{adv}}^{\alpha \beta \gamma' \delta'}(x,z)]
\end{equation}
which are given by the equations
\begin{equation}
G_{\tx{ret}}^{\alpha \beta \gamma' \delta'}(x,z) = 2 \: \Theta[\Sigma(x),z] \: G_{\tx{sym}}^{\alpha \beta \gamma' \delta'}(x,z)
\end{equation}
\begin{equation}
G_{\tx{adv}}^{\alpha \beta \gamma' \delta'}(x,z) = 2 \: \Theta[z,\Sigma(x)] \: G_{\tx{sym}}^{\alpha \beta \gamma' \delta'}(x,z).
\end{equation}
They are solutions of the inhomogeneous differential equation
\begin{equation}
\begin{split}
&\nabla^{2} G_{\tx{ret}}^{\alpha \beta \gamma' \delta'}(x,z) + 2 \: R^{\alpha \: \: \beta}_{\: \: \mu \: \: \nu}(x) \: G_{\tx{ret}}^{\mu \nu \gamma' \delta'}(x,z) = \\
&= \nabla^{2} G_{\tx{adv}}^{\alpha \beta \gamma' \delta'}(x,z) + 2 \: R^{\alpha \: \: \beta}_{\: \: \mu \: \: \nu}(x) \: G
_{\tx{adv}}^{\mu \nu \gamma' \delta'}(x,z) = 
-2 \: \bar{g}^{\gamma' ( \alpha}(x,z) \: \bar{g}^{\beta ) \delta'}(x,z) \: \frac{\delta^{4}(x-z)}{(-g)^\frac{1}{2}} \\
\end{split}
\end{equation}
and have the property that
\begin{equation}
G_{\tx{ret}}^{\alpha \beta \gamma' \delta'}(x,z) = G_{\tx{adv}}^{\gamma' \delta' \alpha \beta}(z,x).
\label{SymGrav}
\end{equation}

The retarded and advanced gravitational fields are given by
\begin{equation}
\bar{h}^{\alpha \beta}_{\tx{ret}} = 16 \pi \int_{spacetime} (-g)^{\frac{1}{2}} \: G^{\alpha \beta \gamma \delta}_{\tx{ret}}(x,y) \: T_{\gamma \delta}(y) \: d^{4}y
\end{equation}
\begin{equation}
\bar{h}^{\alpha \beta}_{\tx{adv}} = 16 \pi \int_{spacetime} (-g)^{\frac{1}{2}} \: G^{\alpha \beta \gamma \delta}_{\tx{adv}}(x,y) \: T_{\gamma \delta}(y) \: d^{4}y
\end{equation}
and $\bar{h}^{\alpha \beta}_{\tx{ret}}$ is the actual trace-reversed gravitational perturbation induced by the moving particle.

The radiative Green's function is defined in terms of the retarded and advanced Green's functions by the equation
\begin{equation}
G_{\tx{rad}}^{\alpha \beta \gamma' \delta'}(x,z) = G_{\tx{ret}}^{\alpha \beta \gamma' \delta'}(x,z) - G_{\tx{adv}}^{\alpha \beta \gamma' \delta'}(x,z)
\end{equation}
and satisfies the homogeneous differential equation
\begin{equation}
\nabla^{2} G_{\tx{rad}}^{\alpha \beta \gamma' \delta'}(x,z) + 2 \: R^{\alpha \: \: \beta}_{\: \: \mu \: \: \nu}(x) \: G_{\tx{rad}}^{\mu \nu \gamma' \delta'}(x,z) = 0.
\end{equation} 
Using the symmetry expressed in Equation (\ref{SymGrav}) for the retarded and advanced Green's functions, a symmetry property for the radiative Green's function can be derived, namely
\begin{equation}
G_{\tx{rad}}^{\alpha \beta \gamma' \delta'}(x,z) = - G_{\tx{rad}}^{\gamma' \delta' \alpha \beta}(z,x).
\end{equation}

Finally, the radiative gravitational perturbation is given by the difference between the retarded and advanced gravitational perturbations
\begin{equation}
\begin{split}
\bar{h}_{\tx{rad}}^{\alpha \beta}(x) &= \bar{h}^{\alpha \beta}_{\tx{ret}}(x) - \bar{h}^{\alpha \beta}_{\tx{adv}}(x) \\ 
&= 16 \pi \int_{spacetime} (-g)^{\frac{1}{2}} \: G_{\tx{rad}}^{\alpha \beta \gamma \delta}(x,y) \: T_{\gamma \delta}(y) \: d^{4}y \\
\end{split}
\end{equation}
and is a solution of the homogeneous differential equation
\begin{equation}
\nabla^{2} \bar{h}_{\tx{rad}}^{\alpha \beta}(x) + 2 \: R^{\alpha \: \: \beta}_{\: \: \mu \: \: \nu} \: \bar{h}_{\tx{rad}}^{\mu \nu}(x) = 0.
\end{equation}

\section{Motion of a Charged Particle in Curved Spacetime}
\label{MotionCPCS}

Following DeWitt and Brehme's analysis, I present the equations of motion and the radiation reaction effects for a particle carrying an electric charge $e$. 
Studying the vector case is useful in comparing the result derived in curved spacetime with Dirac's result for flat spacetime.
Similar analyses can be done for the scalar case and the gravitational case. 
Those analyses can be found in a recent review of the subject by Poisson \cite{PoissonsReview}.

\subsection{Equations of Motion}
\label{EqMotionCPCS}

Let's assume that a particle of bare mass $m_{0}$ and electric charge $e$ is moving in a curved spacetime of metric $g_{\mu \nu}$ on a worldline described by $\Gamma : z(\tau)$, where $\tau$ is the proper time.

In the following, the notation $u^{\alpha'} = (dz^{\alpha'}/d\tau) = \dot{z^{\alpha'}}$ is used.
Also, the symbol $(D/d\tau)$ is used to denote absolute covariant differentiation with respect to the proper time $\tau$ along the worldline.

The particle generates an electromagnetic vector potential $A^{\alpha}(x)$ and an electromagnetic field $F^{\alpha \beta}(x)$ given by Equation (\ref{EMfield}).
The action of the system contains three terms; the first term  comes from the particle, the second from the electromagnetic field and the third from the interaction between the two:
\begin{equation}
S = S_{particle} + S_{field} + S_{interaction},
\end{equation}
where
\begin{equation}
S_{particle} = - m_{0} \: \int_{\Gamma} \: [-g_{\alpha' \beta'}(z) \: u^{\alpha'} \: u^{\beta'} ]^{\frac{1}{2}} \: d\tau
\end{equation}
\begin{equation}
S_{field} = -\frac{1}{16 \pi} \: \int F_{\alpha \beta} \: F^{\alpha \beta} \: \sqrt{-g} \: d^{4}x
\end{equation} 
\begin{equation}
S_{interaction} = e \: \int A_{\alpha'}(z) \: u^{\alpha'} \: d\tau.
\end{equation}

In determining the equations of motion, the stationary action principle can be applied to $S$ for variations in the potential $A^{\alpha}$ and for variations in the worldline $z^{\alpha'}$ separately.
Varying $A^{\alpha}$ gives
\begin{equation}
\nabla_{\beta} F^{\alpha \beta} = 4 \pi J^{\alpha}
\label{F}
\end{equation}
where $J^{\alpha}$ is the current density defined as
\begin{equation}
J^{\alpha}(y) = e \int_{\Gamma} \bar{g}^{\alpha}_{\: \: \beta'}(y,z) \: u^{\beta'} \: \delta^{4}(y,z) \: d\tau.
\end{equation}
The electromagnetic field $F_{\alpha \beta}$ is invariant under a gauge transformation of the form $A_{\alpha} \rightarrow A_{\alpha} + \nabla_{\alpha} \Lambda(x)$, where $\Lambda(x)$ is any scalar function of $x$.
For convenience, $\Lambda(x)$ can be chosen in such a way that the electromagnetic potential satisfies the Lorentz gauge condition
\begin{equation}
\nabla_{\alpha} A^{\alpha} = 0
\end{equation}
in which case Equation (\ref{F}) for the electromagnetic field becomes
\begin{equation}
\nabla^{2} A^{\alpha}(x) - R^{\alpha}_{\: \beta} A^{\beta}(x) = - 4 \pi J^{\alpha}(x).
\label{Motion1}
\end{equation}
Equation (\ref{Motion1}) allows one to determine the electromagnetic field, once the worldline of the charged particle is known.

On the other hand, varying the worldline $z^{\alpha'}$ gives
\begin{equation}
m_{0} \frac{Du^{\alpha'}}{d\tau} = e F^{\alpha'}_{\: \: \: \beta'}(z) \: u^{\beta'}.
\label{Motion2}
\end{equation}
Equation (\ref{Motion2}) allows one to calculate the worldline of the particle, once the electromagnetic field is specified.

As is well-known, the electromagnetic field given by Equation (\ref{Motion1}) is divergent at the location of the particle and cannot be used in Equation (\ref{Motion2}).
An alternative method for studying the motion of the particle is to take advantage of the conservation of energy and momentum for the particle and the field, the same method that was used by Dirac in his analysis of the motion of an electron in flat spacetime.

\subsection{Calculation of the Fields}
\label{CalcOfFields}

It becomes obvious in Section (\ref{ConservEM}) that
in order for the conservation of energy and momentum to be implemented along the worldline of the charged particle, it is necessary for the retarded and advanced fields generated by the particle to be calculated.
A brief description of the calculation is given here; the details can be found in \cite{dewbre}.

\begin{figure}[ht]
\centering
\includegraphics[height=6cm,width=9cm,clip]{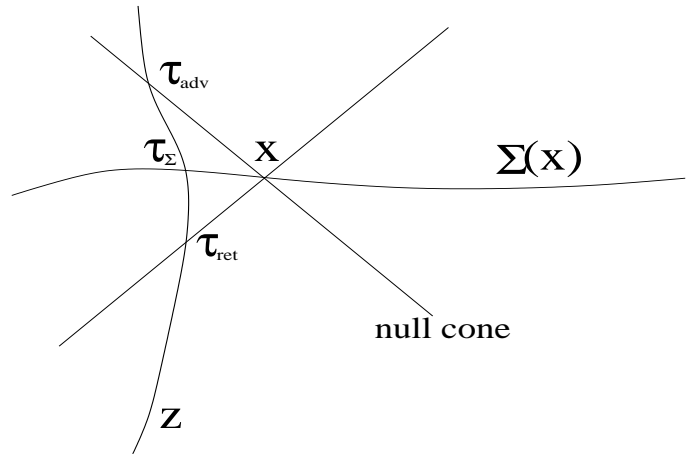}
\caption{Retarded and advanced proper times}
\label{nullcone}
\end{figure}

In the following, $\tau_{\tx{ret}}$ and $\tau_{\tx{adv}}$ are the proper times along the worldline of the particle, where the past and future null cones of $x$ respectively intersect that worldline.
They are referred to as retarded and advanced proper times.
Also, $\tau_{\Sigma}$ is the proper time at the point where the hypersurface $\Sigma(x)$ intersects the worldline of the particle.
All these points are shown in Figure (\ref{nullcone}).

Starting with Equation (\ref{Aret}) for the retarded potential, 
and using the Hadamard expansion for the Green's function, a simple expression for $A_{\alpha}^{\tx{ret}}$ can be obtained
\begin{equation}
\begin{split}
A^{\tx{ret}}_{\alpha}(x) &= 4 \pi \: \int d^{4}y \: (-g)^{\frac{1}{2}} \: G^{\tx{ret}}_{\alpha \beta}(x,y) \: J^{\beta}(y) \\
                         &= 4 \pi e \int d^{4}y(-g)^{\frac{1}{2}} \: 2 \: \Theta[\Sigma(x),y] \: \frac{[U_{\alpha \beta}(x,y) \: \delta(\sigma) - V_{\alpha \beta}(x,y) \: \Theta(-\sigma)]}{8 \pi} \times \\ 
                         & \qquad \qquad \qquad \qquad \qquad \qquad \qquad \qquad \qquad \qquad \: \times \int_{\Gamma} d\tau \: \bar{g}^{\beta}_{\: \: \alpha'}(y,z) \: u^{\alpha'} \: \delta^{4}(y,z) \\
                         &= e \int_{-\infty}^{\tau_{\Sigma}} U_{\alpha \alpha'}(x,z) \: \delta(\sigma) \: u^{\alpha'} d\tau - e \int_{-\infty}^{\tau_{\Sigma}} V_{\alpha \alpha'}(x,z) \: \Theta(-\sigma) \: u^{\alpha'} \: d\tau \\
                         &= e \: \Big [ \frac{U_{\alpha \alpha'} \: u^{\alpha'}}{\nabla_{\beta'}\sigma \: u^{\beta'}} \Big ]_{\tau_{\tx{ret}}} \: - \: e \int_{-\infty}^{\tau_{\tx{ret}}} V_{\alpha \alpha'}(x,z) \: u^{\alpha'} \: d\tau. \\
\end{split}
\label{SimplAret}
\end{equation}
By a similar calculation, starting with Equation (\ref{Aadv}), the advanced potential can be calculated
\begin{equation}
A^{\tx{adv}}_{\alpha}(x) = - e \: \Big [ \frac{U_{\alpha \alpha'} \: u^{\alpha'}}{\nabla_{\beta'}\sigma \: u^{\beta'}} \Big ]_{\tau_{\tx{adv}}} \: - \: e \int_{\tau_{\tx{adv}}}^{+\infty} V_{\alpha \alpha'}(x,z) \: u^{\alpha'} \: d\tau.
\label{SimplAadv}
\end{equation}
Equations (\ref{SimplAret}) and (\ref{SimplAadv}) give the covariant Lienard-Wiechert potentials.
 
The retarded and advanced electromagnetic fields can be calculated by Equation (\ref{EMfield}) if the respective potentials are differentiated. The result is 
\begin{equation}
\begin{split}
F^{\tx{ret/adv}}_{\mu \nu} = &\mp e \bigg \{ (U_{\nu \alpha'} \nabla_{\mu} \sigma - U_{\mu \alpha'} \nabla_{\nu} \sigma) \: u^{\alpha'} \: ( \nabla_{\beta' \gamma'} \sigma \: u^{\beta'} u^{\gamma'} + \nabla_{\beta'} \sigma \frac{Du^{\beta'}}{d\tau}) \: (\nabla_{\delta'} \sigma \: u^{\delta'})^{-3} \\
                       &- [\nabla_{\beta'} (U_{\nu \alpha'} \nabla_{\mu} \sigma - U_{\mu \alpha'} \nabla_{\nu} \sigma) \: u^{\alpha'} u^{\beta'} + (U_{\nu \alpha'} \nabla_{\mu} \sigma - U_{\mu \alpha'} \nabla_{\nu} \sigma) \: \frac{Du^{\alpha'}}{d\tau}] \times \\
                       & \qquad \qquad \qquad \qquad \qquad \qquad \qquad \qquad \qquad \qquad \qquad \qquad  \times (\nabla_{\gamma'} \sigma \: u^{\gamma'})^{-2} \\
                       &+ (\nabla_{\mu} U_{\nu \alpha'} - \nabla_{\nu} U_{\mu \alpha'} + V_{\nu \alpha'} \nabla_{\mu} \sigma - V_{\mu \alpha'} \nabla_{\nu} \sigma) \: u^{\alpha'} \: (\nabla_{\beta'} \sigma \: u^{\beta'})^{-1} \bigg \}_{\tau_{\tx{ret}}/ \tau_{\tx{adv}}} \\
                       &\mp e \int_{\tau_{\tx{ret}}/\tau_{\tx{adv}}}^{\pm \infty} (\nabla_{\mu} V_{\nu \alpha'} - \nabla_{\nu} V_{\mu \alpha'} ) \: u^{\alpha'} \: d\tau. \\
\end{split}
\label{Fretadv1}
\end{equation}
In Equation (\ref{Fretadv1}), the lower of the two signs corresponds to the retarded field and the upper sign corresponds to the advanced field.
This is an important expression because it gives the electromagnetic field at point $x$ in terms of the characteristics of the worldline of the charged particle. 
The significance of this becomes clear when the final equations of motion of the charged particle are written down.

As becomes obvious in Section (\ref{ConservEM}), when calculating the balance of energy and momentum between the particle and the field along the worldline, the values of the fields very close to the worldline need to be known.
Expanding the retarded and advanced fields in powers of $\sigma$ and its derivatives is a tedious but straightforward calculation, which yields a complicated expression for the two fields, Equation (5.12) of \cite{dewbre}.

Following Dirac's ideas, the total field in the vicinity of the charged particle can be written as the sum of the retarded field created by the particle plus the external incoming field
\begin{equation}
F_{\alpha \beta} = F_{\alpha \beta}^{\tx{ret}} + F_{\alpha \beta}^{\tx{in}}
\end{equation}
which, by using Equations (\ref{Fsym}) and (\ref{Frad}), can be written in terms of the radiation and symmetric fields
\begin{equation}
F_{\alpha \beta} = F^{\tx{in}}_{\alpha \beta} + \frac{1}{2} F^{\tx{rad}}_{\alpha \beta} + F^{\tx{sym}}_{\alpha \beta}.
\label{Ftotal}
\end{equation}
This breaking up of the total field is very important because, after implementing the conservation of energy and momentum along the worldline of the particle, only the radiation field $F^{\tx{rad}}_{\alpha \beta}$ and the external incoming field $F^{\tx{in}}_{\alpha \beta}$ show up in the equations of motion of the charged particle.

Since the radiation field is the difference between the retarded and advanced fields, only two terms of Equation (\ref{Fretadv1}) are of interest here.
Specifically, it is sufficient to write
\begin{equation}
\begin{split}
F^{\tx{ret/adv}}_{\mu \nu} = &\: e \: (\bar{g}_{\mu \alpha'} \bar{g}_{\nu \beta'} - \bar{g}_{\nu \alpha'} \bar{g}_{\mu \beta'}) \: \Bigg [ \pm  \frac{2}{3} \: \bigg ( \frac{D^{2}\sigma}{d\tau^{2}} \bigg )^{-2} \: \frac{Du^{\alpha'}}{d\tau} \: u^{\beta'} \\
                           & \qquad \qquad \qquad \qquad \qquad \, \, \pm \frac{1}{2} \int_{\tau}^{\pm\infty} f^{\alpha' \beta'}_{\: \: \: \: \: \: \: \gamma'}(z(\tau),z(\tau')) \: u^{\gamma'} \: d\tau' \Bigg ] \\ 
                         &+ F^{\tx{extra}}_{\mu \nu} + \tx{(higher order terms)} \\ 
\end{split}
\label{Fsigma}
\end{equation}
where, again, the lower sign corresponds to the retarded field and the upper sign corresponds to the advanced field.
In Equation (\ref{Fsigma}), the notation
\begin{equation}
f_{\alpha \beta \alpha'}(x,z) = \nabla_{\beta} V_{\alpha \alpha'}(x,z) - \nabla_{\alpha} V_{\beta \alpha'}(x,z)
\end{equation}
is used.
The term $F^{\tx{extra}}_{\alpha \beta}$ is, in fact, a sum of terms and is the same for the retarded and the advanced fields, so it vanishes when the difference of the two is taken.

\subsection{Conservation of Energy and Momentum}
\label{ConservEM}

The first step, when considering the conservation of energy and momentum, is the construction of a cylindrical surface that surrounds the worldline of the charged particle. 
That surface is referred to as the world-tube from now on.
As in Dirac's analysis, the radius $\epsilon$ of the world-tube is very small, because the balance of energy and momentum between the particle and the field needs to be calculated in the vicinity of the particle's trajectory.
DeWitt and Brehme \cite{dewbre} explain the procedure for the construction of the world-tube in great detail. 
For the needs of the present calculation, the world-tube can be thought of as the three-dimensional surface that is generated by the particle 
if the particle is surrounded by a sphere of a very small radius and is then left to move along its worldline, starting at proper time $\tau_{1}$ and ending at proper time $\tau_{2}$.
In the following, the cylindrical surface is denoted by $\Sigma$, its two end caps are denoted by $\Sigma_{1}$ (corresponding to the earliest proper time $\tau_{1}$) and $\Sigma_{2}$ (corresponding to the latest proper time $\tau_{2}$).
The total volume enclosed by the world-tube is denoted by $V_{\tx{t}}$.

The stress-energy tensor $T^{\alpha \beta}$ of the system consists of one part that comes from the particle and one part that comes from the field
\begin{equation}
T^{\alpha \beta} = T_{\tx{P}}^{\alpha \beta} + T_{\tx{F}}^{\alpha \beta}
\end{equation}
where
\begin{equation}
T_{\tx{P}}^{\alpha \beta} = m_{0} \int \bar{\delta}^{\frac{1}{2}} \: \bar{g}^{\alpha}_{\: \: \alpha'} \bar{g}^{\beta}_{\: \: \beta'} \: u^{\alpha'} u^{\beta'} \: \delta^{4} \: d\tau
\end{equation}
\begin{equation}
T_{\tx{F}}^{\alpha \beta} = \frac{1}{4 \pi} \: (F^{\alpha}_{\: \: \gamma} F^{\beta \gamma} - \frac{1}{4} \: g^{\alpha \beta} F_{\gamma \delta} F^{\gamma \delta})
\end{equation}
where $F_{\alpha \beta}$ is the total field in the vicinity of the particle.
It is the sum of the field generated by the particle plus any external incoming field that is present.

The conservation of energy and momentum for the system can be expressed as
\begin{equation}
\nabla_{\beta} T^{\alpha \beta} = 0.
\label{CONSEnergyMomentum}
\end{equation} 
When integrating Equation (\ref{CONSEnergyMomentum}) over the world-tube, special attention should be given to the fact that the integral $(\int \nabla_{\beta} T^{\alpha \beta} d^{4}x)$ is not a vector in the curved spacetime that is being considered here.
That means that Gauss's theorem cannot be used as usual, to convert the integral over the volume $V_{\tx{t}}$ to an integral over the surface of the world-tube.
The natural way to overcome this difficulty is to consider the integral $\int \bar{g}_{\alpha}^{\: \: \alpha'} \: \nabla_{\beta} T^{\alpha \beta} \: d^{4}x$ instead.
In this integral, the bivector of geodesic parallel displacement is used, so that any contributions to the integral by the point $x$ are referred back to the fixed point $z$, which is assumed to correspond to proper time $\tau$.
That integral is, then, a contravariant vector at point $z$, so Gauss's theorem can be used.
That gives
\begin{equation}
\bigg ( \int_{\Sigma} + \int_{\Sigma_{1}} + \int_{\Sigma_{2}} \bigg ) \bar{g}_{\alpha}^{\: \: \alpha'} \: T^{\alpha \beta} \: d\Sigma_{\beta} - \int_{V_{\tx{t}}} \Big ( \nabla_{\beta} \bar{g}_{\alpha}^{\: \: \alpha'} \Big ) \: T^{\alpha \beta} \: d^4x = 0 
\label{consEM1}
\end{equation}
where $d\Sigma_{\beta}$ is the surface element along the surface of the world-tube.

The limit of the radius of the world-tube going to zero is taken next.
Also the integral over the surface $\Sigma$ of the world-tube is expressed as a double integral, over the proper time and over the solid angle.
In addition, the proper times $\tau_{1}$ and $\tau_{2}$ are let to approach $\tau$. 
If their infinitesimal difference is denoted by $d\tau$, Equation (\ref{consEM1}) becomes
\begin{equation}
m_{0} \frac{Du^{\alpha'}}{d\tau} d\tau + \lim_{\epsilon \to 0} \int_{4 \pi} \bar{g}_{\alpha}^{\: \: \alpha'} T^{\alpha \beta} d\Sigma_{\beta} = 0.
\label{consEM2}
\end{equation}
It is in the calculation of the integral of the second term that the symmetric field and the radiation field mentioned in Equations (\ref{Fsym}), (\ref{Frad}) and (\ref{Ftotal}) are very useful.

As was mentioned earlier, the result contains only the incoming external field, the radiation field and the tail field and is
\begin{equation}
m \frac{Du^{\alpha'}}{d\tau} = e F^{\tx{in} \: \alpha'}_{\: \: \: \: \: \: \: \: \beta'} u^{\beta'} + \frac{1}{2} e F^{\tx{rad} \: \alpha'}_{\: \: \: \: \: \: \: \: \:\: \:  \beta'} u^{\beta'} + \frac{1}{2} e^{2} u^{\beta'} \int_{-\infty}^{\tau} f^{\alpha'}_{\: \: \beta' \gamma'}(z(\tau), z(\tau')) u^{\gamma'}(\tau') d\tau'
\label{resultDB1}
\end{equation}
where $m$ is the renormalized observed mass of the charged particle
\begin{equation}
m = m_{0} + \lim_{\epsilon \to 0} \frac{1}{2} e^{2} \epsilon^{-1}.
\end{equation}

An equivalent expression is derived by using Equation (\ref{Fsigma}) and the definition of the radiation field given in Equation (\ref{Frad}), and it is 
\begin{equation}
\begin{split}
m \frac{Du^{\alpha'}}{d\tau} = \: & e F^{\tx{in} \: \alpha'}_{\: \: \: \: \: \: \: \: \beta'} u^{\beta'} + \frac{2}{3} e^{2} (\frac{D^{2}u^{\alpha'}}{d\tau^{2}} - u^{\alpha'} \frac{Du^{\beta'}}{d\tau} \frac{Du_{\beta'}}{d\tau}) + \\ 
&+ \frac{1}{2} e^{2} u^{\beta'} \int_{-\infty}^{\tau} f^{\alpha'}_{\: \: \beta' \gamma'}(z(\tau), z(\tau')) u^{\gamma'}(\tau') d\tau'.\\
\end{split}
\label{resultDB2}
\end{equation}
Equation (\ref{resultDB2}) contains only the physical characteristics of the worldline of the charged particle and the incoming external field.

\subsection{Radiation Reaction}
\label{RadReacDewBre}

A short discussion about Equations (\ref{resultDB1}) and (\ref{resultDB2}) is in order.
First it is significant to notice that, except for the last term that contains the quantity $f_{\alpha \beta \alpha'}$, these equations are identical to Equations (\ref{resultD1}) and (\ref{resultD2}) derived by Dirac for flat spacetime.
Since that term contains the biscalar $V_{\alpha \alpha'}$, it comes from the tail part of the retarded electromagnetic field.

Also, it is interesting that the integral that shows up in these two equations needs to be calculated from $-\infty$ to the proper time $\tau$, meaning that knowledge of the entire past history of the particle is required.
That is something that renders the use of Equation (\ref{resultDB2}) impractical and is a point that is also discussed in the following chapters.

      \chapter{Self-Force}
\label{UFM}

In general, the self-force can be due to a scalar, an electromagnetic or a gravitational field.
Two different methods for the self-force calculation are presented in this chapter.
The analysis is given in detail for the scalar self-force, due to the simplicity of the notation. 
The results are also given for the electromagnetic and gravitational self-force.

The first method, in which the direct and tail parts of the retarded fields are used, was described by DeWitt and Brehme in \cite{dewbre} for the scalar and electromagnetic fields and by Mino, Sasaki and Tanaka in \cite{mst} for the gravitational field.
An axiomatic approach to this method was also presented by Quinn \cite{q00} for the scalar field and by Quinn and Wald \cite{qw97} for the electromagnetic and gravitational fields.

The second method, in which the singular-source part and the regular remainder of the retarded fields are used, was proposed by Detweiler and Whiting in \cite{detwhi02}, where they described it for the scalar, electromagnetic and gravitational fields.

As in Chapter \ref{RRC}, $z$ denotes a point along the worldline of the moving particle and $x$ any point in spacetime. 
The limit of $x \to \Gamma$ is the coincidence limit.
The retarded and advanced proper times, $\tau_{\tx{ret}}$ and $\tau_{\tx{adv}}$, are the proper times at the points where the null
 cone of $x$ intersects the worldline $\Gamma$ (see Figure (\ref{nullcone})).
Primed indices refer to $z$ and unprimed ones refer to $x$.

\section{Scalar Self-Force}
\label{ScalarUFM}

A particle of scalar charge $q$ is assumed to be moving in a background spacetime, described by the background metric $g_{a b}$.
For simplicity it can be assumed that there is no external scalar field and consequently, if the scalar charge $q$ is small, the lowest order approximation for the worldline $\Gamma : z^{a'}(\tau)$ of the particle is a background geodesic, where $\tau$ is the proper time.
But the scalar field $\psi$ generated by the particle interacts with the particle, inducing a self-force on it.
The self-force gives $\Gamma$ an acceleration, so, to higher order, the particle does not trace a background geodesic.

\subsection{Direct and Tail Fields}
\label{ScalarUFMDT}

The discussion of the scalar field self-force in \cite{dewbre} and \cite{q00} shows that the self-force on the particle can be calculated by
\begin{equation}
{\mathcal{F}}^{a} = q \: \Big [ \nabla^{a} \psi^{\tx{self}} \Big ]_{\tx{particle}}
\label{ForcePsi}
\end{equation}
where $\psi^{\tx{self}}$ is the scalar field that interacts with the particle and is given by
\begin{equation}
\psi^{\tx{self}}(x) = - \Big [ \frac{q \: U(x,z)}{2 \: \dot{\sigma}} \Big ]_{\tau_{\tx{ret}}}^{\tau_{\tx{adv}}} - q \int_{-\infty}^{\tau_{\tx{ret}}} V(x,z(\tau)) \: d\tau.
\label{PsiSelf}
\end{equation}

It can be inferred from the discussion of the scalar Green's functions of Section (\ref{GFS}) that the first term of the right-hand side of Equation (\ref{PsiSelf}) comes from the direct part of the retarded field generated by the moving particle.
This first term is finite and differentiable at the location of the particle and its contribution to the self-force is the curved-spacetime generalization of the Abraham-Lorentz-Dirac force of flat spacetime.
This contribution results from the moving particle's acceleration, if the worldline $\Gamma$ is not a geodesic,
and can be expressed in terms of the acceleration of $\Gamma$ and components of the Riemann tensor \cite{dewbre,mst,q00,qw97}.

The integral term of the right-hand side of Equation (\ref{PsiSelf}) comes from the tail part of the retarded field of the moving particle.
Its contribution to the self-force represents the result of the scattering of the retarded field of the particle, due to the curvature of spacetime.
Taking the derivative of the tail part of $\psi^{\tx{self}}$ gives one term that comes from the implicit dependence of the retarded proper time on $x$ and one term that comes from the dependence of the integrand on $x$, as was calculated by Quinn in \cite{q00}:
\begin{equation}
\begin{split}
\nabla_{a} \Big \{ & - q \int_{-\infty}^{\tau_{\tx{ret}}} V(x,z(\tau)) \: d\tau \Big \}  \\ 
           &= - q \: V(x,z(\tau_{\tx{ret}})) \: [ \nabla_{a}\tau_{\tx{ret}}(x) ] - q \int_{-\infty}^{\tau_{\tx{ret}}} \big \{ \nabla_{a} \: V(x,z(\tau)) \big \} \: d\tau  \\*
           &= - q \: \Big [ V \: \dot{\sigma}^{-1} \: \nabla_{a}\sigma \Big ]_{\tau_{\tx{ret}}} - q \int_{-\infty}^{\tau_{\tx{ret}}} \big \{ \nabla_{a} \: V(x,z(\tau)) \big \} \: d\tau  \\
           &= \Big [ - \frac{q \: R(x)}{12 \: r} (x_{a} - z_{a}) \Big ]_{x \to \Gamma} + O(r) - q \int_{-\infty}^{\tau_{\tx{ret}}} \big \{ \nabla_{a} \: V(x,z(\tau)) \big \} \: d\tau. \\
\end{split}
\label{TailC}
\end{equation} 
In Equation (\ref{TailC}), $r$ is the proper distance from $x$ to $\Gamma$ measured along the spatial geodesic that is orthogonal to $\Gamma$.

The spatial part of the first term of the derivative calculated in Equation (\ref{TailC}) is not well defined when $x$ is on $\Gamma$, unless $R(x) = 0$ there. 
That makes this approach problematic since, for the self-force to be calculated, the derivative must be calculated at the location of the particle.
To overcome this difficulty, $\nabla_{a} \psi^{\tx{self}}$ can be averaged over a small spatial 2-sphere surrounding the particle, thus removing the spatial part.
Then the limit of the radius of the 2-sphere going to zero can be taken and a finite contribution to the self-force is obtained.

Another complication comes from the fact that, in order for the contribution of the tail term to the self-force to be calculated, knowledge of the entire past history of the moving particle is necessary, as is clear from the integral term in Equation (\ref{TailC}). 

It is also important to note that the field $\psi^{\tx{self}}$, just like the field $\psi^{\tx{tail}}$, is not a physically well-understood field.
That is because it is not a solution to any particular differential equation.

\subsection{The S-Field and the R-Field}
\label{ScalarUFMSR}

\begin{figure}[ht]
\centering
\includegraphics[height=10.1cm, width=13.1cm, clip]{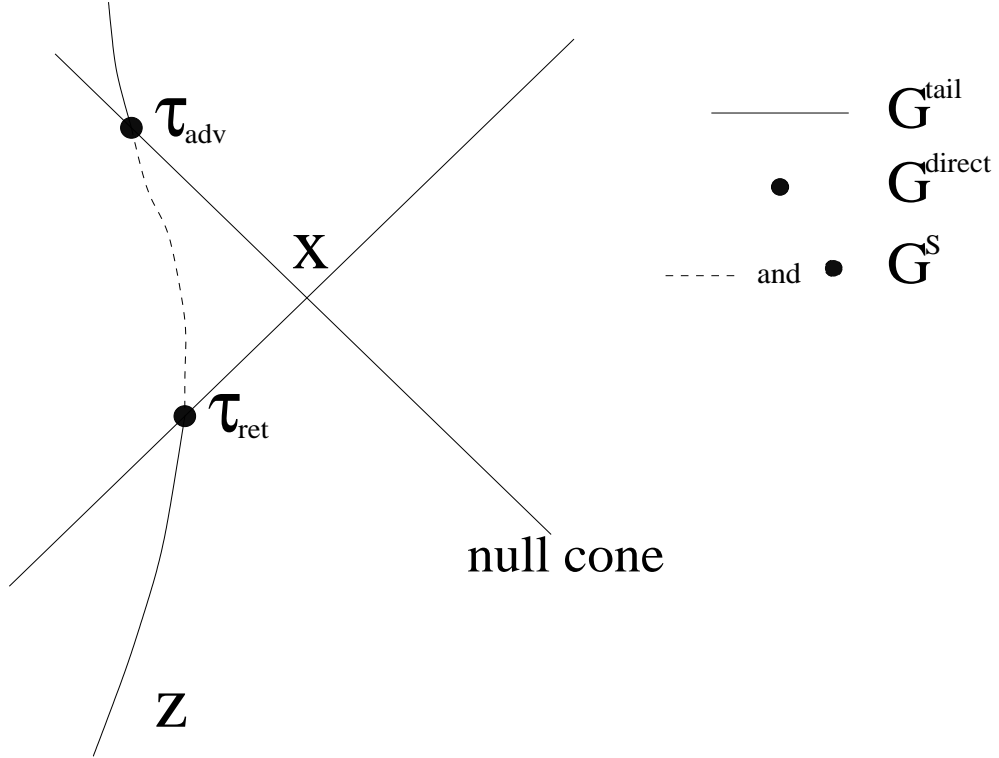}
\caption{Support of $G^{\tx{direct}}$, $G^{\tx{tail}}$ and $G^{\tx{S}}$.}
\label{Ggraph}
\end{figure}

In order to provide an alternative scalar field that gives the self-force on the moving particle, Detweiler and Whiting \cite{detwhi02} took advantage of the fact that adding a homogeneous solution of a differential equation to an inhomogeneous solution of the same differential equation, gives a new inhomogeneous solution.
Specifically, since the biscalar $V(x,z)$ is a symmetric homogeneous solution of the scalar differential Equation (\ref{DelV}), it can be added to the symmetric Green's function $G^{\tx{sym}}(x,z)$ to give a new symmetric inhomogeneous solution $G^{\tx{S}}(x,z)$ of Equation (\ref{dGsym})
\begin{equation}
\begin{split}
G^{\tx{S}}(x,z) &= G^{\tx{sym}}(x,z) + \frac{1}{8 \pi} V(x,z) \\
		&= \frac{1}{8 \pi} [U(x,z) \delta(\sigma) - V(x,z) \Theta(-\sigma) + V(x,z)] \\
		&= \frac{1}{8 \pi} [U(x,z) \delta(\sigma) + V(x,z) \Theta(\sigma)], \\
\end{split}
\label{Gs}
\end{equation} 
which obeys
\begin{equation}
\nabla^{2} G^{\tx{S}}(x,z) = - (-g)^{\frac{1}{2}} \delta^{4}(x-z).
\end{equation}
The superscript $\tx{S}$ is used to indicate that this Green's function obeys the differential equation that contains the Source term.
This new symmetric Green's function has support on the null cone, coming from the term that contains the $\delta$-function, and outside of the null cone, at spacelike separated points, coming from the term that contains $\Theta(\sigma)$.
It has no support within the null cone, as shown in Figure (\ref{Ggraph}).
The local expansion of the biscalar $V(x,z)$ given in Equation (\ref{ExpV}) is sufficient for the purposes of this dissertation, since the singular Green's function 
is only used for points $x$ close to the worldline of the particle.

The Regular-Remainder Green's function is defined in terms of the retarded and singular Green's functions as
\begin{equation}
\begin{split}
G^{\tx{R}}&(x,z) = G^{\tx{\tx{ret}}}(x,z) - G^{\tx{S}}(x,z) = \\
		&= \frac{1}{8 \pi} \Big \{ 2 \Theta[\Sigma(x),z] [U(x,z) \delta(\sigma) - V(x,z) \Theta(-\sigma)] - [U(x,z) \delta(\sigma) + V(x,z) \Theta(\sigma)] \Big \}, \\
\end{split}
\label{RegRem}
\end{equation}
and it is obvious that, by construction, it obeys the homogeneous differential equation
\begin{equation}
\nabla^{2} G^{\tx{R}}(x,z) = 0.
\end{equation}
Clearly, $G^{\tx{R}}(x,z)$ has no support inside the future null cone of $x$.

Using the new Green's functions, the fields $\psi^{\tx{S}}$ and $\psi^{\tx{R}}$ can be defined.
First, the singular field is
\begin{equation}
\begin{split}
\psi^{\tx{S}}(x) &= 4 \pi q \int G^{\tx{S}}[x,z(\tau)] \: d\tau \\
		 &= \Big [ \frac{q \: U(x,z)}{2 \dot{\sigma}} \Big ]_{\tau_{\tx{ret}}} + \Big [ \frac{q \: U(x,z)}{2 \dot{\sigma}} \Big ]_{\tau_{\tx{adv}}} + \frac{q}{2} \int_{\tau_{\tx{ret}}}^{\tau_{\tx{\tx{adv}}}} V(x,z) \: d\tau, \\
\end{split}
\end{equation}
and obeys Poisson's equation
\begin{equation}
\nabla^{2} \psi^{\tx{S}}(x) = - 4 \pi \varrho.
\end{equation}
It is noteworthy that the singular field does not depend on the entire past history of the moving particle, but only on its motion between the retarded and advanced proper times.

The regular-remainder is defined as
\begin{equation}
\begin{split}
\psi^{\tx{R}}(x) &= \psi^{\tx{\tx{ret}}}(x) - \psi^{\tx{S}}(x) \\
		 &= - \Big [ \frac{q \: U(x,z)}{2 \dot{\sigma}} \Big ]_{\tau_{\tx{ret}}}^{\tau_{\tx{\tx{adv}}}} - q \Bigg [ \int_{-\infty}^{\tau_{\tx{ret}}} + \frac{1}{2} \int_{\tau_{\tx{ret}}}^{\tau_{\tx{\tx{adv}}}} \Bigg ] V(x,z) \: d\tau, \\
\end{split}
\label{PsiR}
\end{equation}
and by definition obeys the homogeneous differential equation
\begin{equation}
\nabla^{2} \psi^{\tx{R}}(x) = 0.
\end{equation}
Since the field $\psi^{\tx{R}}$ is a source-free field, it is smooth and differentiable at any point in spacetime and consequently at any point along the worldline of the particle as well. That is the most significant property of it, as far as the calculation of the self-force is concerned.

To compare the scalar fields $\psi^{\tx{self}}$ and $\psi^{\tx{R}}$, the difference of the two is calculated from Equations (\ref{PsiR}) and (\ref{PsiSelf})
\begin{equation}
\psi^{\tx{R}}(x) - \psi^{\tx{self}}(x) = - \frac{q}{2} \int_{\tau_{\tx{ret}}}^{\tau_{\tx{adv}}} V(x,z) \: d\tau.
\end{equation}
Using Equation (\ref{ExpV}) for the biscalar $V(x,z)$ to expand the above integrand in the coincidence limit, the difference of the two fields becomes
\begin{equation}
\begin{split}
\psi^{\tx{R}}(x) - \psi^{\tx{self}}(x) &= - \frac{q}{2} \int_{\tau_{\tx{ret}}}^{\tau_{\tx{adv}}} \big [ -\frac{1}{12} R(x) + O(r) \big  ] d\tau \\
				       &= \frac{1}{12} q r R(x) + O(r^2)  \tx{,} \: \: \tx{ for } \: x \to \Gamma
\end{split}
\end{equation}
where the fact that the difference between the advanced and the retarded proper times is equal to 
\begin{equation}
\tau_{\tx{adv}} - \tau_{\tx{ret}} = 2 r + O(r^2), \tx{  for  } x \to \Gamma
\end{equation}
was used.
Taking the derivative gives
\begin{equation}
\nabla_{a} \psi^{\tx{R}} = \nabla_{a} \psi^{\tx{self}} + \frac{q}{12} \nabla_{a}[r R(x)] + (\tx{terms that vanish as } x \to \Gamma).
\label{DelaPsiR}
\end{equation}
The second term of the right-hand side of Equation (\ref{DelaPsiR}) gives an outwardly pointing spatial unit vector and cancels the first term of the right side of Equation (\ref{TailC}).
Consequently, the fields $\psi^{\tx{R}}$ and $\psi^{\tx{self}}$ give the same self-force.
In addition, the averaging procedure that was necessary for the calculation of the self-force using $\psi^{\tx{self}}$ is not required when $\psi^{\tx{R}}$ is used.

A self-force calculation that has been demonstrated in \cite{detmeswhi} showed an additional benefit of calculating the self-force using the difference of the retarded and singular fields. 
The retarded field can be computed numerically and, since the singular field depends only on the motion of the particle between the retarded and advanced proper times, the knowledge of the entire past history of the particle is not required.
That is a significant advantage of this method compared to the one that involves the direct and tail fields.

Additionally, it is important that all the scalar fields used in this analysis are specific solutions of the homogeneous or inhomogeneous Equation (\ref{scinh}).
That makes them well-defined physical fields, a property that the direct and tail scalar fields do not have.

For those reasons, in this dissertation the self-force is calculated using
\begin{equation}
\mathcal{F}^{a} = q \nabla^a \psi^{\tx{R}}.
\label{ScalarSF1}
\end{equation}

\section{Electromagnetic Self-Force}
\label{EMUFM}

A detailed analysis for the self-force in terms of the direct and tail electromagnetic potentials was done in \cite{dewbre} and was presented in Chapter \ref{RRC}. 
In this section, the calculation of the self-force using the singular-source and the regular remainder fields is presented.

It is assumed that a particle of electric charge $q$ is moving on a worldline $\Gamma : z^{a'}(\tau)$ in a background described by the metric $g_{ab}$.
For the purposes of this section it can be assumed that there is no external electromagnetic field and the lowest order approximation to the  particle's motion is a background geodesic.
The particle creates an electromagnetic potential $A^{a}$ and an electromagnetic field $F^{ab}$, which interact with the particle, causing a self-force to act on it and making its worldline to deviate from a background geodesic to order $q^2$.
In the Lorentz gauge
\begin{equation}
\nabla_{a} A^{a} = 0,
\end{equation}
the electromagnetic potential generated by the particle can be calculated by using Maxwell's equations
\begin{equation}
\nabla^{2} A^{a} - R^{a}_{\: b} A^{b} = - 4 \pi J^{a}
\label{EMPot}
\end{equation}
and the different solutions of interest were described by the Green's functions of Section (\ref{GFV}).

If a potential $A_{a}^{\tx{self}}$, analogous to $\psi^{\tx{self}}$, is used to calculate the self-force in this case, its tail part is
\begin{equation}
A_{a}^{\tx{self (tail)}}(x) = - q \int_{-\infty}^{\tau_{\tx{ret}}} V_{ab'}(x,z(\tau)) \: u^{b'} \: d\tau.
\end{equation}
Taking the derivative of it gives
\begin{equation}
\begin{split}
\nabla_{c} &A_{a}^{\tx{self (tail)}} = - q \Big [ V_{ab'} \: u^{b'} \: \nabla_{c} \tau_{\tx{ret}} \Big ]_{x \to \Gamma} - q \int_{-\infty}^{\tau_{\tx{ret}}} \nabla_{c} (V_{ab'} u^{b'}) d\tau \\
	   &= - q \Big [ \: \frac{1}{2} \: \bar{g}_{a}^{\: \: d'} \: (R_{b'd'} - \frac{1}{6} g_{b'd'} R) \: u^{b'} \: \frac{x_c - z_c}{r} \Big ]_{x \to \Gamma} - q \int_{-\infty}^{\tau_{\tx{ret}}} \nabla_{c} [ V_{ab'} u^{b'} ] \: d\tau, \\
\end{split}
\end{equation}
where Equation (\ref{VR}) for the bivector $V_{ab'}$ was used to derive the last expression.
It is obvious that the same problems that were encountered in the scalar case are encountered here as well.
The first term is not well-defined unless $(R_{a'b'} - \frac{1}{6} g_{a'b'} R) u^{b'}$ is zero at the particle's worldline, the entire past history of the moving particle needs to be known and the potential used in the self-force calculation is not a solution of Maxwell's equations.

To overcome those difficulties, the singular Green's function in the neighborhood of the particle is defined analogously to the scalar case 
\begin{equation}
\begin{split}
G_{aa'}^{\tx{S}}(x,z) &= G_{aa'}^{\tx{sym}}(x,z) + \frac{1}{8 \pi} V_{aa'}(x,z) \\
		      &= \frac{1}{8 \pi} [ U_{aa'}(x,z) \delta(\sigma) + V_{aa'}(x,z) \Theta(\sigma) ]. \\
\end{split}
\end{equation}
It obeys the inhomogeneous differential equation
\begin{equation}
\nabla^{2} G_{aa'}^{\tx{S}}(x,z) - R_{a}^{\: \: b} G_{ba'}^{\tx{S}}(x,z) = - \bar{g}^{-\frac{1}{2}} \bar{g}_{aa'} \delta^{4}(x,z)
\end{equation}
and gives the singular electromagnetic potential
\begin{equation}
A_{a}^{\tx{S}} = 4 \pi \int (-g)^{-\frac{1}{2}} G_{ab}^{\tx{S}}(x,y) J^{b}(y) d^{4}y,
\end{equation}
which is a solution of the inhomogeneous Maxwell's Equations (\ref{EMPot}).
By definition, both $G_{ab'}^{\tx{S}}$ and $A_a^{\tx{S}}$ have no support inside the past and future null cone of $x$.
The regular remainder is defined by
\begin{equation}
A^{\tx{R}}_{a}(x) = A^{\tx{\tx{ret}}}_{a}(x) - A^{\tx{S}}_a(x).
\end{equation}
It has no support within the future null cone of $x$ and it obeys the homogeneous Maxwell's equations. Consequently, it is smooth and differentiable everywhere in space, including every point along the worldline of the particle.
The electromagnetic self-force can be calculated by the equation
\begin{equation}
\mathcal{F}^{a} = q g^{ac}(\nabla_{c}A_{b}^{\tx{R}} - \nabla_{b}A_{c}^{\tx{R}}) \dot{z}^{b}
\label{EMSF1}
\end{equation}
and is well-defined at the particle's location.
It is stressed, again, that both $A_{a}^{\tx{S}}$ and $A_{a}^{\tx{R}}$ are well-defined solutions of Maxwell's equations.

\section{Gravitational Self-Force}
\label{GravUFM}

It is now assumed that a particle of mass $m$ is moving on a worldline $\Gamma : z^{a'}(\tau)$, in a background described by the metric $g_{ab}$.  
The particle causes a metric perturbation $h_{ab}$ on the background metric. 
This perturbation obeys the harmonic gauge
\begin{equation}
\nabla_{a} \bar{h}^{ab} = 0
\end{equation}
where $\bar{h}_{ab}$ is the trace-reversed version of the metric perturbation
\begin{equation}
\bar{h}_{ab} = h_{ab} - \frac{1}{2} g_{ab} h^{c}_{\: \: c}.
\end{equation}
The linearized Einstein equations for it are
\begin{equation}
\nabla^{2} \bar{h}_{ab} + 2 R_{a \: \: b}^{\: \: c \: \: d} \bar{h}_{cd} = - 16 \pi T_{ab}.
\label{Hfield}
\end{equation}
This metric perturbation interacts with the particle, causing a self-force to act on it and forcing its worldline to deviate from a background geodesic.

If $\bar{h}^{\tx{self}}_{ab}$, which is not a solution of Einstein's equations, is used to calculate the self-force, its tail part contains the bitensor $V_{abc'd'}$ given in Equation (\ref{VgR}).
The contribution of this tail term to the derivative comes partly from the dependence of the proper time $\tau_{\tx{ret}}$ on the point $x$.
Specifically
\begin{equation}
\begin{split}
\nabla_{f} \bar{h}^{\tx{self (tail)}}_{ab} &= - m \Big [ V_{abc'd'} u^{c'} u^{d'} \nabla_{f} \tau_{\tx{ret}} \Big ]_{x \to \Gamma} - m \int_{-\infty}^{\tau_{\tx{ret}}} \nabla_{f} \big [ V_{abc'd'} u^{c'} u^{d'}  \big ] d\tau \\
					  &= m \Big [ \bar{g}_{a}^{\: a'} \bar{g}_{b}^{\: b'} R_{c'a'd'b'} \frac{(x_f - z_f)}{r} \Big ]_{x \to \Gamma} - m \int_{-\infty}^{\tau_{\tx{ret}}} \nabla_{f} \big [ V_{abc'd'} u^{c'} u^{d'}  \big ] d\tau. \\
\end{split}
\end{equation}
Just like before, the first term is not well-defined at the particle's worldline, unless
$R_{c'a'd'b'} u^{c'} u^{d'} $ equals zero
along the worldline. 
Also, knowledge of the entire past history of the particle is required in order to calculate the integral term.

The singular Green's function $G_{abc'd'}(x,z)$ in the neighborhood of the particle can be defined as
\begin{equation}
\begin{split}
G^{\tx{S}}_{abc'd'}(x,z) &= G_{abc'd'}^{\tx{sym}}(x,z) + \frac{1}{8 \pi} V_{abc'd'}(x,z) \\
			 &= \frac{1}{8 \pi} [ U_{abc'd'}(x,z) \delta(\sigma) + V_{abc'd'}(x,z) \Theta(\sigma) ]. \\
\end{split}
\end{equation}
It gives the singular field $\bar{h}^{\tx{S}}_{ab}$ which is an inhomogeneous solution of Equation (\ref{Hfield}) and has no support inside the past or future null cone of $x$
\begin{equation}
\bar{h}_{ab}^{\tx{S}} = 16 \pi \int (-g)^{-\frac{1}{2}} G_{abcd}^{\tx{S}}(x,y) T^{cd}(y) d^{4}y.
\end{equation} 
The regular remainder is defined as
\begin{equation}
\bar{h}_{ab}^{\tx{R}}(x) = \bar{h}_{ab}^{\tx{\tx{ret}}}(x) - \bar{h}_{ab}^{\tx{S}}(x).
\end{equation}
It obeys the homogeneous differential equation
\begin{equation}
\nabla^{2} \bar{h}_{ab}^{\tx{R}} + 2 R_{a \: \: b}^{\: \: c \: \: d} \bar{h}^{\tx{R}}_{cd} = 0,
\end{equation}
which means that it is smooth and differentiable everywhere in spacetime.
Also, it has no support within the future null cone of $x$.

Then the self-force can be calculated by
\begin{equation}
\mathcal{F}^{a} = - m (g^{ab} + \dot{z}^{a} \dot{z}^{b}) \dot{z}^{c} \dot{z}^{d} ( \nabla_{c}h_{db}^{\tx{R}} - \frac{1}{2} \nabla_{b}h_{cd}^{\tx{R}})
\label{GravSF1}
\end{equation}
which is well-defined at the location of the particle.
It is important that both $h_{ab}^{\tx{R}}$ and $h_{ab}^{\tx{S}}$ are solutions of specific differential equations.

A significant conclusion can be drawn at this point.
As befits the problem, the self-force is causing the particle to move on a geodesic of the metric $(g_{ab}+h_{ab}^{\tx{R}})$, to order $m^2$.
This metric $(g_{ab}+h_{ab}^{\tx{R}})$ is, itself, a homogeneous solution of the Einstein equations.
An extended discussion of that point is given in \cite{detwhi02} and \cite{det01} and is also presented in Chapter \ref{CONC} of this dissertation.

      \chapter{Singular Field for Schwarzschild Geodesics}
\label{SF}

It was seen in Chapter \ref{UFM} that the singular field is important in calculating the self-force.
It is the part that must be subtracted from the retarded field to give the regular remainder, which is then differentiated to give the self-force.
It is also clear that, since the singular field obeys the inhomogeneous Poisson, Maxwell or Einstein equations, it depends on the worldline of the moving particle and on the background spacetime.

The coordinates that are used in this chapter to caclulate the singular field are the Thorne-Hartle-Zhang coordinates, abbreviated from this point on as THZ coordinates. 
Those coordinates were initially introduced by Thorne and Hartle \cite{TH} and later extended by Zhang \cite{Z}.
A short discussion on them is presented in Section (\ref{THZ}) of this chapter.

The detailed calculation of the singular field of a scalar charge $q$ for geodesics in a Schwarzschild background was presented in \cite{detmeswhi}. 
A brief discussion about that field is given here for completeness, since the regularization parameters associated with it are calculated in Chapter \ref{RP}.
The singular field is also calculated for geodesics in a Schwarzschild background, for a dipole generating a scalar field, for an electric charge, an electric dipole and a magnetic dipole each generating its own electromagnetic field and for a spinning particle generating a gravitational field.
For all these calculations, {\textsc{Maple}} and {\textsc{Grtensor}} were used extensively.
The \textsc{Grtensor} code is explained in Section (\ref{GRTCode}) and presented in Appendix \ref{A0}.
In Sections (\ref{SFD}) through (\ref{SFSP}), only the results that the code gives are presented.

\section{Thorne-Hartle-Zhang Coordinates}
\label{THZ}

The Poisson, Maxwell and Einstein equations for the singular field assume a relatively simple form when they are written in a coordinate system in which the background spacetime looks as flat as possible.
In the following, it is assumed that the particle is moving on a geodesic $\Gamma$ in a vacuum background described by the metric $g_{ab}$. 
Also, $\mathcal{R}$ is a representative length scale of the background geometry, the smallest of the radius of curvature, the scale of inhomogeneities of the background and the time scale of curvature changes along $\Gamma$. 

A normal coordinate system can always be found so that, on the geodesic $\Gamma$, the metric and its first derivatives coincide with the Minkowski metric \cite{mtw}.
Such a normal coordinate system is not unique. 
The one used here is the THZ coordinate system and is used only locally, close to the worldline of the particle.
Specifically it is assumed that the background metric close to the worldline of the particle can be written as
\begin{equation}
\begin{split}
g_{ab} &= \eta_{ab} + H_{ab} \\
	&= \eta_{ab} + {}_{2}H_{ab} + {}_{3}H_{ab} + O(\frac{\rho^4}{\mathcal{R}^4}), \\
\end{split}
\label{TaylorExpandMetric}
\end{equation}
where $\eta_{ab}$ is the flat Minkowski metric in the THZ coordinates $(t,x,y,z)$ and
\begin{equation}
\rho^2 = x^2 + y^2 + z^2. 
\end{equation}
Also
\begin{equation}
\begin{split}
_{2}H_{ab} &dx^a dx^b = - \mathcal{E}_{ij} x^i x^j ( dt^2 + \delta_{kl} dx^k dx^l ) + \frac{4}{3} \epsilon_{kpq} \mathcal{B}^
{q}_{\: i} x^p x^i dt dx^k \\
&- \frac{20}{21} \Big [ \dot{\mathcal{E}}_{ij} x^i x^j x_k - \frac{2}{5} \rho^2 \dot{\mathcal{E}}_{ik}x^i \Big ] dt dx^k + \frac{5}{21} \Big [ x_i \epsilon_{jpq} \dot{\mathcal{B}}^{q}_{\: k} x^p x^k - \frac{1}{5} \rho^2 \epsilon_{pqi} \dot{\mathcal{B}}_{j}^{\: q} x^p \Big ] dx^i dx^j, \\
\end{split}
\label{H2ab}
\end{equation}
\begin{equation}
_{3}H_{ab} dx^a dx^b = O \big (\frac{\rho^3}{\mathcal{R}^3} \big )_{ab} dx^a dx^b + O \big (\frac{\rho^4}{\mathcal{R}^4} \big ) _{ij} dx^i dx^j.
\label{H3ab}
\end{equation}
In this and the following, $a$, $b$, $c$ and $d$ denote spacetime indices.
The indices $i, j, k, l, n, p$ and $q$ are spatial indices and, to the order up to which the calculations are performed, 
they are raised and lowered by the 3-dimensional flat space metric $\delta_{ij}$.
The dot denotes differentiation with respect to the time $t$ along the geodesic.
Also, $\epsilon_{ijk}$ is the 3-dimensional flat space antisymmetric Levi-Civita tensor.

If $H_{ab}$ consists of the terms given in Equation (\ref{H2ab}), the coordinates are second-order THZ coordinates and are well defined up to the addition of arbitrary functions of $O(\frac{\rho^4}{\mathcal{R}^3})$.
If $H_{ab}$ also includes the terms given in Equation (\ref{H3ab}), the coordinates are third-order THZ coordinates and are well defined up to the addition of arbitrary functions of $O(\frac{\rho^5}{\mathcal{R}^4})$.

The tensors $\mathcal{E}$ and $\mathcal{B}$ are spatial, symmetric and trace-free and their components are related to the Riemann tensor on the geodesic $\Gamma$ by
\begin{eqnarray}
\mathcal{E}_{ij} &=& R_{titj} \\
\mathcal{B}_{ij} &=& \frac{1}{2} \epsilon_{i}^{\: \: pq} R_{pqjt}.
\label{EandB}
\end{eqnarray}
They are of $O(\frac{1}{\mathcal{R}^2})$ and their time derivatives are of $O(\frac{1}{\mathcal{R}^3})$.

For the calculation of the singular fields that follows, only the first two terms of Equation (\ref{H2ab}) are included in $g_{ab}$.
The remaining two terms are of $O(\frac{\rho^3}{\mathcal{R}^3})$ and must be included in a higher-order calculation which must also take into account terms coming from the $_{3}H_{ab}$ part of the metric.

The `gothic' form of the metric is also defined as
\begin{equation}
\tx{\bf{g}}^{ab} = \sqrt{-g} g^{ab}
\end{equation}
and its difference from the Minkowski metric is
\begin{equation}
\bar{H}^{ab} = \eta^{ab} - \tx{\bf{g}}^{ab}.
\end{equation}
For the lowest order terms of $\bar{H}_{ab}$ it can be found that
\begin{equation}
\bar{H}^{ab} = {}_{2}\bar{H}^{ab} + {}_{3}\bar{H}^{ab} + O \big (\frac{\rho^4}{\mathcal{R}^4} \big )
\end{equation}
where
\begin{eqnarray}
_{2}\bar{H}^{tt} &=& -2 \mathcal{E}_{ij} x^i x^j \\
_{2}\bar{H}^{tk} &=& -\frac{2}{3} \epsilon^{kpq} \mathcal{B}_{qi} x_{p} x^i + \frac{10}{21} \Big [ \dot{\mathcal{E}}_{ij} x^i x^j x^k - \frac{2}{5} \dot{\mathcal{E}}_{i}^{\: \: k} x^i \rho^2 \Big ] \\
_{2}\bar{H}^{ij} &=& \frac{5}{21} \Big [ x^{(i} \epsilon^{j)pq} \dot{\mathcal{B}}_{qk} x_p x^k -\frac{1}{5} \epsilon^{pq(i} \dot{\mathcal{B}}^{j)}_{\: q} x_p \rho^2 \Big ] \\
_{3}\bar{H}^{ta} &=& O \big ( \frac{\rho^3}{\mathcal{R}^3} \big ), \: \: _{3}\bar{H}^{ij} = O \big ( \frac{\rho^4}{\mathcal{R}^4} \big ).
\label{H2barab}
\end{eqnarray}

Using the simple symmetry properties of the tensors $\mathcal{E}$ and $\mathcal{B}$, the following relationships can be shown for their components and the spatial THZ coordinates $(x,y,z)$:
\begin{eqnarray}
\epsilon_{ijk} \mathcal{E}^{k}_{\: l} + \epsilon_{ikl} \mathcal{E}^k_{\: j} - \epsilon_{jkl} \mathcal{E}^k_{\: i} &=& 0 \\ \label{EBRel1}
\epsilon_{ijk} \mathcal{B}^{k}_{\: l} + \epsilon_{ikl} \mathcal{B}^k_{\: j} - \epsilon_{jkl} \mathcal{B}^k_{\: i} &=& 0 
\label{EBRel2}
\end{eqnarray}
and
\begin{eqnarray}
( -\epsilon_{ijk} \mathcal{E}^k_{\: n} x_l + \epsilon_{ijk} \mathcal{E}_{ln} x^k - \epsilon_{ikl} \mathcal{E}^k_{\: n} x_j + \epsilon_{jkl} \mathcal{E}^k_{\: n} x^i ) x^n x^l &=& 0 \\ \label{EBRel3}
( -\epsilon_{ijk} \mathcal{B}^k_{\: n} x_l + \epsilon_{ijk} \mathcal{B}_{ln} x^k - \epsilon_{ikl} \mathcal{B}^k_{\: n} x_j + \epsilon_{jkl} \mathcal{B}^k_{\: n} x^i ) x^n x^l &=& 0.
\label{EBRel4}
\end{eqnarray}
These relationships are used in the following sections when calculating the singular fields for different sources
to simplify the expressions for those fields.

The calculation of the regularization parameters shown in Chapter \ref{RP} requires one to know the exact relationship between the THZ coordinates and the background Schwarzschild coordinates.
A general procedure for finding that relationship for a given geodesic is to try to satisfy the three basic properties of the second order THZ coordinates:
\begin{itemize}
\item On $\Gamma$: the coordinate $t$ measures the proper time along the geodesic and the spatial coordinates $x, y$ and $z$ are equal to 0. Also, $g_{ab} = \eta_{ab}$ and the first derivatives of the metric vanish there.
\item At linear, stationary order: $\bar{H}^{ij} = O(x^3)$.
\item The coordinates satisfy the harmonic gauge: $\partial_{a} \tx{\bf{g}}^{ab} = O(x^2)$.
\end{itemize}
The details of this procedure are described in Appendix A of \cite{detmeswhi}.

For circular orbits in a Schwarzschild background of mass $M$, the THZ coordinates have been calculated in Appendix B of \cite{detmeswhi} using specific properties of the spherically symmetric background, which is much simpler than following the procedure just described. 
Their functional relationships with the Schwarzschild coordinates are given here in order to facilitate the discussion of the calculation of the singular fields and the regularization parameters.
The Schwarzschild coordinates are $(t_{\tx{s}},r,\theta,\phi)$ and the Schwarzschild metric is
\begin{equation}
ds^2 = -(1-\frac{2M}{r}) dt_{\tx{s}}^2 + (1-\frac{2M}{r})^{-1} dr^2 + r^2 d\theta^2 + r^2 \sin^2\theta d\phi^2.
\end{equation}
The circular orbit given by $\phi = \Omega t_{\tx{s}}$ has orbital frequency equal to $\Omega =  (M r_o^{-3} )^{\frac{1}{2}}$, at Schwarzschild radius $r_o$. 

First, the functions $\mathcal{X, Y, Z}$, which are Lie derived by the Killing vector $\xi^{a} = \frac{\partial}{\partial t_{\tx{s}}} + \Omega \frac{\partial}{\partial \phi}$, are chosen
\begin{eqnarray}
\mathcal{X} &\equiv& \frac{r-r_o}{(1-\frac{2M}{r_o})^{\frac{1}{2}}} \\ \label{THZX}
\mathcal{Y} &\equiv& r_o \sin(\theta) \sin(\phi-\Omega t_{\tx{s}}) \Big ( \frac{r_o-2M}{r_o-3M} \Big )^{\frac{1}{2}} \\ \label{THZY}
\mathcal{Z} &\equiv& r_o \cos(\theta). \label{THZZ}
\end{eqnarray}
Then the functions $\tilde{x}$,  $\tilde{y}$, $z$, also Lie derived by the Killing vector $\xi^a$, and the function $t$ are defined in terms of $\mathcal{X, Y, Z}$
\begin{equation}
\begin{split}
\tilde{x} &= \frac{r \sin\theta \cos(\phi-\Omega t_{\tx{s}}) - r_o}{(1-\frac{2M}{r_o})^{\frac{1}{2}}} \\
	  &+ \frac{M}{r_o^2 (1-\frac{2M}{r_o})^{\frac{1}{2}}} \Big [ -\frac{\mathcal{X}^2}{2} + \mathcal{Y}^2 \Big ( \frac{r_o-3M}{r_o-2M} \Big ) + \mathcal{Z}^2 \Big ] \\
	  &+ \frac{M \mathcal{X}}{2 r_o^3 (r_o-2M) (r_o - 3M)} \big [ -M^2 \mathcal{X}^2 + \mathcal{Y}^2 (r_o-3M)(3r_o-8M) + 3 \mathcal{Z}^2 (r_o - 2M)^2 \big ] \\
	  &+ \frac{M}{r_o^5 (1-\frac{2M}{r_o})^{\frac{1}{2}} (r_o-3M)} \Big [ M \mathcal{X}^4 \frac{(r_o^2 - r_o M + 3 M^2)}{8(r_o - 2M)} \\
	  &+ \frac{\mathcal{X}^2 \mathcal{Y}^2}{28} (28r_o^2 - 114r_oM + 123M^2) + \frac{ \mathcal{X}^2 \mathcal{Z}^2}{14} (14r_o^2 - 48 r_oM + 33M^2) \\
	  &+ \frac{M \mathcal{Y}^4}{56 (r_o-2M)^2} (3r_o^3 - 74 r_o^2M + 337r_o M^2 - 430 M^3) \\
	  &- \frac{M^2 \mathcal{Y}^2 \mathcal{Z}^2 (7r_o -18M)}{4(r_o-2M)} - \frac{M \mathcal{Z}^4}{56} (3 r_o + 22M) \Big], \\ 
\end{split}
\label{THZxtilde}
\end{equation}
\begin{equation}
\begin{split}
\tilde{y} &= r \sin\theta \sin(\phi - \Omega t_{\tx{s}}) \Big ( \frac{r_o-2M}{r_o-3M} \Big )^{\frac{1}{2}} \\
	 &\quad + \frac{M \mathcal{Y}}{2 r_o^3} \Big [ -2 \mathcal{X}^2 + \mathcal{Y}^2 \Big ( \frac{r_o-3M}{r_o-2M} \Big ) + \mathcal{Z}^2 \Big ] \\
	  &\quad + \frac{M \mathcal{X} \mathcal{Y}}{14 r_o^5 (1 - \frac{2M}{r_o})^{\frac{1}{2}} (r_o-3M) } \big [ 2M\mathcal{X}^2 (4r_o-15M) \\ 
	  &\quad +\mathcal{Y}^2(14r_o^2 - 69r_oM + 89 M^2) + 2\mathcal{Z}^2(r_o-2M)(7r_o-24M)],\\
\end{split}
\label{THZytilde}
\end{equation}
\begin{equation}
\begin{split}
z &= r \cos\theta + \frac{M \mathcal{Z}}{2 r_o^3 (r_o-3M)} [ - \mathcal{X}^2 (2r_o-3M) + \mathcal{Y}^2(r_o-3M) + \mathcal{Z}^2(r_o-2M)] \\
  &+ \frac{M \mathcal{X Z}}{14 r_o^5 (1 - \frac{2M}{r_o})^{\frac{1}{2}}(r_o-3M)} [M \mathcal{X}^2 (13r_o-19M) \\
  &+\mathcal{Y}^2 (14r_o^2 - 36r_o M + 9 M^2) + \mathcal{Z}^2 (r_o-2M)(14r_o-15M)],\\
\end{split}
\label{THZz}
\end{equation}
\begin{equation}
\begin{split}
t &= t_{\tx{s}} (1-\frac{3M}{r_o})^{\frac{1}{2}} - \frac{r \Omega \mathcal{Y}}{(1 - \frac{2M}{r_o})^{\frac{1}{2}}} \\
  & + \frac{\Omega M \mathcal{Y}}{r_o^2 (1 - \frac{2M}{r_o})^{\frac{1}{2}} (r_o - 3M)} \Big [ - \frac{\mathcal{X}^2}{2}(r_o-M) + M \mathcal{Y}^2 \frac{r_o - 3M}{3(r_o-2M)} + M\mathcal{Z}^2 \Big ] \\
  &+ \frac{\Omega M \mathcal{XY}}{14 r_o^3 (r_o-2M) (r_o-3M)} \big [ - \mathcal{X}^2 (r_o^2 - 11 r_oM + 11 M^2) \\
  &+\mathcal{Y}^2 (13r_o^2 - 45r_o M + 31 M^2) + \mathcal{Z}^2 (13r_o-5M)(r_o-2M) \big ]. \\
\end{split}
\label{THZt}
\end{equation}
Finally
\begin{equation}
\begin{split}
&x = \tilde{x} \cos (\Omega^{\dagger} t_{\tx{s}}) - \tilde{y} \sin( \Omega^{\dagger} t_{\tx{s}} )\\
&y = \tilde{x} \sin (\Omega^{\dagger} t_{\tx{s}}) + \tilde{y} \cos (\Omega^{\dagger} t_{\tx{s}}) \\
\end{split}
\label{xy}
\end{equation}
where $\Omega^{\dagger} = \Omega \sqrt{1-\frac{3M}{r_o}}$.

There are two coordinate systems of interest. 
The first system is $(t, \tilde{x}, \tilde{y}, z)$ which is a non-inertial coordinate system that co-rotates with the particle, meaning that the $\tilde{x}$ axis always lines up the center of the black hole and the center of the particle.
The $\tilde{y}$ axis is always tangent to the spatially circular orbit and the $z$ axis is always orthogonal to the orbital plane.
As was already mentioned, for the spatial coordinates of this system it holds that: $\mathcal{L}_{\xi} \tilde{x} = \mathcal{L}_{\xi} \tilde{y} = \mathcal{L}_{\xi} z = 0$.

The second system is $(t, x, y, z)$ and is a locally inertial and non-rotating system in the vicinity of $\Gamma$.
However, when viewed far away from $\Gamma$ these coordinates appear to be rotating due to Thomas precession, as is clear from the terms involving the sine and the cosine of $(\Omega^{\dagger} t_{\tx{s}})$ in Equations (\ref{xy}).
It is also noted that: $\mathcal{L}_{\xi} \rho^2 = \mathcal{L}_{\xi}z = \mathcal{L}_{\xi} \nabla_{a} t = 0$, but $\mathcal{L}_{\xi} x$, $\mathcal{L}_{\xi} y$ and $\mathcal{L}_{\xi} t$ are not equal to zero.
This second system is used to calculate the singular fields.
In order to avoid confusion of the field point $x$ and the source point $z$ with the THZ coordinates, the notation needs to be changed.
The field point is denoted as $p$ and its coordinates as $x_p^{a}$ and the source point is denoted as $p'$ and its coordinates as $x_{p'}^{a}$.

\section{Scalar Field of a Charged Particle}
\label{SFCP}

The singular field generated by a particle that carries a scalar charge $q$ and is moving on a geodesic $\Gamma: p'(\tau)$ (where, as usual, $\tau$ is the proper time) of the Schwarzschild background must obey Poisson's  equation
\begin{equation}
\nabla^{2} \psi^{\tx{S}} = - 4 \pi \varrho
\end{equation}
where the $\nabla^{2}$ is written in the THZ coordinates and the source term is
\begin{equation}
\varrho = q \int (-g)^{-\frac{1}{2}} \delta^{4}(p-p'(\tau)) \: d\tau.
\end{equation}
That singular field is derived in \cite{detmeswhi} and is equal to
\begin{equation}
\psi^{\tx{S}} = \frac{q}{\rho} + O(\frac{\rho^3}{\mathcal{R}^{4}}).
\label{ScSField}
\end{equation}
Here, instead of following the exact calculation of the singular field, which is performed in detail in \cite{detmeswhi} following a similar derivation in \cite{TK}, I prove that it does indeed satisfy the scalar field equation to the order specified in Equation (\ref{ScSField}).

The differential operator of the scalar field equation becomes, in THZ coordinates:
\begin{equation}
\begin{split}
\sqrt{-g} \nabla^{a} \nabla_{a} \psi^{\tx{S}} &= \partial_{a} (\eta^{ab} \partial_{b} \psi^{\tx{S}} ) - \partial_{a} (\bar{H}^{ab} \partial_{b} \psi^{\tx{S}}) \\
				&= \eta^{ab} \partial_{a} \partial_{b} \psi^{\tx{S}} - \bar{H}^{ij} \partial_{i} \partial_{j} \psi^{\tx{S}} - 2 \bar{H}^{it} \partial_{(i} \partial_{t)} \psi^{\tx{S}} - \bar{H}^{tt} \partial_{t} \partial_{t} \psi^{\tx{S}}. \\
\end{split}
\end{equation}  
If the field $\psi^{\tx{S}} = (q/\rho)$ is substituted into this equation, the first term gives the expected $\delta$-function singularity and the last two terms vanish since $\rho$ does not depend on the time $t$.
An explicit calculation shows that for the second term, $_{2}\bar{H}^{ij} \times \partial_i \partial_j \psi$ is equal to zero.
The remainder $_{3}\bar{H}^{ij}$ gives a term that scales as $O(\frac{\rho}{\mathcal{R}^{4}})$.
So
\begin{equation}
\sqrt{-g} \nabla^{a} \nabla_{a} \big ( \frac{q}{\rho} \big ) = - 4 \pi q \delta^{3}(x^{i}) + O(\frac{\rho}{\mathcal{R}^4}) \tx{, for } \frac{\rho}{\mathcal{R}} \to 0.
\end{equation}
Consequently, for the remainder to be removed a term of $O(\frac{\rho^3}{\mathcal{R}^4})$ must be added to $q/\rho$.
So $\psi^{\tx{S}} = q/\rho + O(\frac{\rho^3}{\mathcal{R}^4})$ is an inhomogeneous solution of the scalar wave equation and the error in this approximation is $C^2$.

\section{Scalar Field of a Dipole}
\label{SFD}

The calculation of the singular scalar field generated by a dipole moving on a geodesic in a Schwarzschild background is presented in this section.
The dipole moment is assumed to have a random orientation and its THZ components are denoted as $K_{a} = (0, K_x, K_y, K_z)$.

In this and the following sections, the subscripts (or superscripts) (0), (1) and (2) are used to indicate the order of significance of each term or component.
The subscript (0) refers to the most dominant contribution, the subscript (1) refers to the next most significant correction, which is calculated for the singular fields, and the subscript (2) refers to the next correction, the order of which is predicted for the singular fields.
It is important to realize that multiplying two terms of order (1) does not necessarily give a term of order (2), because of the fact that there is no $O(\frac{\rho}{\mathcal{R}})$ correction to the metric (see Equation (\ref{TaylorExpandMetric})) and the fact that the first correction to the metric is of $O(\frac{\rho^2}{\mathcal{R}^2})$ while the second is of $O(\frac{\rho^3}{\mathcal{R}^3})$.

The scalar field of a dipole can be thought of as having the form
\begin{equation}
\Psi^{\tx{S}} = \Psi^{\tx{S}}_{(0)} + \Psi^{\tx{S}}_{(1)} + \Psi^{\tx{S}}_{(2)}
\end{equation}
and it obeys the differential equation
\begin{equation}
\nabla_{(0+1+2)}^{2} \Psi^{\tx{S}} = \nabla_{(0+1+2)}^{2} ( \Psi^{\tx{S}}_{(0)} + \Psi^{\tx{S}}_{(1)} + \Psi^{\tx{S}}_{(2)}) = -4 \pi \varrho
\label{Eq01}
\end{equation}
where the source term is given by
\begin{equation}
\varrho = - \int K_i \nabla^i \big [ \delta^4(p-p'(\tau)) \big ] \sqrt{-g} \: d\tau.
\end{equation}

The zeroth-order term is the scalar field generated by a dipole that is stationary at the origin of a Cartesian coordinate system and it obeys the lowest order differential equation
\begin{equation}
\nabla_{(0)}^{2} \Psi^{\tx{S}}_{(0)} = -4 \pi \varrho.
\end{equation}
It is equal to
\begin{equation}
\Psi^{\tx{S}}_{(0)} = \frac{K_i x^i}{\rho^3} = \frac{K_x x + K_y y + K_z z}{(x^2 + y^2 + z^2)^{3/2}}.
\end{equation}

The first-order term obeys the differential equation derived from Equation (\ref{Eq01})
\begin{equation}
\nabla_{(0)}^{2} \Psi^{\tx{S}}_{(1)} + \nabla_{(1)}^{2} \Psi^{\tx{S}}_{(0)} = 0 \: \: \Rightarrow \: \: 
\nabla_{(0)}^{2} \Psi^{\tx{S}}_{(1)} = - \nabla_{(1)}^{2} \Psi^{\tx{S}}_{(0)}
\label{PsiD1}
\end{equation}
which means that the $\nabla^{2}_{(1)}$ of the zeroth-order part of the field is the source term in the scalar differential equation for the first-order part of the field.
In general, that source term is expected to contain the $\mathcal{E}$'s and the $\mathcal{B}$'s and to give the first-order correction coming from the dipole's motion on the Schwarzschild geodesic.
In this case, Equation (\ref{PsiD1}) gives that
\begin{equation}
\big ( \frac{\partial^2}{\partial x^2} + \frac{\partial^2}{\partial y^2} + \frac{\partial^2}{\partial z^2} \big ) \Psi_{(1)}^{\tx{S}} = 0
\end{equation}
so the first order correction to the field can be set equal to zero.

The source for the next order correction comes from the part of the $\nabla^2$ that is of $O(\frac{\rho^3}{\mathcal{R}^3})$ acting on the zeroth-order scalar field $\Psi^{\tx{S}}_{(0)}$. 
The differential equation is of the form
\begin{equation}
\partial_i \partial_j \Psi^{\tx{S}}_{(2)} = O(\frac{\rho^3}{\mathcal{R}^3}) \times \partial_i \partial_j \Psi^{\tx{S}}_{(0)} \: \sim O(\frac{1}{\rho \mathcal{R}^3}) .
\end{equation}
That means that the next order term must be of $O(\frac{\rho}{\mathcal{R}^3})$.

Finally, the singular scalar field of a dipole moving on a Schwarzschild geodesic is equal to
\begin{equation}
\Psi^{\tx{S}} = \frac{K_x x + K_y y + K_z z}{\rho^3} + O(\frac{\rho}{\mathcal{R}^3}).
\end{equation}

\section{Electromagnetic Potential of a Charged Particle}
\label{SFECP}

In this section, the singular electromagnetic potential generated by a charge $q$ moving on a Schwarzchild geodesic is calculated.
Its general form is
\begin{equation}
A_{\tx{S}}^{a} = A_{\tx{S} (0)}^{a} + A_{\tx{S} (1)}^{a} + A_{\tx{S} (2)}^{a}.
\end{equation}
Since in the vacuum background the Ricci tensor is
\begin{equation}
R_{ab} =0,
\end{equation}
this electromagnetic potential, as well as all those calculated in this chapter, obey the vacuum Maxwell's equations in curved spacetime
\begin{equation}
\nabla^{2}_{(0+1+2)} A_{\tx{S} (0+1+2)}^{a} = - 4 \pi J^{a}.
\label{DelACharge}
\end{equation}
The zeroth-order term obeys the differential equation
\begin{equation}
\nabla_{(0)}^{2}  A_{\tx{S} (0)}^{a} = - 4 \pi J^{a}
\label{DelACharge0}
\end{equation}
the source term being
\begin{equation}
J^{a} = \Big ( q \int (-g)^{-\frac{1}{2}} \delta^{4}[p-p'(\tau)] \: d\tau,\: 0,\:  0,\:  0 \Big ).
\end{equation}
The solution of this differential equation is the well-known Coulomb electromagnetic potential
\begin{equation}
A_{\tx{S} (0)}^{a} = \Big ( \frac{q}{\rho}, \: 0,\: 0, \: 0 \Big ).
\end{equation}

The first-order correction to this electromagnetic potential obeys the differential equation derived from Equation (\ref{DelACharge})
\begin{equation}
\nabla_{(0)}^{2} A_{\tx{S} (1)}^{a} = - \nabla_{(1)}^{2} A_{\tx{S} (0)}^{a}.
\end{equation}
The first-order part of $\nabla^{2}$ (which contains the $
\mathcal{E}$'s and the $\mathcal{B}$'s) acting on the zeroth-order electromagnetic potential is the source term for the first-order correction.
Substituting the THZ components of $A_{\tx{S} (1)}^{a}$
\begin{equation}
A^{a}_{\tx{S} (1)} = ( A^{t}_{\tx{S} (1)}, A^{x}_{\tx{S} (1)}, A^{y}_{\tx{S} (1)}, A^{z}_{\tx{S} (1)} )
\end{equation}
into the differential equation results in four differential equations, one for each one of these components.
Each equation relates a specific sum of second derivatives of a component to a sum of terms of the form:
\begin{equation}
\begin{split}
&\frac{q \mathcal{E}_{..} x^{.} x^{.} x^{.} x^{.}}{\rho^5} \: \: \tx{for the $t$-component,} \\
&\frac{q \mathcal{B}_{..} x^{.} x^{.}}{\rho^3} \: \: \tx{for the spatial components}, \\
\end{split}
\end{equation}
where the dots denote appropriate indices.

Solving these four equations is straightforward, once one notices that the solution should have the form
\begin{equation}
\begin{split}
&\frac{q}{\rho} \mathcal{E}_{ij} x^i x^j \: \quad \quad \tx{for the $t$-component,} \\
&\frac{q}{\rho} \epsilon^{p}_{\: \: ij} B^{i}_{\: k} x^k x^j \: \: \tx{for the p-spatial component,} \\
\end{split}
\end{equation}
each term multiplied by an appropriate algebraic factor.
Substituting these expressions into the differential equations gives a set of simple algebraic equations for these factors, which can be easily solved to give the final expression for the first-order correction:
\begin{equation}
A_{\tx{S} (1)}^{a} = \Big ( - \frac{q}{2 \rho} \mathcal{E}_{ij} x^i x^j, \: \frac{q}{2 \rho} \epsilon^{x}_{\: \: ij} \mathcal{B}^{i}_{\: k} x^j x^k, \: \frac{q}{2 \rho} \epsilon^{y}_{\: \: ij} \mathcal{B}^{i}_{\: k} x^j x^k, \: \frac{q}{2 \rho} \epsilon^{z}_{\: \: ij} \mathcal{B}^{i}_{\: k} x^j x^k \Big ).
\end{equation}

The next order correction comes from the part of the $\nabla^2$ that is of $O(\frac{\rho^3}{\mathcal{R}^3})$ acting on the zeroth-order electomagnetic pothential.
It gives a differential equation for each component of $A_{\tx{S} (2)}^{a}$ of the form
\begin{equation}
\partial_i \partial_j A_{\tx{S} (2)}^{a} = O(\frac{\rho^3}{\mathcal{R}^3}) \times \partial_i \partial_j A_{\tx{S} (0)}^{a} \: \: \sim \: \: O(\frac{1}{\mathcal{R}^3}),
\end{equation}
which indicates that the next order correction must be of $O(\frac{\rho^2}{\mathcal{R}^3})$.
It is noteworthy that the first-order correction $A_{\tx{S} (1)}^{a}$ does not appear in the equation for the second-order correction. 
That is because it only shows up in terms that involve the $O(\frac{\rho^2}{\mathcal{R}^2})$ part of the metric, which are of the form
\begin{equation}
O(\frac{\rho^2}{\mathcal{R}^2}) \times \partial_i \partial_j A_{\tx{S} (1)}^{a} \: \: \sim \: \: O(\frac{\rho}{\mathcal{R}^4})
\end{equation}
and must be included in a higher-order calculation.

Finally, the singular electromagnetic potential of a charge $q$ that is moving on a geodesic in a Schwarzschild background is equal to
\begin{equation}
A_{\tx{S}}^{a} = \frac{q}{\rho} \Big ( [1-\frac{1}{2} \mathcal{E}_{ij} x^i x^j ], \: \frac{1}{2}
\epsilon^{x}_{\: \: ij} \mathcal{B}^{i}_{\: k} x^j x^k, \: \frac{1}{2} \epsilon^{y}
_{ij} \mathcal{B}^{i}_{\: k} x^j x^k, \: \frac{1}{2}  \epsilon^{z}_{\: \: ij} \mathcal{B}^{i}_{\:
 k} x^j x^k \Big ) + O(\frac{\rho^2}{\mathcal{R}^3}).
\end{equation}

\section{Electromagnetic Potential of an Electric Dipole}
\label{SFED}

The calculation of the singular electromagnetic potential of an electric dipole is presented in this section.
The dipole moment is assumed to point at some random direction and its THZ components are $q^{a} = (0, q^x, q^y, q^z)$.

The singular electromagnetic potential can be written as
\begin{equation}
A^{a}_{\tx{S}} = A^{a}_{\tx{S} (0)} + A^{a}_{\tx{S} (1)} + A^{a}_{\tx{S} (2)}
\end{equation}
and obeys the vacuum Maxwell's equations
\begin{equation}
\nabla^{2}_{(0+1+2)} A^{a}_{\tx{S} (0+1+2)} = - 4 \pi J^{a}
\label{VectorAE}
\end{equation}
where the source is
\begin{equation}
J^{a} = \bigg (- \int q_i \nabla^i \big [ \delta^4(p-p'(\tau)) \big ] \sqrt{-g} \: d\tau, \: 0, \: 0, \: 0 \bigg ).
\end{equation}

The zeroth-order term is the electromagnetic potential generated by an electric dipole that is stationary at the origin of the Cartesian coordinate system, so it obeys the differential equation
\begin{equation}
\nabla^{2}_{(0)} A^{a}_{\tx{S} (0)} = - 4 \pi J^{a}.
\end{equation}
The solution to this equation is well-known and has only a $t$-component
\begin{equation}
A^{a}_{\tx{S} (0)} = \bigg ( \frac{q_i x^i}{\rho^{3}}, \: 0, \: 0, \: 0 \bigg ).
\end{equation}

The first-order correction to the electromagnetic potential obeys the differential equation
\begin{equation}
\nabla^2_{(0)} A^{a}_{\tx{S} (1)} = - \nabla^2_{(1)} A^{a}_{\tx{S} (0)}.
\label{OutOfNames}
\end{equation}
If its THZ components are assumed to be
\begin{equation}
A^{a}_{\tx{S} (1)} = ( A^{t}_{\tx{S} (1)}, A^{x}_{\tx{S} (1)}, A^{y}_{\tx{S} (1)}, A^{z}_{\tx{S} (1)})
\end{equation}
and are substituted into Equation (\ref{OutOfNames}),
the differential equation for $A_{\tx{S} (1)}^{a}$ becomes a set of four second-order differential equations for these components.
Each differential equation relates a sum of second derivatives of a component to a sum of terms of the form
\begin{equation}
\begin{split}
&\frac{q_. \mathcal{E}_{..} x^. x^. x^. x^. x^.}{\rho^7} \: \: \tx{for the $t$- component}, \\
&\frac{q_. \mathcal{B}_{..} x^. x^. x^. x^. x^.}{\rho^7} \: \: \tx{for the spatial components}, \\
\end{split}
\end{equation}
where, again, the dots denote appropriate indices.

These differential equations indicate that the solution should be equal to a sum of the terms 
\begin{equation}
\frac{q_i \mathcal{E}_{jk} x^i x^j x^k}{\rho^3} \: \tx{and} \: \frac{q^i \mathcal{E}_{ij} x^j x_k x^k}{\rho^3}
\end{equation}
for the $t$-component and a sum of the terms
\begin{equation}
\begin{split}
&\frac{\epsilon^{p}_{\: ij} q^{i} \mathcal{B}^{j}_{\: l} x^l x_k x^k}{\rho^3}, \: \frac{\epsilon^{p}_{\: ij} q^{i} \mathcal{B}_{kl} x^j x^k x^l}{\rho^3}, \:\frac{\epsilon^{p}_{\: ij} q^{k} \mathcal{B}^{i}_{\: k} x^j x_l x^l}{\rho^3}, \\
&\frac{\epsilon^{p}_{\: ij} q^{k} \mathcal{B}^{i}_{\: l} x^j x^k x^l}{\rho^3}, \: \frac{\epsilon_{ijk} q^{i} \mathcal{B}^{pj} x^k x_l x^l}{\rho^3}, \: \frac{\epsilon^{ijk} q^{i} \mathcal{B}^{j}_{\: l} x^p x^k x^l}{\rho^3},\\
\end{split}
\end{equation}
for the $p$-spatial component,
each term multiplied by an appropriate numerical coefficient.
In fact, using Equations (\ref{EBRel2}) and (\ref{EBRel4}), the last two terms that are expected to show up in the solution for the $p$-component can be eliminated in favor of the remaining four.
Substituting these expressions into the differential equations gives simple abgebraic equations for the coefficients.
The final expressions for the components of $A^{a}_{\tx{S} (1)}$ are
\begin{equation}
\begin{split}
&A^{t}_{\tx{S} (1)} = -\frac{1}{2} \frac{q_i \mathcal{E}_{jk} x^i x^j x^k}{\rho^3}, \\
&A^{p}_{\tx{S} (1)} = \frac{\epsilon^{p}_{\: ij} q^{i} \mathcal{B}^{j}_{\: l} x^l x_k x^k}{2 \rho^3} + \frac{\epsilon^{p}_{\: ij} q^{k} \mathcal{B}^{i}_{\: k} x^j x_l x^l}{2 \rho^3} + \frac{\epsilon^{p}_{\: ij} q^{k} \mathcal{B}^{i}_{\: l} x^j x_k x^l}{ 2 \rho^3}. \\
\end{split}
\end{equation}

The order of the next term in the expansion of  the electromagnetic potential can be predicted.
It is the solution of the differential equation whose source term comes from the $O(\frac{\rho^3}{\mathcal{R}^3})$ part of the $\nabla^2$ acting on $A^{a}_{\tx{S} (0)}$.
That differential equation has the form
\begin{equation}
\partial_i \partial_j A^{a}_{\tx{S} (2)} = O(\frac{\rho^3}{\mathcal{R}^3}) \partial_i \partial_j A^{a}_{\tx{S} (0)} \: \: \sim \: \: O(\frac{1}{\rho \mathcal{R}^3}).
\end{equation}
Consequently, the second-order correction  must be of $O(\frac{\rho}{\mathcal{R}^3})$.
The terms that involve the first-order correction $A_{\tx{S} (1)}^{a}$ do not contribute to the equation for the second-order correction, because they involve the $O(\frac{\rho^2}{\mathcal{R}^2})$ part of the metric and that results in terms of $O(\frac{1}{\mathcal{R}^4})$.

Finally, the singular electromagnetic potential for an electric dipole moving on a Schwarzschild geodesic is equal to
\begin{equation}
\begin{split}
&A^t_{\tx{S}} = \frac{1}{\rho^3} \bigg [ q_i x^i -\frac{1}{2} q_i \mathcal{E}_{jk} x^i x^j x^k \bigg ] + O(\frac{\rho}{\mathcal{R}^3}), \\
&A^p_{\tx{S}} = \frac{1}{2 \rho^3} \bigg [ \epsilon^{p}_{\: ij} q^{i} \mathcal{B}^{j}_{\: l} x^l x_k x^k + \epsilon^{p}_{\: ij} q^{k} \mathcal{B}^{i}_{\: k} x^j x_l x^l + \epsilon^{p}_{\: ij} q^{k} \mathcal{B}^{i}_{\: l} x^j x^k x^l \big  ] + O(\frac{\rho}{\mathcal{R}^3}). \\
\end{split}
\end{equation}

\section{Electromagnetic Potential of a Magnetic Dipole}
\label{SFMD}

In this section, the singular electromagnetic potential of a magnetic dipole moving on a geodesic of the Schwarzschild background is calculated.
The magnetization $m^{a}$ is assumed to point at some random direction and its THZ components are $m^{a} = (0, m^x, m^y, m^z)$.

The singular electromagnetic potential can be written as
\begin{equation}
A^{a}_{\tx{S}} = A^{a}_{\tx{S} (0)} + A^{a}_{\tx{S} (1)} + A^{a}_{\tx{S} (2)}
\end{equation}
and obeys the vacuum Einstein equations
\begin{equation}
\nabla^{2}_{(0+1+2)} A^{a}_{\tx{S} (0+1+2)} = - 4 \pi J^{a} 
\label{VectorAD}
\end{equation}
where the source term is
\begin{equation}
\begin{split}
& J^{0} = 0, \\
&J^{q} = \int \epsilon^{qij} \partial_i \Big [ m_j \delta^4(p - p'(\tau)) \Big ] \sqrt{-g} \: d\tau.\\ 
\end{split}
\end{equation}

The zeroth-order term is the electromagnetic potential generated by a magnetic dipole that is stationary at the origin of a Cartesian coordinate system and is the solution of the differential equation
\begin{equation}
\nabla^{2}_{(0)} A^{a}_{\tx{S} (0)} = - 4 \pi J^{a}.
\end{equation}
Its THZ components are
\begin{equation}
A^{a}_{\tx{S} (0)} = ( 0, \epsilon^x_{\: \: ij} \frac{m^i x^j}{\rho^3}, \epsilon^y_{\: \: ij} \frac{m^i x^j}{\rho^3},\epsilon^z_{\: \: ij} \frac{m^i x^j}{\rho^3}).
\end{equation}

The first-order term obeys the differential equation derived from Maxwell's Equations (\ref{VectorAD})
\begin{equation}
\nabla^{2}_{(0)} A^{a}_{\tx{S} (1)} = - \nabla^{2}_{(1)} A^{a}_{\tx{S} (0)}
\label{AS1Dipole}
\end{equation}
which indicates that the source term for $A^{a}_{\tx{S} (1)}$ is the $\nabla^{2}_{(1)}$ of the zeroth-order part of the potential.
The THZ components of $A^{a}_{\tx{S} (1)}$ are
\begin{equation}
A^{a}_{\tx{S} (1)} = ( A^{t}_{\tx{S} (1)}, A^{x}_{\tx{S} (1)}, A^{y}_{\tx{S} (1)}, A^{z}_{\tx{S} (1)} )
\end{equation}
and when substituted into Equation (\ref{AS1Dipole}) the result is a set of four second-order differential equations, one for each one of those four components.
Each equation relates a sum of second derivatives of one component to the source term which consists of terms of the form:
\begin{equation}
\begin{split}
& \frac{m_{.} \mathcal{B}_{..} x^{.} x^{.} x^{.}}{\rho^5} \: \quad \quad \tx{for the $t$-component}, \\
&\frac{m_{.} \mathcal{E}_{..} x^{.} x^{.} x^{.} x^{.} x^{.}}{\rho^7} \: \: \tx{ for the spatial components}, \\
\end{split}
\end{equation}
where the dots denote the appropriate indices.

Solving these differential equations is tedious but not difficult, because each equation involves only one component of $A^{a}_{\tx{S} (1)}$ and only one of the two tensors $\mathcal{E}$ and $\mathcal{B}$.
A careful look at the equations indicates that the solution should be a sum of the terms
\begin{equation}
\frac{m_{i} \mathcal{B}^{i}_{\: \: j} x^{j} x_{k} x^{k}}{\rho^3} \quad {\tx{and}} \quad \frac{m_{i} \mathcal{B}_{jk} x^{i} x^{j} x^{k}}{\rho^3}
\end{equation}
for the $t$-component and a sum of the terms
\begin{equation}
\begin{split}
&\frac{\epsilon^{p}_{\:ij} m^{i} \mathcal{E}^{j}_{\: \: k} x^{k} x_{l} x^{l}}{\rho^3}, \qquad \frac{\epsilon^{p}_{ \: ij} m^{i} \mathcal{E}_{kl} x^{j} x^{k} x^{l}}{\rho^3}, \qquad \frac{\epsilon^{p}_{\:ij} m^{k} \mathcal{E}^{i}_{\: \: k} x^{j} x_{l} x^{l}}{\rho^3} \\
& \frac{\epsilon^{p}_{\:ij} m^{k} \mathcal{E}^{i}_{\: \: l} x^{j} x_{k} x^{l}}{\rho^3}, \qquad \frac{\epsilon_{ijk} m^{i} \mathcal{E}^{pj} x^{k} x_{l} x^{l}}{\rho^3}, \qquad \frac{\epsilon_{ijk} m^{i} \mathcal{E}^{j}_{\: \: l} x^{k} x^{l} x^{p}}{\rho^3} \\
\end{split}
\end{equation}
for the $p$-spatial component, with appropriate numerical factors in front of each term so that the equations are satisfied.
Using Equations (\ref{EBRel1}) and (\ref{EBRel3}), the first and last terms expected to show up in the final expression for the $p$-component can be eliminated,
since they can be expressed as linear combinations of the remaining four terms.
Substituting these expressions into the differential equations gives a system of four algebraic equations for those factors.
Solving these algebraic equations is trivial.
The result is that the first-order components of the electromagnetic potential are:
\begin{equation}
\begin{split}
&A^{t}_{\tx{S} (1)} = \frac{1}{\rho^3} \big [ \frac{1}{6} m_{i} \mathcal{B}_{jk} x^{i} x^{j} x^{k} - \frac{2}{3} m^{i} \mathcal{B}_{ij} x^{j} x^{k} x_{k} \big ] \\
&A^{p}_{\tx{S} (1)} = \frac{1}{\rho^3} \big [ \epsilon^{p}_{\: \: ij} m^{i} \mathcal{E}_{kl} x^{j} x^{k} x^{l} - \frac{1}{2} \epsilon^{p}_{\: \: ij} m_{k} \mathcal{E}^{i}_{\: \: l} x^{j} x^{k} x^{l} - \frac{1}{2} \epsilon_{ijk} m^{i} \mathcal{E}^{pj} x^k x_l x^l \big ] . \\
\end{split}
\end{equation}

The order of the next correction to the singular electromagnetic potential can be predicted. 
It is the solution of the differential equation that has the $O(\frac{\rho^3}{\mathcal{R}^3})$ part of the $\nabla^2$ acting on the zeroth-order electromagnetic potential as the source term.
Specifically, it looks like
\begin{equation}
\partial_i \partial_j A_{\tx{S} (2)}^{a} = O(\frac{\rho^3}{\mathcal{R}^3}) \times \partial_i \partial_j A_{\tx{S} (0)}^{a} \: \: \sim \: \: O(\frac{1}{\rho \mathcal{R}^3}),
\end{equation}
meaning that the $A_{\tx{S} (2)}^{a}$ correction is of $O(\frac{\rho}{\mathcal{R}^3})$.
As in the previously studied cases, the terms that involve the first-order correction $A_{\tx{S} (1)}^{a}$ do not contribute to this equation, since they involve the $O(\frac{\rho^2}{\mathcal{R}^2})$ part of the metric which results in terms of $O(\frac{1}{\mathcal{R}^4})$.

Finally, the singular electromagnetic potential for a magnetic dipole moving on a Schwarzschild geodesic is equal to
\begin{equation}
\begin{split}
&A^{t}_{\tx{S}} = \frac{1}{\rho^3} \big [ \frac{1}{6} m_{i} \mathcal{B}_{jk} x^{i} x^{j} x^{k} -
 \frac{2}{3} m^{i} \mathcal{B}_{ij} x^{j} x^{k} x_{k} \big ]  + O(\frac{\rho}{\mathcal{R}^3}) \\
&A^{p}_{\tx{S}} =  \epsilon^{p}_{\: ij} \frac{m^i x^j}{\rho^3} + \big [ \epsilon^{p}_{\: \: ij} \frac{m^{i} \mathcal{E}_{kl} x^{j} x
^{k} x^{l}}{\rho^3} - \frac{1}{2} \epsilon^{p}_{\: \: ij} \frac{m_{k} \mathcal{E}^{i}_{\: \: l} x^{j} x^{k} x^{l}}{\rho^3} -
 \frac{1}{2} \epsilon_{ijk} \frac{m^{i} \mathcal{E}^{pj} x^k x_l x^l}{\rho^3} \big ] + O(\frac{\rho}{\mathcal{R}^3}). \\
\end{split}
\end{equation}

\section{Gravitational Field of a Spinning Particle}
\label{SFSP}

The calculation of the singular gravitational field of a spinning particle moving on a Schwarzschild geodesic is presented in this section.
The particle is assumed to have a small angular momentum pointing at some random direction $A^{a} = (0, A^x, A^y, A^z)$ in THZ coordinates.

The singular gravitational field can be written as
\begin{equation}
h_{\tx{S} \: ab} = h^{(0)}_{\tx{S} \: ab} + h^{(1)}_{\tx{S} \: ab} + h^{(2)}_{\tx{S} \: ab}  
\end{equation}
and obeys the linearized Einstein equations
\begin{equation}
\nabla^2_{(0+1+2)} \bar{h}^{(0+1+2)}_{\tx{S} \: ab} + 2 R_{(0+1+2) \: a \: \: b}^{\qquad \quad \: \: c \: \: d} \bar{h}^{(0+1+2)}_{\tx{S} \: cd} = - 16 \pi T_{ab}
\label{LinEinstein}
\end{equation}
where
\begin{equation}
\bar{h}_{\tx{S} \: ab} = h_{\tx{S} \: ab} - \frac{1}{2} g_{ab} h^{\: \: \: \: c}_{\tx{S} \: \: c}
\end{equation}
is the trace-reversed version of $h_{\tx{S} \: ab}$.

The zeroth-order part of the singular field is the gravitational field generated by a particle with angular momentum $A^{a}$ that is stationary at the origin of a Cartesian coordinate system.

For the angular momentum pointing along the z-axis, that is the well-known Kerr solution with the mass set equal to zero.  
Since the angular momentum is assumed to be small and the effects of the mass of the particle are not taken into account, only the terms of the Kerr metric that are linear in the angular momentum need to be considered.
In the Boyer-Lindquist coordinates $(t_{\tx{BL}}, r, \theta, \phi)$ around the spinning particle, $h^{\tx{Z}}_{\tx{S} \: ab}$ (where the superscript Z denotes that this is the part of the zeroth-order gravitational field  coming only from the $z$-component of the angular momentum) is equal to
\begin{equation}
h^{\tx{Z}}_{\tx{S} \: ab} = - \frac{2}{r} A^z \sin^2 \theta
\left(\begin{array}{cccc}
0 & 0 & 0 & 1 \\
0 & 0 & 0 & 0 \\
0 & 0 & 0 & 0 \\
1 & 0 & 0 & 0
\end{array} \right).
\end{equation}
Using \textsc{Grtensor}, this expression can be easily converted into the equivalent expression in THZ coordinates
\begin{equation}
h^{\tx{Z}}_{\tx{S} \: ab} = 2 A^z \frac{1}{\rho^3}
\left(\begin{array}{cccc}
0 & y & - x & 0 \\
y & 0 & 0 & 0 \\
- x & 0 & 0 & 0 \\
0 & 0 & 0 & 0
\end{array} \right).
\end{equation}
The relationships between the Boyer-Lindquist and the THZ coordinates used for this conversion are the usual relationships between the spherical and the Cartesian coordinates, namely
\begin{equation}
\begin{split}
t &= t_{\tx{BL}} \\ 
r &= \sqrt{x^2+y^2+z^2} \\
\theta &= \arctan \frac{\sqrt{x^2+y^2}}{z} \\
\phi &= \arctan \frac{y}{x}. \\
\end{split}
\label{BLTHZz}
\end{equation}
This is sufficient because the corrections to these relationships that involve the angular momentum would give terms of higher order in the angular momentum and must be ignored in this analysis,
since only first-order terms in the angular momentum are kept.

The components of the angular momentum along the $x$ and $y$ axes must be treated separately, 
because of the axial symmetry of the Kerr metric.
The analyses for the angular momentum being along the $x$ and $y$ axes are very similar to that for the angular momentum being along the $z$ axis and the only change comes from the different form that the relationships (\ref{BLTHZz}) have.
Specifically, when the angular momentum points along the $x$ axis, the axial symmetry is around the $x$ axis and the relationships used are
\begin{equation}
\begin{split}
t &= t_{\tx{BL}} \\ 
r &= \sqrt{x^2+y^2+z^2} \\
\theta &= \arctan \frac{\sqrt{y^2+z^2}}{x} \\  
\phi &= \arctan \frac{z}{y}. \\
\end{split}
\label{BLTHZx}
\end{equation}
For the angular momentum pointing along the $y$ axis, the axial symmetry is around the $y$ axis and the relationships are
\begin{equation}
\begin{split}
t &= t_{\tx{BL}} \\ 
r &= \sqrt{x^2+y^2+z^2} \\
\theta &= \arctan \frac{\sqrt{x^2+z^2}}{y} \\ 
\phi &= \arctan \frac{x}{z}. \\
\end{split}
\label{BLTHZy}
\end{equation}
Adding all the contributions that result from this analysis, the zeroth-order singular gravitational field becomes
\begin{equation}
h^{(0)}_{\tx{S} \: ab} = \frac{2}{\rho^3}
\left(\begin{array}{cccc}
0 &  (-A^y z + A^z y) & (A^x z - A^z x) & (-A^x y + A^y x) \\ 
(-A^y z + A^z y) & 0 & 0 & 0 \\
(A^x z - A^z x) & 0 & 0 & 0 \\ 
(-A^x y + A^y x) & 0 & 0 & 0 
\end{array} \right).
\end{equation}

The first-order correction to this gravitational field obeys the differential equation derived from Equation (\ref{LinEinstein}):
\begin{equation}
\nabla^2_{(0)} \bar{h}^{(1)}_{\tx{S} \: ab} + 2 R_{(0) \: a \: \: b}^{\quad \: \:
 c \: \: d} \bar{h}^{(1)}_{\tx{S} \: cd} =
-\nabla^2_{(1)} \bar{h}^{(0)}_{\tx{S} \: ab} - 2 R_{(1) \: a \: \: b}^{\quad \: \: c \: \: d} \bar{h}^{(0)}_{\tx{S} \: cd} 
\end{equation}
so the first-order $\nabla^2$ and the first-order Riemann tensor acting on the zeroth-order solution give the source for the first-order correction.
The first-order correction must be a symmetric tensor so it is assumed to be equal to
\begin{equation}
h^{(1)}_{\tx{S} \: ab} =
\left(\begin{array}{cccc}
h^{(1)}_{tt} & h^{(1)}_{tx} & h^{(1)}_{ty} & h^{(1)}_{tz} \\
h^{(1)}_{tx} & h^{(1)}_{xx} & h^{(1)}_{xy} & h^{(1)}_{xz} \\
h^{(1)}_{ty} & h^{(1)}_{xy} & h^{(1)}_{yy} & h^{(1)}_{yz} \\
h^{(1)}_{tz} & h^{(1)}_{xz} & h^{(1)}_{yz} & h^{(1)}_{zz} 
\end{array} \right).
\end{equation}
Substituting it into the first-order equation results in  10 differential equations.
There is one set of four differential equations for the four diagonal components, each equation containing all four diagonal components.
There is also one differential equation for each one of the $t-i$ components and one differential equation for each one of the $i-j$ components, for $i \neq j$.
In each equation, a sum of second derivatives of components is related to a sum of terms of the form
\begin{equation}
\begin{split}
& \frac{A_{.} \mathcal{B}_{..} x^. x^. x^. x^. x^.}{\rho^7}, \tx{for the $t-t$ and $p-q$ components}, \\
& \frac{A_{.} \mathcal{E}_{..} x^. x^. x^. x^. x^.}{\rho^7}, \tx{for the $t-p$ components}, \\
\end{split}
\end{equation}
where the dots denote the appropriate indices for each term.

Solving the differential equations in this case is slightly more complicated than in the previous cases, mainly because of the fact that four of them involve all diagonal components rather than only one of them.
Still, the process becomes significantly easier if one notices that the $t-t$ component must be an appropriate sum of terms of the form
\begin{equation}
\frac{A_{i} \mathcal{B}_{jk} x^i x^j x^k}{\rho^3} \qquad \tx{and} \qquad \frac{A^i \mathcal{B}_{ij} x^j x_k x^k}{\rho^3},
\end{equation}
each $t-p$ component must be a sum of terms of the form
\begin{equation}
\begin{split}
&\frac{\epsilon_{pij} A^i \mathcal{E}^{j}_{\: l} x^l x_k x^k}{\rho^3}, \: \frac{\epsilon_{pij} A^i \mathcal{E}_{lk} x^j x^l x^k}{\rho^3}, \: \frac{\epsilon_{pij} A^l \mathcal{E}^{i}_{\: l} x^j x_k x^k}{\rho^3}, \\
&\frac{\epsilon_{pij} A_l \mathcal{E}^{i}_{\: k} x^j x^l x^k}{\rho^3}, \: \frac{\epsilon_{ijl} A^i \mathcal{E}^{j}_{\: p} x^l x_k x^k}{\rho^3}, \: \frac{\epsilon_{ijl} A^i \mathcal{E}^{j}_{\: k} x^p x^l x^k}{\rho^3}, \\
\end{split}
\end{equation}
and each $p-q$ spatial component must be a sum of terms of the form
\begin{equation}
\begin{split}
& \frac{A_p \mathcal{B}_{qk} x^k x_l x^l}{\rho^3}, \:  \frac{A_q \mathcal{B}_{pk} x^k x_l x^l}{\rho^3}, \:  \frac{A_p \mathcal{B}_{kl} x_q x^k x^l}{\rho^3}, \:  \frac{A_q \mathcal{B}_{kl} x_p x^k x^l}{\rho^3}, \\ 
& \frac{A_k \mathcal{B}_{pq} x^k x_l x^l}{\rho^3}, \: \frac{A^k \mathcal{B}_{kl} x_p x_q x^l}{\rho^3}, \: \frac{\eta_{pq} A_i \mathcal{B}_{jk} x^i x^j x^k}{\rho^3}, \: \frac{\eta_{pq} A^i \mathcal{B}_{ij} x^j x_k x^k}{\rho^3}, \\
& \frac{A^k \mathcal{B}_{pk} x_q x_l x^l}{\rho^3}, \: \frac{A^k \mathcal{B}_{qk} x_p x_l x^l}{\rho^3},\: \frac{A_k \mathcal{B}_{pl} x_q x^k x^l}{\rho^3}, \: \frac{A_k \mathcal{B}_{ql} x_p x^k x^l}{\rho^3} ,\\
\end{split}
\end{equation}
with appropriate numerical coefficients in front of each term.
Equations (\ref{EBRel1}) and (\ref{EBRel3}) can again be used to eliminate the first and last terms in favor of the remaining four, for the expression for the $t-p$ components.
Substituting these sums into the differential equations gives algebraic equations for the coefficients, which are fairly easy to solve.
The result is that the first-order correction to the gravitational field has components
\begin{equation}
\begin{split}
& h_{\tx{S} \: tt}^{(1)} = - 2 \frac{A_i \mathcal{B}_{jk} x^i x^j x^k}{\rho^3} \\
& h_{\tx{S} \: tp}^{(1)} = \frac{1}{\rho^3} \Big [ \epsilon_{pij} A_l \mathcal{E}^{i}_{\: k} x^j x^l x^k - \epsilon_{pij} A^i \mathcal{E}_{lk} x^j x^l x^k + 3 \epsilon_{ijl} A^i \mathcal{E}_{p}^{\: j} x^l x_k x^k \Big ] \\
& h_{\tx{S} \: pq}^{(1)} = \frac{1}{\rho^3} \Big [ 2 A_i \mathcal{B}_{pq} x^i x_j x^j + \frac{2}{3} A^i \mathcal{B}_{ij} x_p x_q x^j + 2 A^i \mathcal{B}_{i(p} x_{q)} x_j x^j - \frac{2}{3} A_i \mathcal{B}_{j(p} x_{q)} x^i x^j \\
& \qquad \quad \: \: - \frac{10}{3} A_{(p} \mathcal{B}_{q)i} x^i x_j x^j -\frac{2}{3} A_{(p} x_{q)} \mathcal{B}_{ij} x^i x^j + \frac{2}{3} \eta_{pq} \big ( A_i \mathcal{B}_{jk} x^i x^j x^k - A^i \mathcal{B}_{ij} x^j x_k x^k \big ) \Big ]. \\
\end{split}
\end{equation}

The order of the next correction can be predicted.
That correction is the solution of the differential equation
\begin{equation}
\nabla^2_{(0)} \bar{h}^{(2)}_{\tx{S} \: ab} + 2 R_{(0) \: a \: \: b}^{\quad \: \:
 c \: \: d} \bar{h}^{(2)}_{\tx{S} \: cd} =
-\nabla^2_{(2)} \bar{h}^{(0)}_{\tx{S} \: ab} - 2 R_{(2) \: a \: \: b}^{\quad \: \: c \: \: d} \bar{h
}^{(0)}_{\tx{S} \: cd}
\end{equation}
where $\nabla^2_{(2)}$ and $R_{(2) \: a \: \: b}^{\quad \: \: c \: \: d}$ come from the $O(\frac{\rho^3}{\mathcal{R}^3})$ part of the metric.
So the equations look like
\begin{equation}
\partial_i \partial_j \bar{h}^{(2)}_{\tx{S} \: ab} = O(\frac{\rho^3}{\mathcal{R}^3}) \times \partial_i \partial_j \bar{h}^{(0)}_{\tx{S} \: ab} \: \: \sim \: \: O(\frac{1}{\rho \mathcal{R}^3})
\end{equation}
and the second correction to the gravitational field is of $O(\frac{\rho}{\mathcal{R}^3})$.
Again, $\bar{h}^{(1)}_{\tx{S} \: ab}$ does not show up in the equations for the second correction, since it relates to the $O(\frac{\rho^2}{\mathcal{R}^2})$ part of the metric and results in terms of $O(\frac{1}{\mathcal{R}^4})$.

Finally, the singular gravitational field due to a spinning particle moving on a Schwarzschild geodesic is equal to
\begin{equation}
\begin{split}
& h_{\tx{S} \: tt} = - 2 \frac{A_i \mathcal{B}_{jk} x^i x^j x^k}{\rho^3} + O(\frac{\rho}{\mathcal{R}^3}) \\
& h_{\tx{S} \: tp} = -2 \frac{\epsilon_{pij} A^i x^j}{\rho^3} 
 + \frac{1}{\rho^3} \Big [ \epsilon_{pij} A_l \mathcal{E}^{i}_{\: k} x^j x^l
 x^k - \epsilon_{pij} A^i \mathcal{E}_{lk} x^j x^l x^k + 3 \epsilon_{ijl} A^i \mathcal{E}_{p}^{\: j}
 x^l x_k x^k \Big ] \\
&\qquad \quad + O(\frac{\rho}{\mathcal{R}^3}) \\
& h_{\tx{S} \: pq} = \frac{1}{\rho^3} \Big [ 2 A_i \mathcal{B}_{pq} x^i x_j x^j + \frac{2}{3}
A^i \mathcal{B}_{ij} x_p x_q x^j + 2 A^i \mathcal{B}_{i(p} x_{q)} x_j x^j - \frac{2}{3} A_i \mathcal{B}_{j(p} x_{q)} x^i x^j \\
& \quad \qquad \: \: - \frac{10}{3} A_{(p} \mathcal{B}_{q)i} x^i x_j x^j -\frac{2}{3} A_{(p} x_{q)}
\mathcal{B}_{ij} x^i x^j + \frac{2}{3} \eta_{pq} \big ( A_i \mathcal{B}_{jk} x^i x^j x^k - A^i \mathcal{B}_{ij} x^j x_k x^k \big ) \Big ] \\
& \qquad \quad + O(\frac{\rho}{\mathcal{R}^3}). \\
\end{split}
\end{equation}

\section{\textsc{Grtensor} Code}
\label{GRTCode}

The \textsc{Grtensor} code (running under \textsc{Maple}) used to derive the differential equations for the first order correction to the singular field is given in Appendix \ref{A0}.
Since the case of the gravitational field is the most complicated one,
the analysis is presented for the gravitational field generated by a spinning particle with the angular momentum pointing along the THZ $z$-axis.
The analyses for the scalar fields and the electromagnetic potentials are very similar, and can be easily deduced from that for the gravitational field.

An effort was made to keep the symbols in the code in accordance with the ones used in this chapter for the various quantities.
In the situations where that is not the case, the comments in the code should make the notation clear enough for the reader to follow.

The parameter \verb+e+ is used to keep track of the order of each term in the components of the tensors $\mathcal{E}$ and $\mathcal{B}$ and is set equal to $1$ at the end.
Throughout the calculation, only first order terms are kept.
Specifically, the Christoffel symbols and the components of the Riemann and Ricci tensors are calculated first and all their terms that are of order higher than 1 are set equal to zero.
Doing that makes the subsequent analysis significantly simpler and the running time of the code significantly shorter.

After the various quantities associated with the problem are calculated, the \emph{test} tensor \verb+hbartest(a,b)+, 
whose exact dependence on the THZ coordinates is not specified, 
is used as a trial solution in the linearized Einstein equations and the source term coming from the zeroth-order solution \verb+h0(a,b)+ is examined.
That helps identify the terms that should be expected to show up in each component of the solution \verb+hbar(a,b)+. 
Specifically, it helps determine which tensor's components, $\mathcal{E}$ (denoted as \verb+EE+ in the code) or $\mathcal{B}$ (denoted as \verb+BB+), should show up in each component of \verb+hbar(a,b)+ and gives an idea of how they should be contracted to the spatial THZ coordinates $x, y,z$. 
The terms that result from this analysis
are multiplied by algebraic factors and the appropriate sum is substituted into the linearized Einstein equations.
The result is a simple system of algebraic equations which can be easily solved to give the values of the algebraic factors. That completes the solution.

The last component of this analysis is a simple confirmation performed for the gravitational field that was calculated, that the algebraic coefficients obtained do, indeed, give the required solution.
The confirmation is simply done by replacing the initially unknown algebraic coefficients with their exact values in the expression for the solution and substituting that expression into the Einstein equations.
Despite the fact that it was not explicitly mentioned in Sections (\ref{SFD})-(\ref{SFSP}), that confirmation was performed for all singular fields and potentials that were calculated.

      \chapter{Regularization Parameters for the Scalar Field}
\label{RP}

A mode-sum regularizarion procedure for the scalar singular field is presented and implemented in this chapter.
This regularization procedure was first proposed by Barack and Ori in \cite{bo0}, where they described the calculation of the regularization parameters for the direct part of the self-force on a particle carrying a scalar charge.
The procedure  was later implemented by different groups for the calculation of the regularization parameters for the direct part of the self-force on a scalar charge on different geodesics \cite{mns,bo0,bo1,burko,BarBur,barack1} and also for non-geodesic motion \cite{barack1,burko} around a Schwarzschild black hole.
The calculation of the regularization parameters has also been performed for the direct part of the electromagnetic self-force \cite{bo2} and for the direct part of the gravitational self-force \cite{barack2,bo2,BMetal}, for arbitrary geodesics around a Schwarzschild black hole.

Even though the regularization procedure was initially described for the contribution of the direct part of the scalar field to the self-force, it can be used equally successfully for the contribution of the singular scalar field to the self-force, as was demonstrated in \cite{detmeswhi}.
In that paper, the regularization parameters for the self-force on a scalar charge in circular orbit around a Schwarzschild black hole were calculated and the self-force results ended up being in excellent agreement with the results that were derived using the direct scalar field \cite{bo0,mns,bo1,burko}.

This chapter begins with an outline of the regularization procedure for the scalar self-force.
The description closely follows that given by Barack and Ori \cite{bo0} for the direct self-force but is presented here for the singular self-force instead.
Then, the regularization parameters are calculated for the singular scalar field (rather than the scalar self-force) of a charged particle that moves on an equatorial  circular orbit in a Schwarzschild background and the results of \cite{detmeswhi} for the scalar self-force are reproduced.
Finally, the regularization parameters for the first derivative of the singular part of the self-force are also calculated.

\section{Regularization Procedure}
\label{RegProc}

As was shown in Chapter \ref{UFM}, the self-force on a particle that carries a scalar charge $q$ can be calculated from the equation
\begin{equation}
\mathcal{F}_{a}^{\tx{R}} = q \lim_{p \to p'} \nabla_{a} \psi^{\tx{R}} 
	= q \lim_{p \to p'} \nabla_{a} \big ( \psi^{\tx{ret}} - \psi^{\tx{S}} \big ) 
\label{SelfForceR}
\end{equation}
where $p'$ is the point along the worldline of the charged particle on which the self-force needs to be calculated and $p$ is a point in the vicinity of $p'$.

It is assumed that the charge $q$ is moving in a Schwarzschild background of mass $M$ and the Schwarzschild coordinates are $(t_{\tx{s}},r,\theta,\phi)$.
For the calculation of the retarded field, the source term in Poisson's Equation (\ref{scinh}) can be decomposed in terms of spherical harmonics and 
the retarded field can be written as
\begin{equation}
\psi^{\tx{ret}}(t_{\tx{s}},r,\theta,\phi) = \sum_{l=0}^{\infty} \sum_{m=-l}^{l} \psi^{\tx{ret}}_{lm}(r,t_{\tx{s}}) Y_{lm}(\theta, \phi).
\end{equation}
Then, the $lm$-components of $\psi^{\tx{ret}}$ can be calculated numerically.
That calculation is discussed in great detail in Chapter \ref{RSF}.
Here it is just noted that the
important property of the $\psi^{\tx{ret}}_{lm}$'s and of their first $r$-derivatives is that they are finite at the location of the particle, even though $\psi^{\tx{ret}}$ is singular there.

If the spherical harmonic decomposition of the singular field is also considered, Equation (\ref{SelfForceR}) becomes
\begin{equation}
\mathcal{F}_{a}^{\tx{R}} = q \lim_{p \to p'} \nabla_{a} \sum_{l=0}^{\infty} \sum_{m=-l}^{l} \big ( \psi_{lm}^{\tx{ret}} - \psi_{lm}^{\tx{S}} \big ) Y_{lm} = q \lim_{p \to p'} \nabla_{a} \sum_{l,m} \big ( \psi_{lm}^{\tx{ret}} - \psi_{lm}^{\tx{S}} \big ) Y_{lm}.
\label{SelfForceR2}
\end{equation}
It is helpful to define the multipole $l$-modes of the two contributions to the self-force, which result after performing the $m$-summation of each term individually in Equation (\ref{SelfForceR2}).
Specifically
\begin{equation}
\begin{split}
\mathcal{F}_{la}^{\tx{ret}} &= q \nabla_{a} \sum_{m} \psi^{\tx{ret}}_{lm} Y_{lm}, \\
\mathcal{F}_{la}^{\tx{S}} &= q \nabla_{a} \sum_{m} \psi^{\tx{S}}_{lm} Y_{lm}, \\
\end{split}
\end{equation}
which gives for the self-force
\begin{equation}
\mathcal{F}_{a}^{\tx{R}} =   \lim_{p \to p'} \sum_{l} \big ( \mathcal{F}_{la}^{\tx{ret}} - \mathcal{F}_{la}^{\tx{S}} \big ).
\label{FRlmodes}
\end{equation}
In Equation (\ref{FRlmodes}), the difference in the multipole $l$-modes must be taken before the summation over $l$ is performed.

From this point on, the discussion of the regularization procedure becomes specific to the problem of the scalar field $(q/\rho)$, since more detailed results are available for this case.
However, a similar analysis can be done for any other scalar field.
The goal is to find a function $h_{la}$ such that the series
\begin{equation}
\sum_{l} ( \mathcal{F}_{la}^{\tx{ret}} - h_{la} )
\end{equation}
converges.
When such a function is found, the self-force can be written as
\begin{equation}
\mathcal{F}_{a}^{\tx{R}} = \sum_{l} ( \mathcal{F}_{la}^{\tx{ret}} - h_{la} ) - E_{a}
\end{equation}
where
\begin{equation}
E_{a} \equiv \lim_{p \to p'} \sum_{l} (\mathcal{F}_{la}^{\tx{S}} - h_{la}).
\end{equation}
Because of its definition, the function $h_{la}$ should be calculated by investigating the asymptotic expansion of $\mathcal{F}_{la}^{\tx{ret}}$ for large $l$.
On the other hand, because the self-force is known to be well-defined, $\mathcal{F}_{la}^{\tx{ret}}$ and $\mathcal{F}_{la}^{\tx{S}}$ are expected to have the same large-$l$ behavior, so $h_{la}$ can be determined by the asymptotic behavior of $\mathcal{F}_{la}^{\tx{S}}$ instead.

The singular part $\mathcal{F}_{a}^{\tx{S}}$ of the self-force consists of terms of different order in the limit $p \to p'$ and it has been shown \cite{detmeswhi} that,
in principle, only the first three of those terms are expected to give non-zero contributions, for the field $(q/\rho)$.
However, for reasons that will become clear shortly, the next order terms are included and
\begin{equation}
\mathcal{F}_{a}^{\tx{S}} = \mathcal{F}_{a}^{\tx{S}(A)} + \mathcal{F}_{a}^{\tx{S}(B)} + \mathcal{F}_{a}^{\tx{S}(C)} + \mathcal{F}_{a}^{\tx{S}(D)} + \mathcal{F}_{a}^{\tx{S}(E)}. 
\end{equation}
The superscripts $A, B, C, D$ indicate the different orders,
$A$ coming from the most dominant term, $B$ from the next more dominant and so on.
The superscript $E$ refers to all terms of order higher than that of $\mathcal{F}_{a}^{\tx{S}(D)}$.
The mode-sum regularization procedure amounts to performing the spherical harmonic decomposition of each such term, which results in an expression of the form
\begin{equation}
\mathcal{F}_{a}^{\tx{S}} = \sum_{l,m} ( A_{a}^{lm} + B_{a}^{lm} + C_{a}^{lm} + D_{a}^{lm} + E_{a}^{lm}) Y_{lm}
\label{FaSseries}
\end{equation}
with $A_{a}^{lm}$ corresponding to $\mathcal{F}_{a}^{\tx{S}(A)}$, etc.
For the simple case of a scalar charge moving in a Schwarzschild background and
generating the scalar field $(q/\rho)$, the parameters $A_{a}^{lm}, B_{a}^{lm}, C_{a}^{lm}$ and $D_{a}^{lm}$ have been shown \cite{detmeswhi} to have a very simple form so that, when the explicit expression for the spherical harmonics $Y_{lm}$ is substituted into Equation (\ref{FaSseries}),
the summation over $m$ can be performed and the result is an expression of the form
\begin{equation}
\mathcal{F}_{a}^{\tx{S}} = \sum_{l} \big [ A_{a} (l+\frac{1}{2}) + B_{a}  + C_{a} \frac{1}{(l+\frac{1}{2})} + \frac{D_a}{(2l-1)(2l+3)} + E_{a} O(l^{-4}) \big ] 
\end{equation}
where the regularization parameters $A_{a}, B_{a}, C_{a}, D_{a} \tx{ and } E_{a}$ are $l$-independent quantities which do depend on the background geometry and the characteristics of the orbit.

Finally, the self-force can be calculated by
\begin{equation}
\begin{split}
\mathcal{F}_{a}^{\tx{R}} = \sum_{l=0}^{\infty} \big [ &\lim_{p \to p'} \mathcal{F}_{la}^{\tx{ret}} \\
					&- A_{a} (l+\frac{1}{2}) - B_{a} - C_{a} \frac{1}{(l+\frac{1}{2})} -\frac{D_a}{(2l-1)(2l+3)} - E_{a} O(l^{-4}) \big ].
\end{split}
\label{SelfForceFinal}
\end{equation}
One important point that should be made is that the infinite sum over $l$ must be performed, in order for the self-force to be calculated.
Notice, however, that the contributions for large $l$ get less significant for the terms containing $D_a$ and $E_a$.
The reason for including these terms is now clear.
Even though the sum over $l$ of each of these two terms is exactly equal to 0, including these terms improves the convergence of the sum.
An additional benefit of including these terms is that the approximation to $\mathcal{F}_{a}^{\tx{R}}$ becomes more differentiable, as is explained in \cite{detmeswhi}.

\section{Order Calculation of the Scalar Field}
\label{OrderCalculation}

It should be obvious from the analysis of Section (\ref{RegProc}) that,
in order to calculate the regularization parameters for the singular field generated by a charge $q$, it is necessary to have an expression of $(1/\rho)$ in which the order of each term is known.
The derivation of such an expression is presented in this section for an equatorial circular orbit of radius $r_o$ in the Schwarzschild background. 
It is noted that the results were derived using \textsc{Maple} extensively.
It is also noted that the derivation was presented in \cite{detmeswhi} where the results of it for the radial derivative $\partial_r \big ( \frac{1}{\rho} \big ) $ were given.
For simplicity, the scalar charge $q$ is set to 1.

In order for the calculation of the self-force regularization parameters to be made easier, the Schwarzschild coordinates can be rotated, as explained in \cite{bo1}. 
Specifically, new angles $\Theta$ and $\Phi$ can be defined in terms of the usual Schwarzschild angles by the equations
\begin{eqnarray} \label{NewAngles}
\sin \theta \: \cos(\phi - \Omega t_{\tx{s}}) & = & \cos \Theta \\
\sin \theta \: \sin(\phi - \Omega t_{\tx{s}}) & = & \sin\Theta \: \cos\Phi \\
\cos \theta & = & \sin\Theta \: \sin\Phi
\end{eqnarray}
so that the coordinate location of the particle is moved from the equatorial plane, where $\theta = \frac{\pi}{2}$, to a location where $\sin \Theta = 0$, for a specific $t_{\tx{s}}$.
Such a coordinate rotation preserves the index $l$ of any spherical harmonic $Y_{lm}(\theta, \phi)$.
That means that any $Y_{lm}(\theta, \phi)$ is mapped  into a linear combination of spherical harmonics $Y_{lm'}(\Theta, \Phi)$, where $m' = -l, \ldots, l$.
Consequently, each $l$-multipole mode of the field or the self-force that results after summation over $m$ is the same, regardless of which angles, $(\theta,\phi)$ or $(\Theta,\Phi)$, are used for calculating it.

The benefit of this coordinate rotation for the calculation of the regularization parameters for the singular part of the self-force can be understood if one remembers that in the limit $p \to p'$ the angle $\Theta$ is equal to 0.
That means that $Y_{lm}(0,\Phi)$ has to be used, for which
\begin{eqnarray}
Y_{lm}(0,\Phi) &=& 0, \tx{ for } m \neq 0 \\
Y_{lm}(0,\Phi) &=& \sqrt{\frac{2l+1}{4 \pi}}, \tx{ for } m = 0.
\end{eqnarray}
So the sum over $m$ can, after this coordinate transformation, be replaced with just the $m=0$ term.
However, for the regularization parameters of the singular field, the limit $p \to p'$ must not be taken, so the $Y_{lm}$'s for all $m$'s must be taken into consideration.
The $m=0$ spherical harmonic must be considered only when the regularization parameters for the self-force are derived from the ones for the scalar field and when the regularization parameters for the first $r$-derivative of the self-force are calculated.

A comment on the order of each term needs to be made at this point.
The parameter $\epsilon^n$ is used to indicate a term of order $x^n$ in the coincidence limit $p \to p'$.
In that limit, $r \to r_o$ and $\Theta \to 0$.
That means that the factor $(r - r_o)$ is of order $\epsilon$ and the factor $(1 - \cos \Theta)$ is of order $\epsilon^2$.
At the end of the calculation, the parameter $\epsilon$ can be set equal to 1.

The relationships between the Schwarzschild coordinates $(t_{\tx{s}}, r, \theta, \phi)$ and the THZ coordinates $(t,x,y,z)$ for a circular orbit on the equatorial plane of a Schwarzschild background that were given in Section (\ref{THZ}) are used for this calculation.
As was mentioned earlier, $\rho^2 = x^2 + y^2 + z^2$ in terms of the spatial THZ coordinates $x$, $y$ and $z$.
However, it is clear from Equation (\ref{xy}) that the relationship
\begin{equation} 
x^2 + y^2 = \tilde{x}^2 + \tilde{y}^2
\end{equation}
holds between $\{x,y\}$ and $\{\tilde{x},\tilde{y}\}$, so the sum $(\tilde{x}^2 + \tilde{y}^2)$ is used to calculate $\rho^2$.
The Equations (\ref{THZxtilde}), (\ref{THZytilde}) and (\ref{THZz}) are substituted into the expression for $\rho^2$, with the Schwarzschild angles $\theta$ and $\phi$ replaced by the new angles $\Theta$ and $\Phi$.
That substitution gives that the lowest order term for $\rho^2$, denoted by $\tilde{\rho}^2$, is
\begin{equation}
\tilde{\rho}^2 \equiv \frac{r_o \Delta^2}{r_o - 2M} + 2 r_o^2 \frac{r_o -2M}{r_o -3M} \chi (1-\cos\Theta),
\label{elim1}
\end{equation}
where
\begin{equation}
\Delta \equiv r - r_o 
\label{elim2}
\end{equation}
and
\begin{equation}
\chi \equiv 1 - \frac{M}{r_o - 2M} \sin^2 \Phi.
\label{elim3}
\end{equation}
Clearly, $\tilde{\rho}^2$ is of order $\epsilon^2$ in the coincidence limit, as should have been expected.
Next the variables $\Theta$, $\Phi$ and $r$ are eliminated from the order expression of $\rho^2$ in favor of the variables $\Delta$, $\tilde{\rho}$ and $\chi$, by using Equations (\ref{elim1}), (\ref{elim2}) and (\ref{elim3}). 
Finally, the result is inverted and the square root is taken, in order to obtain the order expression of $(1/\rho)$.

The result of this calculation is a very long expression.
Here, I only give the terms of this expression that are necessary to calculate the regularization parameters of the singular field, the singular self-force along the radial direction and the $r$-derivative of the singular self-force along the radial direction.
Exactly how that is determined will become clear shortly, when the general term $\Delta^{\alpha} \tilde{\rho}^{p} \chi^{-q}$, where $\alpha$, $p$ and $q$ stand for integers, will be discussed.

The terms of interest are
\begin{equation}
\begin{split}
\frac{1}{\rho} &= \epsilon^{-1} \frac{1}{\tilde{\rho}} + \\
		& + \epsilon^{0} \bigg \{ \bigg [ \frac{r_o-3M}{2 r_o (r_o-2M)} \frac{1}{\chi} - \frac{1}{r_o} \bigg ] \frac{\Delta}{\tilde{\rho}} +  \bigg [ \frac{2 r_o-3M}{2 r_o (r_o-2M)} - \frac{r_o-3M}{2 r_o (r_o-2M)} \frac{1}{\chi} \bigg ] \frac{\Delta^3}{\tilde{\rho}^3}  \bigg \} \\
		& + \epsilon^{1} \bigg \{ \frac{r_o-3M}{8 r_o^2 (r_o-2M)} \bigg [ \frac{1}{\chi} - \frac{r_o +M}{r_o} \frac{1}{\chi^2} \bigg ] \tilde{\rho} \\
		& \qquad + \bigg [ \frac{2r_o-3M}{2 r_o^2 (r_o-2M)} - \frac{5r_o^2-22r_oM+21M^2}{4(r_o-2M)^2 r_o^2} \frac{1}{\chi} + \frac{5r_o^2-22r_oM+21M^2}{8 r_o^2 (r_o-2M)^2} \frac{1}{\chi^2}  \bigg ] \frac{\Delta^2}{\tilde{\rho}}  \\
		& \qquad + O \big (  \frac{\Delta^4}{\tilde{\rho}^3}, \frac{\Delta^6}{\tilde{\rho}^5} \big ) \bigg \} \\
		& + \epsilon^{2} \bigg \{ \bigg [ - \frac{M (r_o-2M)}{2 r_o^4 (r_o-3M)} - \frac{(r_o-M)(r_o-4M)}{8 (r_o-2M) r_o^4} \frac{1}{\chi} \\ 
		& \qquad + \frac{(r_o-3M)(5r_o^2-7r_oM-14M^2)}{16 r_o^4 (r_o-2M)^2} \frac{1}{\chi^2} - \frac{3 (r_o+M) (r_o-3M)^2}{16 r_o^4 (r_o-2M)^2} \frac{1}{\chi^3} \bigg ]
 \Delta \tilde{\rho} \\
		& \qquad + O \big ( \frac{\Delta^3}{\tilde{\rho}} , \frac{\Delta^5}{\tilde{\rho}^3}, \frac{\Delta^7}{\tilde{\rho}^5}, \frac{\Delta^9}{\tilde{\rho}^7} \big ) \bigg \} \\
		& + O(\epsilon^3). \\
\end{split}
\label{OrderRho}
\end{equation}

\section{Scalar Monopole Field}
\label{ScalarMonopole}

As is clear from the previous calculation of $(1/\rho)$, the angular dependence of every term shows up in factors of the form $\tilde{\rho}^p \chi^{-q}$, where $p$  is an odd integer and $q = 0, 1, 2, \ldots$.
The spherical harmonic decomposition of that factor, which is necessary when calculating the regularization parameters for the scalar field, is given in detail in Appendix \ref{A1}.
The result is
\begin{equation}
\tilde{\rho}^p \chi^{-q} = \bigg [ \frac{2 r_o^2 (r_o-2M)}{r_o - 3M} \bigg ]^{p/2} \sum_{l=0}^{\infty} \sum_{m=-l}^{l} E_{l,m}^{p,q}(\gamma^2, \frac{M}{r_o-2M}) Y_{lm}(\Theta, \Phi)
\label{rhochidec}
\end{equation}
where
\begin{equation}
\gamma^2 = \frac{\Delta^2 (r_o-3M)}{2 r_o (r_o-2M)^2}
\end{equation}
and the coefficients $E_{l,m}^{p,q}(\gamma^2, \frac{M}{r_o-2M})$ are given in Equation (\ref{Efinal}).
It is stressed  that all the $r$-dependence of $\tilde{\rho}^p \chi^{-q}$ resides in the sum $\sum_{s=0}^{\infty} \gamma^{2s}$ and the term $(\gamma^2)^{\frac{p}{2} +n +1}$, in Equation (\ref{Efinal}) for $E_{l,m}^{p,q}$, and is always proportional to powers of $\Delta = (r-r_o)$.

At this point, a note on the term $(\gamma^2)^{\frac{\beta}{2}}$ for an odd integer $\beta$ is in order.
For that term, the square root of $\gamma^2$ must be considered.
One might think that
\begin{equation}
(\gamma^2)^{\frac{1}{2}} = \bigg [ \frac{r_o -3M}{2 r_o (r_o-2M)^2} \Delta^2 \bigg ]^{\frac{1}{2}} = \bigg [ \frac{r_o-3M}{2 r_o (r_o-2M)^2} \bigg ]^{\frac{1}{2}} \Delta.
\end{equation}
But that implies that being on the equatorial plane with $\Theta = 0$ and approaching the particle by taking the limit $r \to r_o$, would give for the leading term $\tilde{\rho}$ in the expansion of $\rho$:
\begin{equation}
\begin{split}
\tilde{\rho} &= \bigg [ 2 r_o^2 \frac{r_o-2M}{r_o-3M} \bigg ]^{\frac{1}{2}} \chi^{\frac{1}{2}} \bigg ( \frac{\gamma^2}{\chi} + 1 - \cos 0 \bigg )^{\frac{1}{2}}  \\
	&=\bigg [ 2 r_o^2 \frac{r_o-2M}{r_o-3M} \bigg ]^{\frac{1}{2}} (\gamma^2)^{\frac{1}{2}} \\
	&= \bigg ( \frac{r_o}{r_o-2M} \bigg )^{\frac{1}{2}} \Delta. \\
\end{split}
\end{equation}
According to this, the sign of the leading order term of $\rho$ could be either positive or negative, depending on whether $r > r_o$ or $r < r_o$.
That is clearly not correct, since by definition $\rho = \sqrt{x^2 + y^2 + z^2}$, which is always positive.
For that reason, taking the square root of $\gamma^2$ always implies
\begin{equation}
(\gamma^2)^{\frac{1}{2}} = \bigg [ \frac{r_o-3M}{2 r_o (r_o-2M)^2} \bigg ]^{\frac{1}{2}} |\Delta|.
\end{equation}

For the self-force along the radial direction to be calculated, the $r$-derivative of $(1/\rho)$ and the limit $r \to r_o$ have to  be taken.
That means that any term of order $(r-r_o)^2$ or higher gives, after the limit is taken, no contribution to the self-force.
However, as has already been mentioned, it is desired to calculate the first derivative with respect to $r$ of the self-force, which is the second derivative with respect to $r$ of the field, in the limit $r \to r_o$.
Consequently, terms of order $(r-r_o)^2$ have to be retained, because they do give a contribution at that limit, while any terms of order $(r-r_o)^{\frac{5}{2}}$ or higher can be disregarded.

As is clear from Equation (\ref{Efinal}) of Appendix \ref{A1}, the general term $\Delta^{\alpha} \tilde{\rho}^p \chi^{-q}$ in Equation (\ref{OrderRho}) has two pieces that contain the  $r$-dependence.
The first piece comes from the sum over $s$ in $E_{l,m}^{p,q}$ and gives terms  proportional to $\gamma^{\alpha+2s}$.
Such a term should be kept only for $\alpha + 2s < \frac{5}{2}$ or
\begin{equation}
s < \frac{5}{4} - \frac{\alpha}{2}. 
\label{skeep}
\end{equation}
Since $s \geq 0$, the $\gamma^{\alpha+2s}$ contribution can be immediately disregarded for terms with $\alpha \geq \frac{5}{2}$, while for terms with  $\alpha < \frac{5}{2}$ a limited number of values of $s$ have to be retained in the sum. 
The second piece that contains $r$-dependence comes from the term $(\gamma^2)^{\frac{p}{2}+n+1}$ and is proportional to $\gamma^{p + \alpha +2n +2}$.
Such a term should be kept only for $p + \alpha +2n+2 < \frac{5}{2}$ or
\begin{equation}
n < \frac{1}{4} - \frac{p+\alpha}{2}. 
\label{nkeep}
\end{equation}
Since $n \geq 0$, all terms of Equation (\ref{OrderRho}) for which $p + \alpha \geq \frac{1}{2}$ do not contribute through the $\gamma^{p+\alpha+2n+2}$ piece.
Combinations of $p$ and $\alpha$ for which $ p + \alpha < \frac{1}{2}$ should be examined individually and a specific number of $n$'s must be retained.
The terms of the expansion (\ref{OrderRho}) for $(1/\rho)$ for which just the order (and not the explicit expression) is given are terms that fall in both categories that according to this analysis can be ignored. 

I present now the calculation of the regularization parameters coming from the terms of different order, for the singular field $(1/\rho)$.
The abbreviation used for the hypergeometric function is
\begin{equation}
F_{a}^{\lambda,m} \equiv {}_{2}F_{1} (a,\lambda+\frac{1}{2};\frac{|m|}{2} + 1;\frac{M}{r_o-2M}),
\end{equation}
and it is also noted that the hypergeometric function
\begin{equation}
F_{a}^{0,0} = {}_{2}F_{1}(a, \frac{1}{2};1;\frac{M}{r_o-2M})
\end{equation}
is denoted by $F_{a}$, as is done in \cite{detmeswhi}.
Also, for the three sums over $k$, $\nu$ and $\lambda$ the abbreviation
\begin{equation}
\begin{split}
\bigg ( \sum_{k,\nu,\lambda} & \bigg ) = \sqrt{\frac{2l+1}{\pi}} \frac{\Gamma(l-|m|+1)}{[\Gamma(l-m+1) \Gamma(l+m+1)]^{\frac{1}{2}}} \times \\
				& \sum_{k=0}^{[\frac{l}{2}]} (-1)^k 2^{l-2k} \frac{\Gamma(l-k+\frac{1}{2})}{\Gamma(k+1) \Gamma(l-2k-|m|+1)} \sum_{\nu=0}^{\frac{|m|}{2}} \frac{\Gamma(\frac{|m|}{2} +1)}{\Gamma(\nu+1) \Gamma(\frac{|m|}{2}-\nu+1)} \times \\
				& \sum_{\lambda = 0}^{\frac{|m|}{2}} (-1)^{\lambda} \frac{\Gamma(\frac{|m|+1}{2})}{\Gamma(\lambda+1) \Gamma(\frac{|m|}{2}-\lambda+1)}
\end{split}
\end{equation}
that is used in appendix \ref{A1} is also used here.
In addition, the upper limit of the $n$-summation is denoted by
\begin{equation}
N = l -|m| - 2k + 2\nu.
\end{equation}

For the lowest order term of Equation (\ref{OrderRho})
\begin{equation}
T_{(-1)} \equiv \epsilon^{-1} \frac{1}{\tilde{\rho}}
\end{equation}
the exponents of $\Delta^{\alpha} \tilde{\rho}^p \chi^{-q}$ are $\alpha=0, p=-1, q=0$.
Equations (\ref{skeep}) and (\ref{nkeep}) give that only the $s=0$ and $s=1$ terms of the first piece and only the $n=0$ term of the second piece of $E_{l,m}^{-1,q}$ need to be kept.
Consequently
\begin{equation}
T_{(-1)} = \epsilon^{-1} \sum_{l,m} A_{lm} Y_{lm}(\Theta, \Phi)
\label{T_1}
\end{equation}
where, using Equation (\ref{rhochidec})
\begin{equation}
\begin{split}
A_{lm} &= \bigg [\frac{r_o-3M}{2 r_o^2 (r_o-2M)} \bigg ]^{\frac{1}{2}} E_{l,m}^{-1,0} \\
	&= \bigg [ \frac{r_o-3M}{2 r_o^2 (r_o-2M)} \bigg ]^{\frac{1}{2}} \bigg ( \sum_{k,\nu,\lambda} \bigg ) \bigg \{ - 2 (\gamma^2)^{\frac{1}{2}} F_{1}^{\lambda,m} \\
	& \quad \: + \sum_{n=0}^{N} \frac{(-1)^{l+n} 2^{2n+\frac{1}{2}}}{(2n+1)!!} \frac{\Gamma(N+1)}{\Gamma(N-n+1)} 
	\Big [ 2 F_{1/2}^{\lambda,m} + (n-\frac{1}{2}) F_{3/2}^{\lambda,m} \gamma^2 \Big ] \bigg \} \\
\end{split}
\end{equation}
or, when the explicit expression for $\gamma^2$ and its square root are substituted
\begin{equation}
\begin{split}
A_{lm} &= \bigg [ \frac{r_o-3M}{2 r_o^2 (r_o-2M)} \bigg ]^{\frac{1}{2}} \bigg ( \sum_{k,\nu,\lambda} \bigg ) \bigg \{ - 2 \Big [ \frac{r_o-3M}{2 r_o (r_o-2M)^2} \Big ]^{\frac{1}{2}} |\Delta| F_{1}^{\lambda,m} \\
	&+ \sum_{n=0}^{N} \frac{(-1)^{l+n} 2^{2n+\frac{1}{2}}}{(2n+1)!!} \frac{\Gamma(N+1)}{\Gamma(N-n+1)} 
	\Big [ 2 F_{1/2}^{\lambda,m} + (n-\frac{1}{2} ) F_{3/2}^{\lambda,m} \frac{r_o-3M}{2 r_o (r_o-2M)^2} \Delta^2 \Big ] \bigg \}. \\
\end{split}
\label{Alm}
\end{equation}
As is explained in Appendix \ref{A1}, only the even $m$'s should be included in the sum of Equation (\ref{T_1}).

The regularization parameter $A_r$ given in \cite{detmeswhi} for the self-force can be easily obtained from this result.
First the $m=0$ parameter is considered, for reasons that were explained earlier.
That makes the sums over $\lambda$ and over $\nu$ equivalent to the terms with $\lambda =0$ and $\nu=0$.
For the hypergeometric function $F_{1}^{0,0}$ it can easily be proven that
\begin{equation}
F_{1}^{0,0} = F_1 = \sqrt{\frac{r_o-2M}{r_o-3M}} 
\label{hypergeomF1}
\end{equation}
using Equations 15.3.3 of \cite{HMF}.
Then the first derivative of $A_{lm}$ with respect to $r$ is taken.
That makes the term proportional to $F_{1/2}^{\lambda,m}$ in Equation (\ref{Alm}) vanish, while the term proportional to $F_{3/2}^{\lambda,m}$ gives a factor of $\Delta$ which vanishes when the coincidence limit is taken.
For the first term the derivative of $|\Delta|$ gives sgn$(\Delta)$.
Finally, the spherical harmonic $Y_{l0}(0 , \Phi)$ is substituted with $\sqrt{\frac{2l+1}{4 \pi}}$.
This procedure gives
\begin{equation}
\begin{split}
\frac{\partial}{\partial r} T_{(-1)} \bigg |_{p \to p'} &= \epsilon^{-2} \sum_{l=0}^{\infty} \frac{d A_{l0}}{dr} \bigg |_{r = r_o} \sqrt{\frac{2l+1}{4 \pi}}  \\
					&= \epsilon^{-2} (- \tx{sgn}(\Delta)) \frac{[r_o (r_o-3M)]^{\frac{1}{2}}}{r_o^2 (r_o-2M)} \frac{1}{\sqrt{\pi}} \sum_{l=0}^{\infty} \bigg \{ (l+\frac{1}{2}) \times \\
					& \qquad \qquad \qquad \sum_{k=0}^{[\frac{l}{2}]} (-1)^k 2^{l-2k} \frac{\Gamma(l-k+\frac{1}{2})}{\Gamma(k+1) \Gamma(l-2k+1)} \bigg \} . \\ 
\end{split}
\end{equation}
The sum over $k$ can be easily calculated for any value of $l$, using {\textsc{Maple}}.
A general proof that it is equal to $\sqrt{\pi}$ for any $l$ cannot be given.
However, for every value of $l$ that was tried, the sum ended up being $\sqrt{\pi}$, which gives
\begin{equation}
\frac{\partial}{\partial r} T_{(-1)} \bigg |_{p \to p'} = \epsilon^{-2} ( - \tx{sgn}(\Delta)) \frac{[r_o (r_o-3M)]^{\frac{1}{2}}}{r_o^2 (r_o-2M)} \sum_{l=0}^{\infty} (l+\frac{1}{2}) 
\end{equation}
which is the result of \cite{detmeswhi}.

Now, the second derivative of the term $T_{(-1)}$ with respect to $r$ is taken, which is equal to the contribution of that term to the first derivative of the self-force.
The $m=0$ component is considered since the coincidence limit must be taken, so $\lambda =0$ and $\nu = 0$ as well.
Only the term proportional to $F_{3/2}^{\lambda,m}$ survives the differentiation, since it is proportional to $\Delta^2$.
The result is
\begin{equation}
\begin{split}
\frac{ \partial^2 T_{(-1)}}{\partial r^2} \bigg |_{p \to p'} &= \epsilon^{-3} \sum_{l=0}^{\infty} \frac{d^2 A_{l0}}{dr^2} \bigg|_{r=r_o} \sqrt{\frac{2l+1}{4 \pi}} \\
					&= \epsilon^{-3} \frac{(r_o-3M)^{\frac{3}{2}}}{r_o^2 (r_o-2M)^{\frac{5}{2}}} \frac{1}{\sqrt{\pi}} F_{3/2} \sum_{l=0}^{\infty} (l+\frac{1}{2}) \times \\
					& \quad \sum_{k=0}^{[\frac{l}{2}]} (-1)^k 2^{l-2k} \frac{\Gamma(l-k+\frac{1}{2})}{\Gamma(k+1)} \sum_{n=0}^{l-2k} (-1)^{l+n} \frac{2^{2n} (n-\frac{1}{2})}{\Gamma(l-2k-n+1) (2n+1)!!}. \\
\end{split}
\label{SecDerivT(-1)}
\end{equation}

The zeroth-order term contribution that is considered in Equation (\ref{OrderRho}) is
\begin{equation}
T_{(0)} \equiv \epsilon^0 \bigg \{ \Big [ \frac{r_o-3M}{2 r_o (r_o-2M)} \frac{1}{\chi} - \frac{1}{r_o} \bigg ] \frac{\Delta}{\tilde{\rho}} + \bigg [ \frac{2r_o-3M}{2 r_o (r_o-2M)} - \frac{r_o-3M}{2 r_o (r_o-2M)} \frac{1}{\chi} \Big ] \frac{\Delta^3}{\tilde{\rho}^3} \bigg \}.
\end{equation}
For the piece that is proportional to $\Delta \tilde{\rho}^{-1}$ the exponents are $\alpha = 1, p=-1, q=1$ for the term that contains $\chi$ and $\alpha=1, p=-1, q=0$ for the term that does not contain $\chi$.
In both cases, Equations (\ref{skeep}) and (\ref{nkeep}) indicate that the $s=0$ term of the first piece and the $n=0$ term of the second piece of $E_{l,m}^{-1,q}$ should be kept.
For the piece that is proportional to $\Delta^3 \tilde{\rho}^{-3}$ the exponents are $\alpha =3, p=-3, \chi=0$ for the term that does not contain $\chi$ and $\alpha =3, p=-3, \chi=1$ for the term that does contain $\chi$.
In this case, Equations (\ref{skeep}) and (\ref{nkeep}) give that none of the $s$-terms and only the $n=0$ term of the second piece of $E_{l,m}^{-3,q}$ need to be kept. 
That gives
\begin{equation}
T_{(0)} = \epsilon^{0} \sum_{l,m} B_{lm} Y_{lm}(\Theta, \Phi)
\label{T0}
\end{equation}
where, from Equation (\ref{rhochidec})
\begin{equation}
\begin{split}
B_{lm} &= \Delta \bigg [ \frac{r_o-3M}{2 r_o^2 (r_o-2M)} \bigg ]^{\frac{1}{2}} \bigg [ \frac{r_o-3M}{2 r_o (r_o-2M)} E_{l,m}^{-1,1} - \frac{1}{r_o} E_{l,m}^{-1,0} \bigg ] \\ 
	&\: \quad + \Delta^3 \bigg [ \frac{r_o-3M}{2 r_o^2 (r_o-2M)} \bigg ]^{\frac{3}{2}} \bigg [ \frac{2 r_o-3M}{2 r_o (r_o-2M)} E_{l,m}^{-3,0} - \frac{r_o-3M}{2 r_o (r_o-2M)} E_{l,m}^{-3,1} \bigg ]. \\
\end{split}
\label{BLM1}
\end{equation}
Substituing the explicit expressions of the coefficients $E_{l,m}^{p,q}$ into Equation (\ref{BLM1}) for $B_{lm}$ results in
\begin{equation}
\begin{split}
B_{lm}
	&= \sqrt{\frac{r_o-3M}{2 r_o^2 (r_o-2M)}}  \Big ( \sum_{k,\nu,\lambda} \Big ) \times \\
	&\qquad \Bigg \{ \: \Delta \sum_{n=0}^{N} \frac{(-1)^{l+n} 2^{2n+\frac{3}{2}}}{(2n+1)!!} 
\frac{\Gamma(N+1)}{\Gamma(N-n+1)} \bigg ( \frac{r_o-3M}{2 r_o (r_o-2M)} F_{3/2}^{\lambda,m} - \frac{1}{r_o} F_{1/2}^{\lambda,m} \bigg ) \\
	&\qquad \quad -\Delta |\Delta| \frac{\sqrt{2} (r_o-3M)^{\frac{1}{2}}}{r_o^{\frac{1}{2}} (r_o-2M)} \Big ( \frac{r_o-3M}{2 r_o(r_o-2M)} F_{2}^{\lambda,m} - \frac{1}{r_o} F_{1}^{\lambda,m} \Big ) \\
	&\qquad \quad +\frac{\Delta^3}{|\Delta|} \frac{(r_o-3M)^{\frac{3}{2}}}{\sqrt{2} r_o^{\frac{5}{2}} (r_o-2M)} \Big [ \frac{2r_o-3M}{r_o-3M} F_{1}^{\lambda,m} - F_{2}^{\lambda,m} \Big ] \Bigg \}.\\
\end{split}
\end{equation}
It is easy to recognize that
\begin{equation}
\Delta |\Delta| =  \frac{\Delta^3}{|\Delta|}
\end{equation}
so the last two terms in the last expression for $B_{lm}$ can be combined to give a significantly simpler expression, namely
\begin{equation}
\begin{split}
B_{lm} &= \sqrt{\frac{r_o-3M}{2 r_o^2 (r_o-2M)}}  \Big ( \sum_{k,\nu,\lambda} \Big ) \times \\
	&\qquad \Bigg \{ \: \Delta \sum_{n=0}^{N} \frac{(-1)^{l+n} 2^{2n+\frac{3}{2}}}{(2n+1)!!} \frac{\Gamma(N+1)}{\Gamma(N-n+1)} \bigg ( \frac{r_o-3M}{2 r_o (r_o-2M)} F_{3/2}^{\lambda,m} - \frac{1}{r_o} F_{1/2}^{\lambda,m} \bigg ) + \\
	&\qquad \quad \Delta |\Delta| \frac{ (r_o-3M)^{\frac{1}{2}}}{ \sqrt{2} r_o^{\frac{5}{2}} (r_o-2M)} \Big [ -\frac{2 (r_o-3M)(r_o-M)}{(r_o-2M)} F_{2}^{\lambda,m} + (4 r_o-3M) F_{1}^{\lambda,m} \Big ] \Bigg \}. 
\end{split}
\label{Blm}
\end{equation}
As is the case for $A_{lm}$, only the coefficients for which $m$ is even contribute to the sum in Equation (\ref{T0}).

The regularization parameter $B_r$ calculated in \cite{detmeswhi} can be easily derived from the general expression given in Equation (\ref{Blm}).
First, the $m=0$ coefficient is considered, which also makes $\lambda=0$ and $\nu=0$.
The first derivative of the factor $\Delta |\Delta|$ with respect to $r$ in Equation (\ref{Blm}) gives zero in the coincidence limit $r \to r_o$.
So the only term that contributes is the one that is proportional to $\Delta$.
If the spherical harmonic $Y_{l0}(0,\Phi)$ is again replaced with $\sqrt{\frac{2l+1}{4 \pi}}$
the result is
\begin{equation}
\begin{split}
\frac{\partial}{\partial r} T_{(0)} \bigg |_{p \to p'} &= \epsilon^{-1} \sum_{l=0}^{\infty} \frac{\partial B_{l0}}{\partial r} \bigg |_{r=r_o} \sqrt{\frac{2l+1}{4 \pi}} \\
				&= \epsilon^{-1} \bigg [ \frac{r_o-3M}{r_o^4 (r_o-2M)} \bigg ]^{\frac{1}{2}} \Big [ \frac{r_o-3M}{2 (r_o-2M)} F_{3/2} - F_{1/2} \Big ] \times \\
				& \qquad \quad \sum_{l=0}^{\infty}  
				\frac{2l+1}{2 \sqrt{2 \pi}} \times \\
				& \qquad \quad \sum_{k=0}^{[\frac{l}{2}]} (-1)^k 2^{l-2k} \frac{\Gamma(l-k+\frac{1}{2})}{\Gamma(k+1)} \sum_{n=0}^{l-2k} \frac{2^{2n+\frac{3}{2}}}{(2n+1)!!} \frac{(-1)^{l+n}}{\Gamma(l-2k-n+1)}. \\
\end{split}
\label{dT0dr}
\end{equation}
The two finite sums over $k$ and over $n$ can be easily calculated for any value of $l$ by using \textsc{Maple}.
For any value of $l$ that was tried, the result was equal to $\frac{2 \sqrt{2 \pi}}{2l+1}$, but, again, there is no general proof for that.
Substituting this result for the double sum into Equation (\ref{dT0dr}) gives that
\begin{equation}
\frac{\partial}{\partial r} T_{(0)} \bigg |_{p \to p'}= \epsilon^{-1} \bigg [ \frac{r_o-3M}{r_o^4 (r_o-2M)} \bigg ]^{\frac{1}{2}} \Big [ \frac{r_o-3M}{2 (r_o-2M)} F_{3/2} - F_{1/2} \Big ] \sum_{l=0}^{\infty} 1 
\end{equation}
which is the same as the result given in \cite{detmeswhi}.

The contribution of $T_{(0)}$ to the derivative of the self-force can now be calculated.
The second derivative of Equation (\ref{Blm}) has to be taken, and the $m=0$ component to be considered.
Thus, the term proportional to $\Delta$ vanishes but the second derivative of the term $\Delta |\Delta|$ gives
\begin{equation}
\frac{d^2}{dr^2} (\Delta |\Delta|) = 2 \tx{sgn}(\Delta) 
\end{equation}
which does give a contribution in the coincidence limit.
The result is
\begin{equation}
\begin{split}
\frac{\partial^2}{\partial r^2} T_{(0)} \bigg |_{p \to p'} &= \epsilon^{-2} \sum_{l=0}^{\infty} \frac{d^2 B_{l0}}{dr^2} \bigg |_{r=r_o} \sqrt{\frac{2l+1}{4 \pi}} \\
				&= \epsilon^{-2} \frac{(r_o-3M)}{r_o^{\frac{5}{2}} (r_o-2M)^{\frac{3}{2}}} \frac{\tx{sgn}(\Delta)}{\sqrt{\pi}} \sum_{l=0}^{\infty} (l+\frac{1}{2}) \sum_{k=0}^{[\frac{l}{2}]} \frac{(-1)^k 2^{l-2k} \Gamma(l-k+\frac{1}{2})}{\Gamma(k+1) \Gamma(l-2k+1)} \times \\
				& \qquad \bigg [ -2 \frac{(r_o-M)(r_o-3M)}{r_o (r_o-2M)} F_2 + \frac{4 r_o-3M}{r_o} F_1 \bigg ]. \\
\end{split}
\end{equation}
Using Equation 15.2.13 of \cite{HMF} and the previously derived result for the hypergeometric function $F_1$ (Equation (\ref{hypergeomF1})), a simple expression for the hypergeometric function $F_2$ can also be obtained and it is
\begin{equation}
F_2 = \frac{1}{2} \sqrt{\frac{r_o-2M}{r_o-3M}} \Big ( \frac{2r_o-5M}{r_o-3M} \Big ).
\label{hypergeomF2}
\end{equation}
The final expression for the contribution of $B_{lm}$ to the second derivative of the scalar field is
\begin{equation}
\begin{split}
\frac{\partial^2}{\partial r^2} T_{(0)} \bigg |_{p \to p'} &= \epsilon^{-2} 
 \frac{\tx{sgn}(\Delta)}{\sqrt{\pi}} \sum_{l=0}^{\infty} (l+\frac{1}{2}) \sum_{k=0}^{[\frac{l}{2}]} (-1)^k 
\frac{2^{l-2k} \; \Gamma(l-k+\frac{1}{2})}{\Gamma(k+1) \Gamma(l-2k+1)} \times \\
                                & \qquad \frac{(r_o-3M)^{\frac{1}{2}}}{r_o^{\frac{7}{2}} (r_o-2M)^{2}}  (2r_o^2 - 4r_oM+M^2). \\ 
\end{split}
\label{SecDerivT0}
\end{equation}

The term of order $\epsilon^1$ in Equation (\ref{OrderRho}) is considered next.
The part of it that needs to be kept is
\begin{equation}
\begin{split}
T_{(1)} &\equiv \epsilon^1 \bigg \{ \frac{r_o-3M}{8 r_o^2 (r_o-2M)} \bigg [ \frac{1}{\chi} - \frac{r_o+M}{r_o} \frac{1}{\chi^2} \bigg ] \tilde{\rho} \\
		&\quad + \bigg [ \frac{2r_o-3M}{2 r_o^2 (r_o-2M)} - \frac{5r_o^2-22r_oM+21M^2}{4(r_o-2M)^2 r_o^2} \frac{1}{\chi} \\
		&\qquad + \frac{5r_o^2-22r_oM+21M^2}{8r_o^2(r_o-2M)^2} \frac{1}{\chi^2} \bigg ] \frac{\Delta^2}{\tilde{\rho}} \bigg \}. \\
\end{split}
\end{equation}
For the part that is proportional to $\tilde{\rho}$ the exponents are $\alpha=0, p=1, q=1$ for the term containing $(1/\chi)$ and $\alpha=0, p=1, q=2$ for the term containing $(1/\chi^2)$.
For both these terms, Equations (\ref{skeep}) and (\ref{nkeep}) indicate that only the $s=0$ and $s=1$ terms of the first piece and none of the values of $n$ of the second piece of $E_{l,m}^{1,q}$ should be kept.
For the part that is proportional to $\Delta^2 \tilde{\rho}^{-1}$ the exponents are $\alpha=2, p=-1, q=0$ for the term not containing $\chi$ and $\alpha=2, p=-1, q=1$ and $\alpha=2, p=-1, q=2$ for the other two terms.
In this case, only the $s=0$ term should be kept in $E_{l,m}^{-1,q}$.
Consequently
\begin{equation}
T_{(1)} = \epsilon^{1} \sum_{l,m} C_{lm} Y_{lm}(\Theta, \Phi)
\end{equation}
where only the even $m$'s contribute to the above sum and, from Equation (\ref{rhochidec}), the coefficients $C_{lm}$ are:
\begin{equation}
\begin{split}
C_{lm}  = & \bigg [ \frac{2 r_o^2 (r_o-2M)}{r_o-3M} \bigg ]^{\frac{1}{2}} \frac{r_o-3M}{8 r_o^2 (r_o-2M)} \bigg [ E_{l,m}^{1,1} - \frac{r_o+M}{r_o} E_{l,m}^{1,2} \bigg ]  \\
	&+ \Delta^2 \Big [ \frac{r_o-3M}{2 r_o^2 (r_o-2M)} \Big ]^{\frac{1}{2}} \times \\
	&\quad \bigg [ \frac{2r_o-3M}{2 r_o^2 (r_o-2M)} E_{l,m}^{-1,0} 
	- \frac{5r_o^2 -22r_oM+21M^2}{4r_o^2 (r_o-2M)^2} E_{l,m}^{-1,1} \\
	& \quad  + \frac{5r_o^2-22r_oM+21M^2}{8r_o^2(r_o-2M)^2} E_{l,m}^{-1,2} \bigg ]. \\
\end{split}
\end{equation}
After the explicit expressions for the coefficients $E_{l,m}^{p,q}$ and for $\gamma^2$ are substituted into the expression for $C_{lm}$, the result is
\begin{equation}
\begin{split}
C_{lm} &=  \frac{1}{8 r_o} \sqrt{\frac{r_o-3M}{r_o-2M}} \bigg ( \sum_{k,\nu,\lambda} \bigg ) 
	\sum_{n=0}^{N} \frac{(-1)^{l+n} 2^{2n+\frac{3}{2}}}{(2n+1)!!} \frac{\Gamma(N+1)}{\Gamma(N-n+1)} \times \\
	&\quad \bigg \{ \frac{2\sqrt{2}}{2n+3} \Big ( F_{1/2}^{\lambda,m} - \frac{r_o+M}{r_o} F_{3/2}^{\lambda,m} \Big ) \\
	&\quad + \frac{\Delta^2}{\sqrt{2} r_o (r_o-2M)^2} \Big [ \frac{2n+1}{2n+3} \frac{(r_o-3M)}{2} \big ( F_{3/2}^{\lambda,m} - \frac{r_o+M}{r_o} F_{5/2}^{\lambda,m} \big ) \\
	&+ \frac{4(r_o-2M)(2r_o-3M)}{r_o} F_{1/2}^{\lambda,m} 
	- \frac{2 (5r_o^2 - 22r_oM+21M^2)}{r_o} F_{3/2}^{\lambda,m} \\
	&+ \frac{5r_o^2-22r_oM+21M^2}{r_o} F_{5/2}^{\lambda,m} \Big ] \bigg \}. \\
\end{split}
\label{CLMF}
\end{equation}

Obtaining the result of \cite{detmeswhi} for the regularization parameter $C_r$ for the self-force is trivial in this case. 
That is because taking the first derivative of $C_{lm}$ with respect to $r$ immediately makes the term with no $\Delta$ dependence vanish and the terms proportional to $\Delta^2$ end up having a factor of $\Delta$ which also vanishes in the coincidence limit.
That means that the $O(\epsilon)$ term does not contibute to the self-force on the scalar charge.

Things are different for the first derivative of the self-force.
Taking the second derivative of Equation (\ref{CLMF}) makes the term with no $\Delta$ dependence vanish, but the term proportional to $\Delta^2$ does give a contribution which must be calculated.
Since the coincidence limit is taken, $m=0$ and
\begin{equation}
\frac{\partial^2}{\partial r^2} T_{(1)} \bigg |_{p \to p'} = \epsilon^{-1} \sum_{l=0}^{\infty} \frac{d^2 C_{l0}}{dr^2} \sqrt{\frac{2l+1}{4 \pi}} 
\end{equation}
which, after a short calculation, gives
\begin{equation}
\begin{split}
\frac{\partial^2}{\partial r^2} T_{(1)} \bigg |_{p \to p'} &= \epsilon^{-1} \frac{1}{\sqrt{\pi}} \sqrt{\frac{r_o-3M}{r_o-2M}} \frac{1}{2 r_o^2 (r_o-2M)^2} \sum_{l=0}^{\infty} (l+\frac{1}{2}) \times \\
				&\: \sum_{k=0}^{[\frac{l}{2}]} (-1)^k 2^{l-2k} \frac{\Gamma(l-k+\frac{1}{2})}{\Gamma(k+1)} \sum_{n=0}^{l-2k} \frac{2^{2n} (-1)^{l+n}}{\Gamma(l-2k-n+1) (2n+1)!!} \times \\
				& \: \bigg [ \frac{r_o-3M}{2} \frac{2n+1}{2n+3} \Big (F_{3/2} - \frac{r_o+M}{r_o} F_{5/2} \Big ) + \frac{4 (r_o-2M)(2r_o-3M)}{r_o} F_{1/2} \\
				& \: - \frac{2(5r_o^2-22r_oM+21M^2)}{r_o} F_{3/2} + \frac{5r_o^2-22r_oM+21M^2}{r_o} F_{5/2} \Big ].\\
\end{split}
\label{SecDerivT1}
\end{equation}

The last term of Equation (\ref{OrderRho}) that is examined is the term of order $\epsilon^2$.
The part of it that must be considered is
\begin{equation}
\begin{split}
T_{(2)} \equiv \epsilon^2 \bigg [ &-\frac{M(r_o-2M)}{2 r_o^4 (r_o-3M)} - \frac{(r_o-M)(r_o-4M)}{8 (r_o-2M) r_o^4} \frac{1}{\chi} \\
			&+ \frac{(r_o-3M)(5r_o^2-7r_oM-14M^2)}{16 r_o^4 (r_o-2M)^2} \frac{1}{\chi^2} - \frac{3 (r_o+M) (r_o-3M)^2}{16 r_o^4(r_o-2M)^2} \frac{1}{\chi^3} \bigg ] \Delta \tilde{\rho}.\\
\end{split}
\end{equation}
In that term, the exponents for $\Delta$ and $\tilde{\rho}$ are $\alpha = 1, p=1$, while the exponent of $\chi$ takes the values $q=0,1,2,3$ for each term.
Equations (\ref{skeep}) and (\ref{nkeep}) indicate that only the $s=0$ term of the first piece containing the $r$-dependence of each $E_{l,m}^{1,q}$ needs to be kept and the second piece can be disregarded.
Then, the $O(\epsilon^2)$ term can be written as
\begin{equation}
T_{(2)} = \epsilon^2 \sum_{l,m} D_{lm} Y_{lm}(\Theta, \Phi)
\end{equation}
where, according to Equation (\ref{rhochidec})
\begin{equation}
\begin{split}
D_{lm} = \Delta \Big [ &\frac{2 r_o^2 (r_o-2M)}{r_o-3M} \Big ]^{\frac{1}{2}} \bigg [ -\frac{M(r_o-2M)}{2 r_o^4(r_o-3M)} E_{l,m}^{1,0} - \frac{(r_o-M)(r_o-4M)}{8 (r_o-2M) r_o^4} E_{l,m}^{1,1} \\
		&\frac{(r_o-3M)(5r_o^2-7r_oM-14M^2)}{16 r_o^4 (r_o-2M)^2} E_{l,m}^{1,2} - \frac{3(r_o+M)(r_o-3M)^2}{16 r_o^4(r_o-2M)^2} E_{l,m}^{1,3} \bigg ]. \\
\end{split}
\label{DLME1}
\end{equation}
Substituting of the coefficients $E_{l,m}^{p,q}$ into Equation (\ref{DLME1}) gives
\begin{equation}
\begin{split}
D_{lm} = &\Delta \Big [ \frac{2 r_o^2 (r_o-2M)}{r_o-3M} \Big ]^{\frac{1}{2}} \bigg ( \sum_{k,\nu,\lambda} \bigg ) \sum_{n=0}^{N} \frac{\Gamma(N+1)}{\Gamma(N-n+1)} 
			\frac{(-1)^{l+n} 2^{2n+\frac{5}{2}}}{(2n+3)!!} \times \\
	&\bigg [ - \frac{M(r_o-2M)}{2 r_o^4 (r_o-3M)} F_{-1/2}^{\lambda,m} - \frac{(r_o-M)(r_o-4M)}{8 r_o^4 (r_o-2M)} F_{1/2}^{\lambda,m} \\
			& + \frac{(r_o-3M)(5r_o^2-7r_oM-14M^2)}{16 r_o^4 (r_o-2M)^2} F_{3/2}^{\lambda,m} - \frac{3 (r_o+M)(r_o-3M)^2}{16 r_o^4 (r_o-2M)^2} F_{5/2}^{\lambda,m} \bigg ]. \\
\end{split}
\end{equation}

Reproducing the result of \cite{detmeswhi} for $D_r$ can be done by following the same method that was used for the regularization parameters $A_r$ and $B_r$.
First, the $m=0$ coefficient is considered, which makes $\lambda=0$ and $\nu=0$.
The first derivative with respect to $r$ is then taken and that simply makes the factor $\Delta$ vanish.
Finally the expression of $Y_{l0}$ for $\Theta=0$ is used.
The result is
\begin{equation}
\begin{split}
\frac{\partial}{\partial r} T_{(2)} \bigg |_{p \to p'} &= \epsilon^{1} \sum_{l=0}^{\infty} \frac{\partial D_{l0}}{\partial r} \sqrt{\frac{2l+1}{4 \pi}} \\
				&= \epsilon^1 \bigg [ \frac{2 r_o^2 (r_o-2M)}{r_o-3M} \bigg ]^{\frac{1}{2}} \bigg [ -\frac{M(r_o-2M)}{2 r_o^4 (r_o-3M)} F_{-1/2} - \frac{(r_o-M)(r_o-4M)}{8 r_o^4 (r_o-2M)} F_{1/2} \\
				&+ \frac{(r_o-3M)(5r_o^2-7r_oM-14M^2)}{16 r_o^4 (r_o-2M)^2} F_{3/2} - \frac{3(r_o+M)(r_o-3M)^2}{16 r_o^4(r_o-2M)^2} F_{5/2} \bigg ] \times  \\
				& \sum_{l=0}^{\infty} \frac{ - 2 \sqrt{2}}{(2l-1)(2l+3)} \\
\end{split}
\end{equation}
which is the same as the result of \cite{detmeswhi}.
To derive this final expression, \textsc{Maple} was used to evaluate the double sum:
\begin{equation}
\begin{split}
\sum_{k=0}^{[\frac{l}{2}]} (-1)^k 2^{l-2k} \frac{\Gamma(l-k+\frac{1}{2})}{\Gamma(k+1)} \sum_{n=0}^{l-2k} & \frac{(-1)^{l+n} 2^{2n+\frac{5}{2}}}{(2n+3)!! \Gamma(l-2k-n+1)} = \\
		& - \frac{4 \sqrt{2 \pi}}{(2l-1)(2l+1)(2l+3)}\\
\end{split}
\end{equation}
for different values of $l$.
Even though a general proof that this equality holds for every $l$ cannot be provided, this equation was verified to hold true for any value of $l$ that was tried.

As far as the contribution of this term to the derivative of the self-force is concerned, it can be easily seen to equal zero since there are no terms that can survive after the second derivative of $D_{lm}$ with respect to $r$ is taken.
Consequently
\begin{equation}
\frac{\partial^2}{\partial r^2} T_{(2)} \bigg |_{p \to p'} = 0.
\end{equation}

One final comment should be made about Equations (\ref{SecDerivT(-1)}), (\ref{SecDerivT0}) and (\ref{SecDerivT1}).
Because these equations involve complicated sums of $k$ and $n$ it may be thought that they are not particularly handy for the calculation of the contributions of the corresponding terms to the first derivative of the self-force. 
Nonetheless, all the sums involved are finite and can always be calculated using mathematical software or numerical code, which renders these equations relatively easy to use.

      \chapter{Calculation of the Retarded Field}
\label{RSF}

As was explained in Chapter 4, when calculating the self-force on a scalar charge $q$ one needs to know the retarded and the singular part of the scalar field.
In order to be able to produce results for the self-force,
I present here the calculation of the retarded field generated by a scalar charge moving on a circular orbit in a Schwarzschild background, as well as its contribution to the self-force.
The method used is similar to that presented by Burko in \cite{burko} and the calculation was outlined in appendix E of \cite{detmeswhi}.

\section{Analytical Work}
\label{AnalytWork}

The scalar field $\psi^{\tx{ret}}$ at a point $p$ generated by a scalar charge $q$ moving on a circular orbit of radius $r_o$ on the equatorial plane of a Schwarzschild background of mass $M$ obeys Poisson's equation
\begin{equation}
\nabla^2 \psi^{\tx{ret}}(p(x^a)) = - 4 \pi \varrho
\end{equation}
where $x^a$ are the Schwarzschild coordinates of point $p$ and the source term is given by the integral
\begin{equation}
\begin{split}
\varrho &= q \int (-g)^{-\frac{1}{2}} \delta^4(p-p') d\tau \\
	&= q \int (-g)^{-\frac{1}{2}} \frac{\delta_(r-r_o)}{r_o^2} \delta(\theta - \frac{\pi}{2}) \delta(\phi-\Omega t_{\tx{s}}) \delta(t_{\tx{s}}-t_{\tx{s}}(\tau)) d\tau \\
	&= q \frac{\delta(r-r_o)}{r_o^2} \delta(\theta-\frac{\pi}{2}) \delta(\phi-\Omega t_{\tx{s}}) \Big ( \frac{dt_{\tx{s}}}{d\tau} \Big )^{-1}. \\
\end{split}
\end{equation}
In this expression $p'$ is the location of the charge $q$, $\tau$ is the proper time along the circular orbit of the charge $q$ and the relationship between the proper time $\tau$ and the Schwarzschild time $t_{\tx{s}}$ gives
\begin{equation}
\frac{dt_{\tx{s}}}{d\tau} = \frac{1}{\sqrt{1-\frac{3M}{r_o}}}.
\end{equation}

Then, by writing Poisson's equation in Scwarzschild coordinates one gets
\begin{equation}
\begin{split}
\nabla^2 \psi^{\tx{ret}} &= \Big [ - \Big ( 1-\frac{2M}{r} \Big )^{-1} \frac{\partial^2 }{\partial t_{\tx{s}}^2} + \frac{2r-2M}{r^2} \frac{\partial }{\partial r} + \Big (1-\frac{2M}{r} \Big ) \frac{\partial^2 }{\partial r^2} \\
			&\: \: + \frac{\cos\theta}{r^2 \sin\theta} \frac{\partial}{\partial \theta} + \frac{1}{r^2} \frac{\partial^2 }{\partial \theta^2} + \frac{1}{r^2 \sin^2 \theta} \frac{\partial^2 }{\partial \phi^2} \Big ] \psi^{\tx{ret}}\\
			&= - 4 \pi q \frac{\delta(r-r_o)}{r_o^2} \delta(\theta-\frac{\pi}{2}) \delta(\phi-\Omega t_{\tx{s}}) \sqrt{1-\frac{3M}{r_o}}.
\end{split}
\label{Del2PsiSchw}
\end{equation}
As was mentioned in Chapter 6, both the source and the field can be decomposed into spherical harmonics.
Specifically, the spherical harmonic decomposition of the source can be done using the spherical harmonic decomposition of the 3-dimensional $\delta$ function which is given in \cite{Jackson}
\begin{equation}
\begin{split}
\varrho
	&= q \: \frac{\delta(r-r_o)}{r_o^2} \sum_{l=0}^{\infty} \sum_{m=-l}^{l} \: \frac{1}{\sin\frac{\pi}{2}} \: Y^{\ast}_{lm}(\frac{\pi}{2},\Omega t_{\tx{s}}) \: Y_{lm}(\theta,\phi) \sqrt{1-\frac{3M}{r_o}} \\
	&= q \: \frac{\delta(r-r_o)}{r_o^2} \sum_{l,m} \tx{e}^{- i m \Omega t_{\tx{s}}} \: Y^{\ast}_{lm}(\frac{\pi}{2},0) \: Y_{lm}(\theta,\phi) \sqrt{1-\frac{3M}{r_o}}\\
	&= \sum_{l,m} \frac{\delta(r-r_o)}{4 \pi r_o} \: q_{lm} \: \tx{e}^{i \omega_m t_{\tx{s}}} \:  Y_{lm}(\theta,\phi)
\end{split}
\end{equation}
where
\begin{equation}
\omega_m = -  m \Omega
\end{equation}
and
\begin{equation}
q_{lm} = \frac{4 \pi q}{r_o} Y^{\ast}_{lm}(\frac{\pi}{2},0) \sqrt{1-\frac{3M}{r_o}}.
\end{equation}
The retarded field can be written as
\begin{equation}
\psi^{\tx{ret}}(t_{\tx{s}},r,
\theta,\phi) = \sum_{l,m} \psi_{lm}(r) \tx{e}^{i \omega_m t_{\tx{s}}} Y_{lm}(\theta,\phi)
\end{equation}
where the time dependence is the one implied by the spherical harmonic decomposition of the $\delta$ function.

Substituting the expressions for $q_{lm}$ and $\psi^{\tx{ret}}$ into Equation (\ref{Del2PsiSchw}) results in a differential equation for the radial part $\psi_{lm}(r)$ only:
\begin{equation}
\frac{d^2 \psi_{lm}}{dr^2} + \frac{2 (r-M)}{r (r-2M)} \frac{d \psi_{lm}}{dr} + \Big [ \frac{\omega_m^2 r^2}{(r-2M)^2} - \frac{l(l+1)}{r (r-2M)} \Big ] \psi_{lm} = - \frac{q_{lm}}{r_o-2M} \delta(r-r_o).
\label{DifEqPsilm}
\end{equation}
For points with $r \neq r_o$ the right hand side of Equation (\ref{DifEqPsilm}) is equal to zero and the differential equation can be solved numerically.

A variable that is used in the following is $r_{\ast}$, which is defined as
\begin{equation}
r_{\ast} = r + 2M  \ln(\frac{r}{2M}-1),
\end{equation}
and for which
\begin{equation}
\frac{dr_{\ast}}{dr} = \Big ( 1 - \frac{2M}{r} \Big )^{-1}.
\end{equation}

The behavior of $\psi_{lm}$ at the two boundaries needs to be considered first.
For large distances away from the orbit, i.e. in the limit $r \to \infty$, the $\psi_{lm}(r)$ is expected to behave as an outgoing wave, so it is assumed that
\begin{equation}
\psi_{lm}^{\tx{INF}}(r) = \frac{\tx{e}^{-i \omega_m r_{\ast}}}{r} \sum_{n=0}^{\infty} \frac{a_n}{r^n}.
\label{PsilmInfty}
\end{equation}
This expression must satisfy Equation (\ref{DifEqPsilm}).
The coefficients $a_n$ can be found by asserting exactly that.
First, the first and second derivatives of $\psi^{\tx{INF}}_{lm}(r)$ with respect to $r$ are calculated.
These derivatives are
\begin{equation}
\frac{d \psi_{lm}^{\tx{INF}}}{dr} = - \tx{e}^{- i \omega_m r_{\ast}} \bigg [ \frac{i \omega_m}{r-2M} \sum_{n=0}^{\infty} \frac{a_n}{r^n} + \sum_{n=0}^{\infty} (n+1) \frac{a_n}{r^{n+2}} \bigg ]
\end{equation}
and
\begin{equation}
\begin{split}
\frac{d^2 \psi_{lm}^{\tx{INF}}}{dr^2} = \tx{e}^{- i \omega_m r_{\ast}} \bigg [ & -\frac{\omega_m^2}{(r-2M)^2} \sum_{n=0}^{\infty} \frac{a_n}{r^{n-1}} 
+ \frac{2 i \omega_m}{(r-2M)} \sum_{n=0}^{\infty} \Big [ \frac{M}{r-2M} + (n+1) \Big ] \frac{a_n}{r^{n+1}} \\
		&+ \sum_{n=0}^{\infty} (n+1)(n+2) \frac{a_n}{r^{n+3}} \bigg ]. \\
\end{split}
\end{equation}
Substituting Equation (\ref{PsilmInfty}) and its two derivatives into Equation (\ref{DifEqPsilm})  results to an equation for the coefficients $a_n$.
That equation needs to be multiplied by $(r-2M)^2$ in order to get rid of the $(r-2M)$ and $(r-2M)^{2}$ factors that appear in some denominators.
Any factors of $(r-2M)$ that show up in the numerators can be absorbed into the sums of $r^n$ so that the final expression contains sums with powers of $r$ only.
By setting the factor of each $r^{n}$ term equal to zero, a recursion relation for the $a_n$'s is obtained and it is
\begin{equation}
\begin{split}
a_n = &\frac{l(l+1) - n(n-1) + 4 i \omega M}{2 i \omega n} a_{n-1} - i M \frac{(n-1)(2n-3)-l(l+1)}{\omega n} a_{n-2} \\
      &+ i 2 M^2 \frac{(n-2)^2}{\omega n} a_{n-3}.
\end{split}
\label{an}
\end{equation}
For this recursion relation it is assumed that $a_0 = 1$ and that $a_n =0$ for $n<0$.

The behavior of $\psi_{lm}(r)$ at the event horizon $r=2M$ is now considered.
There, $\psi_{lm}(r)$ should behave as an ingoing wave, so it can be assumed to be
\begin{equation}
\psi_{lm}^{\tx{HOR}}(r) = \frac{\tx{e}^{i \omega_m r_{\ast}}}{r} \sum_{n=0}^{\infty} b_n (r-2M)^n.
\label{PsilmHor}
\end{equation}
This expression must also be a solution of Equation (\ref{DifEqPsilm}).
As in the previous case, the derivatives are calculated first
\begin{equation}
\frac{d \psi_{lm}^{\tx{HOR}}}{dr} = \tx{e}^{i \omega_m r_{\ast}} \bigg [ \sum_{n=0}^{\infty} \big ( i \omega_m + \frac{n}{r} \big ) b_n (r-2M)^{n-1} - \frac{1}{r^2} \sum_{n=0}^{\infty} b_n (r-2M)^n \bigg ]
\end{equation}
and
\begin{equation}
\begin{split}
\frac{d^2 \psi_{lm}^{\tx{HOR}}}{dr^2} = &\tx{e}^{i \omega_m r_{\ast}} \bigg \{  \sum_{n=0}^{\infty} \Big [ -\frac{2 i \omega_m M}{r} - \omega_m^2 r + 2 i \omega_m n + \frac{n(n+1)}{r} \Big ] b_n (r-2M)^{n-2} \\
	& \qquad - \frac{2}{r} \sum_{n=0}^{\infty} \Big [i \omega_m + \frac{n}{r} \Big ] b_n (r-2M)^{n-1} + \frac{2}{r^3} \sum_{n=0}^{\infty} b_n (r-2M)^{n} \bigg \}.  \\
\end{split}
\end{equation}
Substituting Equation (\ref{PsilmHor}) and these derivatives into Equation (\ref{DifEqPsilm}) gives an equation for the coefficients $b_n$.
This equation must first be multiplied by $r^3$ so that the factors $r^3$, $r^2$ and $r$ that appear in some denominators vanish.
After that, any factors of $r^2$ and $r$ that show up in the numerators can be dealt with by the substitutions
\begin{eqnarray}
r^2 &=& (r-2M)^2 + 4M(r-2M) + 4 M^2 \\
r &=& (r-2M) + 2M
\end{eqnarray}
so that the final equation contains only sums of powers of $(r-2M)$.
By setting the coefficients of each power of $(r-2M)$ equal to zero, 
the recursion relation for the coefficients $b_n$ is obtained
\begin{equation}
\begin{split}
b_n = &- \frac{12 i \omega_m M (n-1) + (2n-3)(n-1) - (l^2+l+1)}{2M(4 i \omega_m Mn+n^2)} b_{n-1} \\
	&- \frac{12 i \omega_m M (n-2) + (n-2)(n-3) - l(l+1)}{4M^2 (4 i \omega_m n + n^2)} b_{n-2} \\
	& - \frac{i \omega_m (n-3)}{2 M^2 (4 i \omega_m M n+ n^2)} b_{n-3} \\
\end{split}
\label{bn}
\end{equation}
where $b_0 =1$ and $b_n=0$ for $n<0$.

\begin{figure}[ht]
\centering
\includegraphics[height=9.2cm,width=13.2cm,clip]{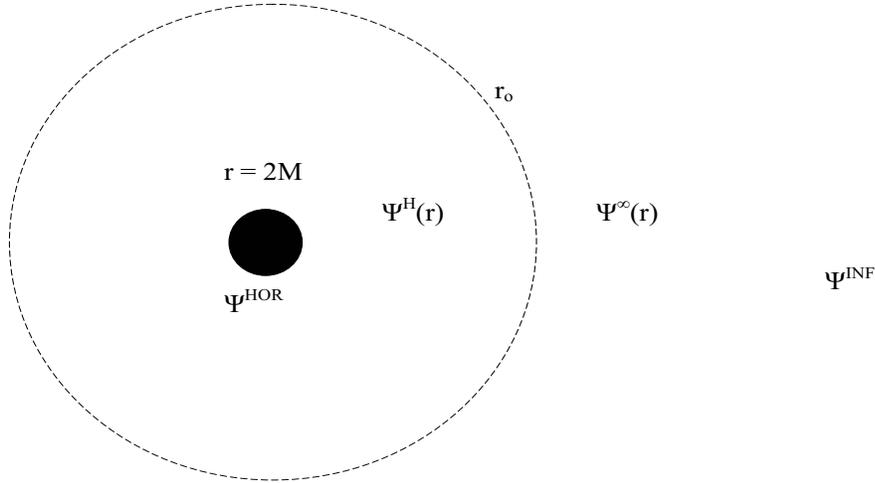}
\caption{Solutions obtained with the numerical code}
\label{RetSolutions}
\end{figure}

The expression (\ref{PsilmInfty}) with the coefficients (\ref{an}) is used as the starting solution for the numerical integration of Equation (\ref{DifEqPsilm}) from infinity to the radius of the orbit $r_o$.
This expression is an asymptotic expansion so it is important that the integration starts at a value of $r$ which is large enough for the terms of the sum to reach machine accuracy before the series  begins to diverge.
The result of this first integration is a homogeneous solution of Equation (\ref{DifEqPsilm}) for the region $r > r_o$, denoted here by $\psi_{lm}^{\infty}(r)$. 
Then, the expression (\ref{PsilmHor}) with the coefficients (\ref{bn}) is used as the starting solution for the numerical integration of Equation (\ref{DifEqPsilm}) from the event horizon $r=2M$ to the radius of the orbit $r_o$.
In this case, the integration starts at a radius $r$ that is close enough to $2M$ for the terms of the sum  to reach machine accuracy before the series starts to diverge.
The result of this integration is a homogeneous solution of Equation (\ref{DifEqPsilm}) for the region $r< r_o$, denoted here as $\psi_{lm}^{\tx{H}}(r)$.
All these solutions are shown in Figure (\ref{RetSolutions}).

Once the two solutions are found, they have to be matched at the location of the particle $r = r_o$.
If $\psi_{lm}(r)$ stands for the final normalized solution, which results after multiplying $\psi_{lm}^{\tx{H}}(r)$ and $\psi_{lm}^{\infty}(r)$ by the normalization factors $\hat{A}_{lm}$ and $\hat{B}_{lm}$ respectively, then
\begin{equation}
\psi_{lm}(r) = \left\{ \begin{array}{ll}
\hat{A}_{lm} \: \psi_{lm}^{\tx{H}}(r), \quad \tx{for } r<r_o \\
\hat{B}_{lm} \: \psi_{lm}^{\infty}(r), \quad \tx{for } r > r_o 
\end{array} \right.
\end{equation}
the factors $\hat{A}_{lm}$ and $\hat{B}_{lm}$ can be calculated.
The $\hat{}$ is used to avoid confusion with the regularization parameters $A_{lm}$ and $B_{lm}$ calculated in Chapter \ref{RP}.

The solution itself should be continuous at $r=r_o$:
\begin{equation}
\hat{A}_{lm} \:  \psi_{lm}^{\tx{H}}(r_o) = \hat{B}_{lm} \:  \psi_{lm}^{\infty}(r_o)
\label{ABEqn1}
\end{equation}
but its first derivative with respect to $r$ , also provided by the numerical integration, should be discontinuous.
The discontinuity is determined by the differential Equation (\ref{DifEqPsilm}) when the right-hand 
side term that contains the $\delta$ function is not set equal to zero.
Integrating Equation (\ref{DifEqPsilm}) from $r_o^-$ to $r_o^+$ gives
\begin{displaymath}
\begin{split}
&\int_{r_o^-}^{r_o^+} \frac{d}{dr} \Big ( \frac{d \psi_{lm}}{dr} \Big ) dr + \int_{r_o^-}^{r_o^+} \frac{2(r-M)}{r(r-2M)} \frac{d \psi_{lm}}{dr} dr + \int_{r_o^-}^{r_o^+} \frac{ \omega_m^2 r^2}{(r-2M)^2} \psi_{lm} dr \\
&- \int_{r_o^-}^{r_o^+} \frac{l(l+1)}{r(r-2M)} \psi_{lm} dr = - \frac{q_{lm}}{r_o-2M} \int_{r_o^-}^{r_o^+} \delta(r-r_o) dr \: \Longrightarrow \\
&\bigg [ \frac{d \psi_{lm}}{dr} \bigg ]_{r_o^-}^{r_o^+} + \bigg [ \frac{2 (r-M)}{r(r-2M)} \psi_{lm} \bigg ]_{r_o^-}^{r_o^+} - \int_{r_o^-}^{r_o^+} 2 \frac{r^2-2Mr-(r-M)(2r-2M)}{r^2 (r-2M)^2} \psi_{lm} dr \\
&\int_{r_o^-}^{r_o^+} \frac{ \omega_m^2 r^2}{(r-2M)^2} \psi_{lm} dr - \int_{r_o^-}^{r_o^+} \frac{l(l+1)}{r(r-2M)} \psi_{lm} dr = - \frac{q_{lm}}{r_o-2M},  \\
\end{split}
\end{displaymath}
or, taking into account the fact that for all the terms of the left-hand side other than the first one the integrand is continuous at $r = r_o$ and the integrals give no contribution
\begin{equation}
\hat{B}_{lm} \frac{d\psi_{lm}^{\infty}}{dr} \bigg |_{r_o} - \hat{A}_{lm} \frac{d \psi_{lm}^{\tx{H}}}{dr} \bigg |_{r_o} = - \frac{q_{lm}}{r_o-2M}.
\label{ABEqn2}
\end{equation}
Equations (\ref{ABEqn1}) and (\ref{ABEqn2}) can be easily solved for the factors $\hat{A}_{lm}$ and $\hat{B}_{lm}$ and the result is
\begin{eqnarray}
\hat{A}_{lm} &=& -\frac{q_{lm}}{r_o-2M} \: \frac{\psi_{lm}^{\infty}(r_o)}{\psi_{lm}^{\tx{H}}(r_o) \: \psi_{lm}^{\infty \; '}(r_o) - \psi_{lm}^{\infty}(r_o) \: \psi_{lm}^{\tx{H} \; '}(r_o)} \\ \label{Anorm}
\hat{B}_{lm} &=& -\frac{q_{lm}}{r_o-2M} \: \frac{\psi_{lm}^{\tx{H}}(r_o)}{\psi_{lm}^{\tx{H}}(r_o) \: \psi_{lm}^{\infty \; '}(r_o) - \psi_{lm}^{\infty}(r_o) \: \psi_{lm}^{\tx{H} \; '}(r_o)}  \label{Bnorm}
\end{eqnarray}
where the primes denote derivatives with respect to $r$.

After the retarded field is calculated, its contribution to the self-force along the radial direction can be obtained.
In fact, every $lm$-component as well as every $l$-component can be calculated by
\begin{eqnarray}
\mathcal{F}_{lm \: r}^{\tx{ret}} &=& \frac{d \psi_{lm}^{\tx{ret}}}{dr} \bigg |_{r_o} \\
\mathcal{F}_{l \: r}^{\tx{ret}} &=& \sum_{m=-l}^{l} \frac{d \psi_{lm}^{\tx{ret}}}{dr} \bigg |_{r_o}.
\end{eqnarray}
Finally, the total contribution from the retarded field is
\begin{equation}
\mathcal{F}_{r}^{\tx{ret}} = \sum_{l=0}^{\infty} \mathcal{F}_{l \: r}^{\tx{ret}}.
\end{equation}

A note on the angles $\theta$ and $\phi$ is necessary at this point.
The usual Schwarzschild angles $\theta$ and $\phi$ were used for the calculation of the $l$-modes of the retarded field in this chapter, while the rotated angles $\Theta$ and $\Phi$ were used for the calculation of the singular field and its regularization parameters in Chapter \ref{RP}.
Questions may arise from the fact that the two results must be subtracted for the calculation of the self-force.
But, as was mentioned in Chapter \ref{RP}, the rotation $(\theta,\phi) \to (\Theta,\Phi)$ preserves the spherical harmonic index $l$ so, as long as the summation over $m$ precedes the subtraction of the $l$-modes, there is no problem with the method.

\section{Numerical Code}
\label{NumCode}

The numerical code that was written for the calculation of the radial part of the retarded field and its contribution to the self-force is given in Appendix \ref{A2}.
The language used is C.
For the integration of Equation (\ref{DifEqPsilm}) the adaptive stepsize Runge-Kutta method is used, which is described in \cite{NR}.
The functions of \cite{NR} that were used for the integration are: \verb+odeint(), rkqs(), rkck()+.
Also, the function \verb+plgndr()+ for the calculation of the Associated Legendre polynomials, also available in \cite{NR}, was used.
The details for all these functions are given in \cite{NR} and 
for that reason they are not presented here.

An effort was made to keep the symbols used in the code for the different variables identical to the ones used in the analysis of this chapter. 
In the cases where that was not possible, the comments on the code should make the notation clear enough for the reader to follow.

      \chapter{Applications and Conclusions}
\label{CONC}

Some applications of the self-force caclulation are presented in this chapter and some important conclusions are drawn.
The effects of the scalar self-force are presented in detail.
The case of the gravitational self-force is also discussed.

\section{Equations of Motion}
\label{CONCEM}

In Chapter \ref{UFM} it was explained how the self-force on different particles moving in curved spacetime
can be calculated.
Since the ultimate goal is the determination of the motion of those particles, it is important to know how the self-force affects the equations of motion as well.

Assume that a point particle that carries a finite charge $q$ (which can be a scalar charge, an electric charge or the mass of the particle)
is moving in a known background spacetime which is characterized by the metric $g_{ab}$.
For simplicity also assume that no other (external to the particle) scalar, electromagnetic or gravitational fields exist.
If the effects of the charge $q$ are ignored, the point particle moves on a background geodesic.
If the geodesic is described by $z^a(s)$, where $s$ is the proper time and $u^a$ is the tangent to the geodesic, the geodesic equation is
\begin{equation}
u^a \nabla_a u^b = 0
\end{equation}
where $\nabla_a$ is the covariant derivative with respect to the background metric $g_{ab}$. 
The geodesic is assumed to be parametrized so that
\begin{equation}
u^a u_a = -1.
\label{uaua}
\end{equation}

If the interaction of the particle with its own fields is to be taken into account, the self-force $\mathcal{F}_{\tx{self}}^{a}$ should be included into the equations of motion.
Specifically, if $m$ is the mass of the particle, the equation that holds is
\begin{equation}
m u^a \nabla_a u^b = \mathcal{F}_{\tx{self}}^b
\label{SFGeodesic}
\end{equation}
where $\mathcal{F}_{\tx{self}}^a$ is given by Equations (\ref{ScalarSF1}), (\ref{EMSF1}) and (\ref{GravSF1}) for the scalar, electromagnetic and gravitational field respectively and is of $O(q^2)$.

\section{Effects of the Scalar Self-Force}
\label{EffectSF}

In general, Equation (\ref{SFGeodesic}) gives the effect of the self-force on each component 
of the 4-velocity of a scalar particle moving in spacetime, when the correct expression for the scalar self-force is substituted.
If the spherically symmetric Schwarzschild background is considered, it is easily deduced that only 3 components of the 4-velocity are affected by the self-force.
Indeed, as is explained in \cite{Wald}, the Schwarzschild metric has parity reflection symmetry $\theta \to \pi - \theta$,
so if the initial position $x^a$ and the initial tangent $u^a$ both lie on the equatorial plane $(\theta=\frac{\pi}{2})$ then the entire path must lie on the equatorial plane as well.
In addition, due to the spherical symmetry of the Schwarzschild background, every path can be brought by a rotation to the equatorial plane.
Consequently one can only consider equatorial orbits, without imposing any real restriction on the path of the particle.
The analysis that follows will be presented in \cite{dwdm} in more detail.

Assume that a trajectory of the Schwarzschild background is described by the coordinates
\begin{equation}
x^a : \{ t_{\tx{s}} = \tau(s), r = R(s), \theta=\frac{\pi}{2}, \phi=\Phi(s) \}.
\end{equation}
If $E$ is the energy per unit mass and $J$ the angular momentum per unit mass of the particle, it can easily be shown that
\begin{equation}
\begin{split}
u^a &= \frac{dx^a}{ds} = \Big ( \frac{ER}{(R-2M)}, \: \dot{R},\:  0, \: \frac{J}{R^2} \Big ) \\
u_a &= g_{ab} u^a = \Big ( -E, \: \frac{\dot{R}R}{(R-2M)},\: 0,\: J \Big ).\\
\end{split}
\label{SchPath}
\end{equation}
The normalization condition given in Equation (\ref{uaua}) for the 4-velocity gives a relationship between $E$, $J$ and $\dot{R}$, specifically
\begin{equation}
E^2- \dot{R}^2 = \Big ( 1+\frac{J^2}{R^2} \Big ) \Big ( 1 - \frac{2M}{R} \Big ).
\label{Normal}
\end{equation}
If the interaction of the scalar particle with its own field is to be taken into account, the geodesic equation is modified to
\begin{equation}
u^a \nabla_a u_b = q \nabla_b \psi^{\tx{R}}. 
\label{GeodesicPlusSF}
\end{equation}
Calculating the non-vanishing components of Equation (\ref{GeodesicPlusSF}) gives three separate equations, namely
\begin{eqnarray}
-\frac{du_t}{ds} &=& \frac{dE}{ds} = -q \partial_t \psi^{\tx{R}} \\ \label{dEds}
\frac{du_{\phi}}{ds} &=& \frac{dJ}{ds} = q \partial_{\phi} \psi^{\tx{R}} \\ \label{dphids}
\frac{d\dot{R}}{ds} &=& -\frac{M}{R^2} + \frac{R-3M}{R^4} J^2 + \frac{R-2M}{R} q \partial_r \psi^{\tx{R}}.  
\label{dRdotds}
\end{eqnarray}

In order to predict the effect of the self-force on circular orbits, one can assume that $\dot{R} =0$ and $\ddot{R}=0$ at some instant.
Then Equation (\ref{dRdotds}) gives the simple expression
\begin{equation}
J^2 = \frac{R^4}{R-3M} \Big ( \frac{M}{R^2} - \frac{R-2M}{R} q \partial_r \psi^{\tx{R}} \Big )
\label{CircOrb1}
\end{equation}
which gives the effect of the self-force on the angular momentum per unit mass of the scalar particle.

Also, Equation (\ref{Normal}) gives
\begin{equation}
E^2 = \Big ( 1-\frac{2M}{R} \Big ) \Big ( 1+\frac{J^2}{R^2} \Big ).
\label{CircOrb2}
\end{equation}
Combining Equations (\ref{CircOrb1}) and (\ref{CircOrb2}) results in
\begin{equation}
E^2 = \frac{(R-2M)^2}{R (R-3M)} (1-qR\partial_r \psi^{\tx{R}})
\label{EnergySelfForce}
\end{equation}
which shows the effect of the self-force on the energy of the scalar particle.
All the results mentioned up to this point are exact.

Since the effect of the scalar charge $q$ can be assumed to be small, the approximation 
that $(R \partial_r \psi^{\tx{R}}) << 1$ can be made, 
and Equation (\ref{EnergySelfForce}) for the energy $E$ can be written as
\begin{equation}
E = \frac{R-2M}{\sqrt{R(R-3M)}} (1-\frac{1}{2} q R\partial_r \psi^{\tx{R}}).
\label{CircOrb3}
\end{equation}

The orbital frequency is defined as
\begin{equation}
\Omega' \equiv \frac{d\phi}{dt_{\tx{s}}} = \frac{J (R-2M)}{E R^3},
\label{OrbFreq}
\end{equation}
from Equation (\ref{SchPath}).
Substituting Equations (\ref{CircOrb1}) and (\ref{CircOrb3}) into Equation (\ref{OrbFreq}), and using again the approximation $(R \partial_r \psi^{\tx{R}}) << 1$, a simple expression for the effect of the self-force on the orbital frequency is obtained:
\begin{equation}
\Omega' = \sqrt{\frac{M}{R^3}} \Big [ 1 - \frac{1}{2} \frac{R(R-3M)}{M} q \partial_r \psi^{\tx{R}} \Big ].
\label{OrbFreqSF}
\end{equation}
As expected, if the interaction of the particle with its own scalar field is ignored, the well-known result $\Omega = \sqrt{MR^{-3}}$ is recovered.

\subsection{Calculation of the Self-Force}
\label{EffectSF1}

A sample self-force calculation is presented in this section.
The scalar charge is assumed to be $q=1$ and to be moving on a circular orbit of radius $R=10M$ in a Schwarzschild background of $M=1$.

\begin{figure}[ht]
\centering
\includegraphics[height=10.1cm,width=13.1cm,clip]{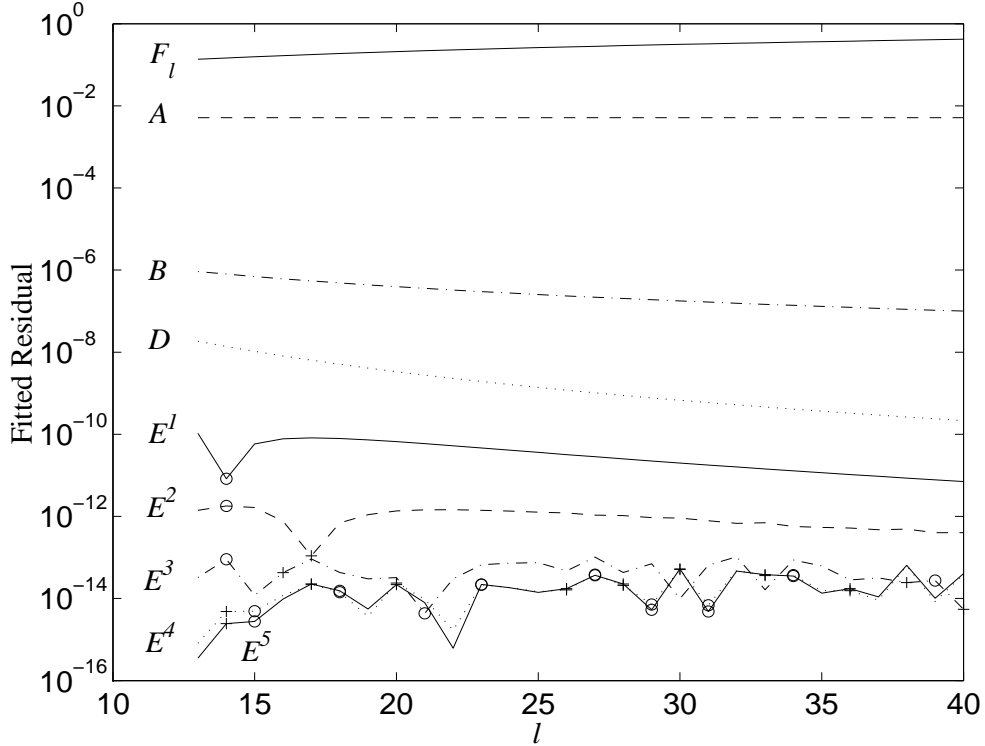}
\caption{Scalar Self-Force}
\label{fig8_1}
\end{figure}

The $lm$-components $\psi_{lm}(r)$ of the retarded field generated by the scalar charge and their contributions $\mathcal{F}_{l \: r}^{\tx{ret}}$ to the $l$-multipole modes of the retarded part of the self-force
are first calculated for $l=0,1, \ldots ,40$,
using the numerical code described in Chapter \ref{RSF} and given in Appendix \ref{A2}.
Then, the non-zero regularization parameters $A_r$, $B_r$ and $D_r$ are used \footnote{Remember that it was proven that $C_r=0$ for circular Schwarzschild orbits, in Chapter \ref{RP}.}
to subtract off the contributions of the singular-field self-force up to $O(l^{-2})$, according to Equation (\ref{SelfForceFinal}).
These results are shown in Figure (\ref{fig8_1}).
In that figure, the curve marked $F_l$ is the contribution $\mathcal{F}^{\tx{ret}}_{l\:r}$ as a function of $l$.
The curves marked $A$, $B$ and $D$ show the terms $(\mathcal{F}^{\tx{ret}}_{l\:r} - \mathcal{F}^{\tx{S}}_{l\:r})$ as functions of $l$, where $\mathcal{F}^{\tx{S}}_{l\:r}$ successively includes the contributions from the regularization parameters $A_r$, $B_r$ and $D_r$ respectively.
It is clear that the series for the self-force converges when the terms containing the regularization parameter $B_r$ are included.
However, the convergence becomes much faster the terms containing the parameter $D_r$ are also included.

As is rigorously explained in \cite{detmeswhi},
the contributions from the $E_a$ terms in Equation (\ref{SelfForceFinal}) are successively
of order $l^{-4}$, $l^{-6}$, $\ldots$.
Specifically, the $O(l^{-4})$ term is associated with the parameter $E_r^1$,
the $O(l^{-6})$ term is associated with the parameter $E_r^2$ and,
in general, the $O(l^{-2k})$ term is associated with the parameter $E_r^{k-1}$.
The parameters $E_r^k$ are $l$-independent \cite{detmeswhi} and can be determined as follows.
The contribution from each term is proven \cite{detmeswhi} to be
\begin{equation}
E_r^{k} \frac{(2l+1)}{[ (2l-2k-1)(2l-2k+1) \ldots (3l+2k+1)(2k+2k+3)]} \mathcal{P}_{k+1/2}
\label{Residual}
\end{equation}
where
\begin{equation}
\mathcal{P}_{k+1/2} = \frac{(-1)^{k+1} 2^{k+\frac{3}{2}} }{[(2k+1)!!]^2}.
\end{equation}
Assume that the residual $(\mathcal{F}^{\tx{ret}}_{l\:r} - \mathcal{F}^{\tx{S}}_{l\:r})$ as a function of $l$,
resulting after the subtraction of the contribution of $A_r$, $B_r$ and $D_r$, is calculated.
The coefficients $E_r^k$ can be determined numerically by fitting that residual,
to a linear combination of terms of the form (\ref{Residual}) and successively removing their contributions. 

The curves marked $E^1, \ldots, E^5$ in Figure (\ref{fig8_1}) are the residuals resulting after the numerical fit of the parameters $E_r^1, \ldots, E^5_r$ respectively.
It was noticed that fitting more than four of the coefficients $E_r^k$ did not improve the residual, which had already reached machine accuracy by the fourth fitting.

\subsection{Change in Orbital Frequency}
\label{EffectSF2}

\begin{figure}[h!]
\centering
\includegraphics[height=10.1cm,width=13.1cm,clip]{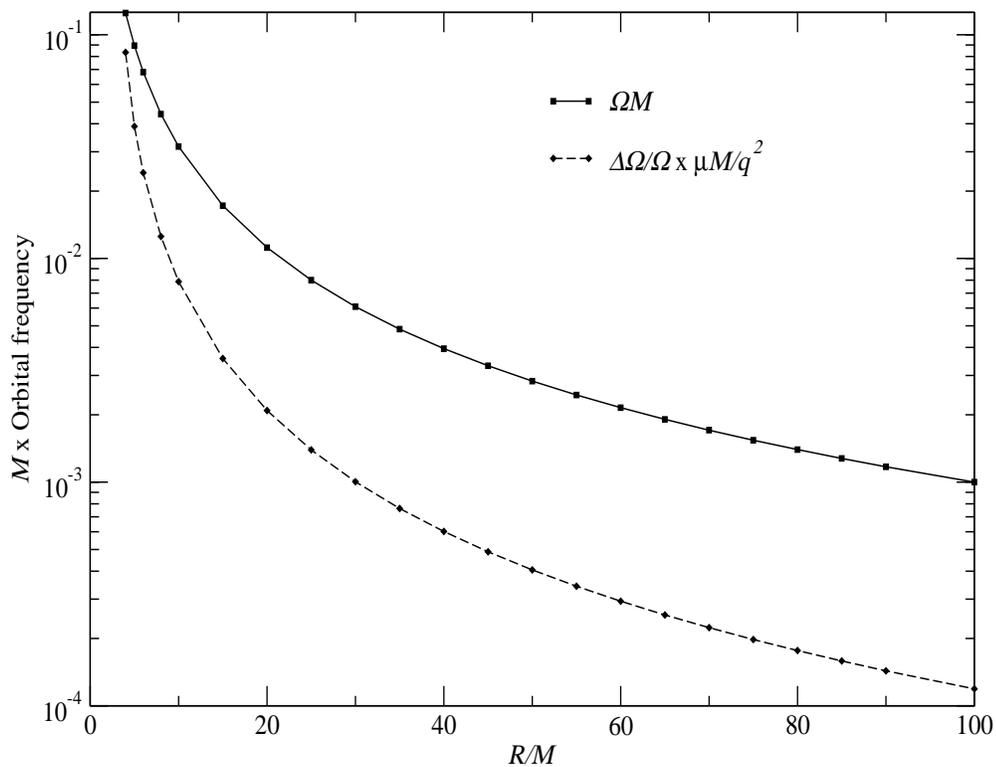}
\caption{Change in Orbital Frequency}
\label{fig8_2}
\end{figure}

It was shown in Section (\ref{EffectSF}) that the orbital frequency of the circular orbit 
is affected by the self-force.
Using Equation (\ref{OrbFreqSF}) the difference in frequency can be calculated; it is
\begin{equation}
\frac{\Omega'-\Omega}{\Omega} = \frac{\Delta \Omega}{\Omega} = - \frac{1}{2} \frac{R(R-3M)}{M} q \partial_r \psi^{\tx{R}}.
\end{equation}
The effect of the self-force on the orbital frequency is shown in Figure (\ref{fig8_2}), where the ratio $(\Delta \Omega/\Omega)$ is plotted as a function of $(R/M)$.
The orbital frequency $\Omega$ before the self-force effects are included is also plotted for comparison. 

%
%

\section{Effects of the Gravitational Self-Force}
\label{CONCGrav}

Assume that a particle of mass $m$ is moving in a background spacetime described by the metric $g_{ab}$, with no other extrernal fields present.
Equation (\ref{SFGeodesic}) gives the worldline of the particle that includes the self-force effect
if $\mathcal{F}^a_{\tx{self}}$ is substituted from Equation (\ref{GravSF1}).
Clearly, given the initial position and 4-velocity of the particle, that worldline is determined exclusively by the background metric $g_{ab}$ and the metric perturbation $h_{ab}^{\tx{R}}$ induced by the particle.
Both of these contributions are solutions of the homogeneous Einstein equations.

An observer making local measurements of the metric in the vicinity of the particle measures two contributions to it.
The first contribution comes from the background metric $g_{ab}$ combined with the $h_{ab}^{\tx{R}}$-part of the actual metric perturbation generated by the particle.
Since both these terms obey the homogeneous Einstein equations, the observer sees no source for either one, so
no local measurement can allow him to distinguish $h_{ab}^{\tx{R}}$ from the background metric $g_{ab}$.
The second contribution is $h_{ab}^{\tx{S}}$, a metric perturbation also generated by the particle which, according to the observer, is sourced by the particle.
The local observer sees the particle move in a \emph{background} geodesic of the metric $(g_{ab}+h_{ab}^{\tx{R}})$.
Because he only makes local measurements, he observes no radiation generated by the moving particle and consequently no effect that he could describe as radiation reaction \cite{detwhi02}.

      \appendix		
      \chapter{\textsc{GRTensor} Code for the Singular Fields}
\label{A0}
  
\begin{singlespace}
\begin{verbatim}
restart; 
 
readlib(grii);
grtensor();
 
qload(thz);
 
e := epsilon;
 
ord := [t, x, y, z, Bxx, Byy];
 
EpsEqn := {e^10=0, e^9=0, e^8=0, e^7=0, e^6=0, e^5=0, e^4=0, e^3=0, 
	   e^2=0};
Traces := {Ezz=-Exx-Eyy, Bzz=-Bxx-Byy};



 # The Ricci Tensor Should Be Zero #

grcalc(detg);
grmap(detg, series, 'x', e, 4);
grmap(detg, convert, 'x', polynom);
grmap(detg, expand, 'x');
grmap(detg, subs, EpsEqn, 'x'); 
grmap(detg, collect, 'x', ord, distributed, factor);
grmap(detg, subs, Traces, 'x');
grmap(detg, collect, 'x', ord, distributed,factor);
grdisplay(detg);
 
grcalc(g(up,up));
grmap(g(up,up), series, 'x', e, 4);  
grmap(g(up,up), convert, 'x', polynom);  
grmap(g(up,up), expand, 'x');  
grmap(g(up,up), subs, EpsEqn, 'x');  
grmap(g(up,up), collect, 'x', ord, distributed, factor);  
grmap(g(up,up), subs, Traces, 'x');  
grmap(g(up,up), collect, 'x', ord, distributed, factor);  
grdisplay(g(up,up));
 
grcalc(Chr(dn,dn,dn));
grmap(Chr(dn,dn,dn), series, 'x', e, 4);  
grmap(Chr(dn,dn,dn),convert, 'x', polynom);  
grmap(Chr(dn,dn,dn), expand, 'x'); 
grmap(Chr(dn,dn,dn), subs, EpsEqn, 'x'); 
grmap(Chr(dn,dn,dn), collect, 'x', ord, distributed, factor);  
grmap(Chr(dn,dn,dn), subs, Traces, 'x');  
grmap(Chr(dn,dn,dn), collect, 'x', ord, distributed, factor);  
grdisplay(Chr(dn,dn,dn));
 
grcalc(Chr(dn,dn,up));
grmap(Chr(dn,dn,up), series, 'x', e, 4);  
grmap(Chr(dn,dn,up), convert, 'x', polynom);  
grmap(Chr(dn,dn,up), expand, 'x'); 
grmap(Chr(dn,dn,up), subs, EpsEqn, 'x');  
grmap(Chr(dn,dn,up), collect, 'x', ord, distributed, factor);  
grmap(Chr(dn,dn,up), subs, Traces, 'x');  
grmap(Chr(dn,dn,up), collect, 'x', ord, distributed, factor);
grdisplay(Chr(dn,dn,up));
 
grcalc(Chr(dn,up,up));
grmap(Chr(dn,up,up), series, 'x', e, 4);  
grmap(Chr(dn,up,up), convert, 'x', polynom);  
grmap(Chr(dn,up,up), expand, 'x'); 
grmap(Chr(dn,up,up), subs, EpsEqn, 'x');  
grmap(Chr(dn,up,up), collect, 'x', ord, distributed, factor);  
grmap(Chr(dn,up,up), subs, Traces, 'x');  
grmap(Chr(dn,up,up), collect, 'x', ord, distributed, factor);
grdisplay(Chr(dn,up,up));
 
grcalc(Chr(dn,up,dn));
grmap(Chr(dn,up,dn), series, 'x', e, 4);  
grmap(Chr(dn,up,dn), convert, 'x', polynom);  
grmap(Chr(dn,up,dn), expand, 'x'); 
grmap(Chr(dn,up,dn), subs, EpsEqn, 'x');  
grmap(Chr(dn,up,dn), collect, 'x', ord, distributed, factor);  
grmap(Chr(dn,up,dn), subs, Traces, 'x');  
grmap(Chr(dn,up,dn), collect, 'x', ord, distributed, factor);  
grdisplay(Chr(dn,up,dn));
  
grcalc(R(dn,dn,dn,dn));
grmap(R(dn,dn,dn,dn), series, 'x', e, 4);  
grmap(R(dn,dn,dn,dn), convert, 'x', polynom);  
grmap(R(dn,dn,dn,dn), expand, 'x'); 
grmap(R(dn,dn,dn,dn), subs, EpsEqn, 'x');  
grmap(R(dn,dn,dn,dn), collect, 'x', ord, distributed, factor);  
grmap(R(dn,dn,dn,dn), subs, Traces, 'x');  
grmap(R(dn,dn,dn,dn), collect, 'x', ord, distributed, factor);  
grdisplay(R(dn,dn,dn,dn));
 
grcalc(R(up,up,dn,dn));
grmap(R(up,up,dn,dn), series, 'x', e, 4);  
grmap(R(up,up,dn,dn), convert, 'x', polynom);  
grmap(R(up,up,dn,dn), expand, 'x'); 
grmap(R(up,up,dn,dn), subs, EpsEqn, 'x');  
grmap(R(up,up,dn,dn), collect, 'x', ord, distributed, factor);  
grmap(R(up,up,dn,dn), subs, Traces, 'x');  
grmap(R(up,up,dn,dn), collect, 'x', ord, distributed, factor);  
grdisplay(R(up,up,dn,dn));
 
grcalc(R(up,dn,dn,dn));
grmap(R(up,dn,dn,dn), series, 'x', e, 4);  
grmap(R(up,dn,dn,dn), convert, 'x', polynom);  
grmap(R(up,dn,dn,dn), expand, 'x'); 
grmap(R(up,dn,dn,dn), subs, EpsEqn, 'x');  
grmap(R(up,dn,dn,dn), collect, 'x', ord, distributed, factor);  
grmap(R(up,dn,dn,dn), subs, Traces, 'x');  
grmap(R(up,dn,dn,dn), collect, 'x', ord, distributed, factor);
grdisplay(R(up,dn,dn,dn));
 
grcalc(R(dn,up,dn,up));
grmap(R(dn,up,dn,up), series, 'x', e, 4);  
grmap(R(dn,up,dn,up), convert, 'x', polynom);  
grmap(R(dn,up,dn,up), expand, 'x'); 
grmap(R(dn,up,dn,up), subs, EpsEqn, 'x');  
grmap(R(dn,up,dn,up), collect, 'x', ord, distributed, factor); 
grmap(R(dn,up,dn,up), subs, Traces, 'x');  
grmap(R(dn,up,dn,up), collect, 'x', ord, distributed, factor);  
grdisplay(R(dn,up,dn,up));
 
grcalc(R(dn,dn));
grmap(R(dn,dn), series, 'x', e, 4);  
grmap(R(dn,dn), convert, 'x', polynom);  
grmap(R(dn,dn), expand, 'x');  grmap(R(dn,dn), subs, EpsEqn, 'x');  
grmap(R(dn,dn), collect, 'x', ord, distributed, factor); 
grmap(R(dn,dn), subs, Traces, 'x');  
grmap(R(dn,dn), collect, 'x', ord, distributed, factor);  
grdisplay(R(dn,dn));
 



 # Useful Quantities #
 
grdef(`etadn{(a b)} := -kdelta{^$t a}*kdelta{^$t b}
	+kdelta{^$x a}*kdelta{^$x b} + kdelta{^$y a}*kdelta{^$y b}
	+kdelta{^$z a}*kdelta{^$z b}`);
grcalc(etadn(dn,dn));
grdisplay(etadn(dn,dn));
 
grdef(`etaup{(^a ^b)} := -kdelta{$t ^a}*kdelta{$t ^b}
	+kdelta{$x ^a}*kdelta{$x ^b} + kdelta{$y ^a}*kdelta{$y ^b}
	+kdelta{$z ^a}*kdelta{$z ^b}`);
grcalc(etaup(up,up));
grdisplay(etaup(up,up));
 
grdef(`eps{[a b c]} := LevCS{d a b c}*kdelta{$t ^d}`);
grcalc(eps(dn,dn,dn));
grdisplay(eps(dn,dn,dn));


grdef(`EE{(a b)} := kdelta{^$x a}*kdelta{^$x b}*e*Exx
		+ kdelta{^$y a}*kdelta{^$y b}*e*Eyy 
		+ kdelta{^$z a}*kdelta{^$z b}*e*Ezz 
		+ kdelta{^$x a}*kdelta{^$y b}*e*Exy
   		+ kdelta{^$x a}*kdelta{^$z b}*e*Exz 
		+ kdelta{^$y a}*kdelta{^$z b}*e*Eyz`);
grcalc(EE(dn,dn)); 
grdisplay(EE(dn,dn));
 
grdef(`BB{(a b)} := kdelta{^$x a}*kdelta{^$x b}*e*Bxx
		+ kdelta{^$y a}*kdelta{^$y b}*e*Byy
		+ kdelta{^$z a}*kdelta{^$z b}*e*Bzz 
		+ kdelta{^$x a}*kdelta{^$y b}*e*Bxy
		+ kdelta{^$x a}*kdelta{^$z b}*e*Bxz 
		+ kdelta{^$y a}*kdelta{^$z b}*e*Byz`);
grcalc(BB(dn,dn)); 
grdisplay(BB(dn,dn));
 
 
grdef(`h0{(b c)} := 
      kdelta{^$t b}*kdelta{^$x c}*2*A*mu*y/((x^2+y^2+z^2)^(3/2))
    + kdelta{^$t b}*kdelta{^$y c}*(-2)*A*mu*x/((x^2+y^2+z^2)^(3/2))`);
grcalc(h0(dn,dn)); 
grdisplay(h0(dn,dn));
 
 
grdef(`epsx{[a b]} := eps{c a b}*kdelta{$x ^c}`);
grcalc(epsx(dn,dn));
grdisplay(epsx(dn,dn));
 
grdef(`epsy{[a b]} := eps{c a b}*kdelta{$y ^c}`);
grcalc(epsy(dn,dn));
grdisplay(epsy(dn,dn));
 
grdef(`epsz{[a b]} := eps{c a b}*kdelta{$z ^c}`);
grcalc(epsz(dn,dn));
grdisplay(epsz(dn,dn));
 
 
 
 # Chech What The Equations Look Like: #

grdef(`h1test{(a b)} := kdelta{^$t a}*kdelta{^$t b}*e*h1tt 
		+ kdelta{^$t a}*kdelta{^$x b}*e*h1tx 
		+ kdelta{^$t a}*kdelta{^$y b}*e*h1ty 
		+ kdelta{^$t a}*kdelta{^$z b}*e*h1tz 
		+ kdelta{^$x a}*kdelta{^$x b}*e*h1xx 
		+ kdelta{^$x a}*kdelta{^$y b}*e*h1xy 
		+ kdelta{^$x a}*kdelta{^$z b}*e*h1xz 
		+ kdelta{^$y a}*kdelta{^$y b}*e*h1yy 
		+ kdelta{^$y a}*kdelta{^$z b}*e*h1yz 
		+ kdelta{^$z a}*kdelta{^$z b}*e*h1zz`);
grcalc(h1test(dn,dn)); 
grdisplay(h1test(dn,dn));
 
grdef(`htest{(a b)} := h0{a b} + h1test{a b}` );
grcalc(htest(dn,dn)); 
grdisplay(htest(dn,dn));
 
grdef(`htesttrace := htest{^c c}`); 
grcalc(htesttrace);
grmap(htesttrace, series, 'x', e, 4);  
grmap(htesttrace, convert, 'x', polynom);  
grmap(htesttrace, expand, 'x');  grmap(htesttrace, subs, EpsEqn, 'x');  
grmap(htesttrace, collect, 'x', ord, distributed, factor);  
grmap(htesttrace, subs, Traces, 'x');  
grmap(htesttrace, collect, 'x', ord, distributed, factor);  
grdisplay(htesttrace);
 
grdef(`hbartest{(a b)} := htest{a b} - (1/2)*g{a b}*htesttrace`);
grcalc(hbartest(dn,dn)); 
grmap(hbartest(dn,dn), series, 'x', e, 4); 
grmap(hbartest(dn,dn), convert, 'x', polynom);  
grmap(hbartest(dn,dn), expand, 'x');  
grmap(hbartest(dn,dn), subs, EpsEqn, 'x'); 
grmap(hbartest(dn,dn), collect, 'x', ord, distributed, factor); 
grmap(hbartest(dn,dn), subs, Traces, 'x');  
grmap(hbartest(dn,dn), collect, 'x', ord, distributed, factor); 
grdisplay(hbartest(dn,dn));   
                                   
 

 # Left-Hand Side Of The Equation #
 
grcalc(hbartest(dn,dn,cdn)); 
grmap(hbartest(dn,dn,cdn), series,'x', e, 4);  
grmap(hbartest(dn,dn,cdn), convert, 'x', polynom); 
grmap(hbartest(dn,dn,cdn), expand, 'x');  
grmap(hbartest(dn,dn,cdn), subs, EpsEqn, 'x');  
grmap(hbartest(dn,dn,cdn), collect, 'x', ord, distributed, factor);  
grmap(hbartest(dn,dn,cdn), subs, Traces, 'x'); 
grmap(hbartest(dn,dn,cdn), collect, 'x', ord, distributed, factor);
grdisplay(hbartest(dn,dn,cdn));
 
grdef(`Dhbartest{a b ^c} := hbartest{a b ;^c}`);
grcalc(Dhbartest(dn,dn,up)); 
grmap(Dhbartest(dn,dn,up), series, 'x', e, 4); 
grmap(Dhbartest(dn,dn,up), convert, 'x', polynom); 
grmap(Dhbartest(dn,dn,up), expand, 'x');  
grmap(Dhbartest(dn,dn,up), subs, EpsEqn, 'x');  
grmap(Dhbartest(dn,dn,up), collect, 'x', ord, distributed, factor);  
grmap(Dhbartest(dn,dn,up), subs, Traces, 'x'); 
grmap(Dhbartest(dn,dn,up), collect, 'x', ord, distributed, factor);
grdisplay(Dhbartest(dn,dn,up));
 
grdef(`DDhbartest{a b} := Dhbartest{a b ^c ;c}`);
grcalc(DDhbartest(dn,dn)); 
grmap(DDhbartest(dn,dn), series, 'x', e, 4);  
grmap(DDhbartest(dn,dn), convert, 'x', polynom); 
grmap(DDhbartest(dn,dn), expand, 'x');  
grmap(DDhbartest(dn,dn), subs, EpsEqn, 'x');  
grmap(DDhbartest(dn,dn), collect, 'x', ord, distributed, factor);  
grmap(DDhbartest(dn,dn), subs, Traces, 'x'); 
grmap(DDhbartest(dn,dn), collect, 'x', ord, distributed, factor);
grdisplay(DDhbartest(dn,dn));
 
grdef(`Rhbartest{a b} := 2*R{a ^c b ^d}*hbartest{c d}`);
grcalc(Rhbartest(dn,dn)); 
grmap(Rhbartest(dn,dn), series, 'x', e, 4);  
grmap(Rhbartest(dn,dn), convert, 'x', polynom); 
grmap(Rhbartest(dn,dn), expand, 'x');  
grmap(Rhbartest(dn,dn), subs, EpsEqn, 'x');  
grmap(Rhbartest(dn,dn), collect, 'x', ord, distributed, factor);  
grmap(Rhbartest(dn,dn), subs, Traces, 'x'); 
grmap(Rhbartest(dn,dn), collect, 'x', ord, distributed, factor); 
grdisplay(Rhbartest(dn,dn));
 
grdef(`LHStest{(a b)} := DDhbartest{a b} + Rhbartest{a b}`);
grcalc(LHStest(dn,dn)); 
grmap(LHStest(dn,dn), series, 'x', e, 4); 
grmap(LHStest(dn,dn), convert, 'x', polynom);  
grmap(LHStest(dn,dn), expand, 'x');  
grmap(LHStest(dn,dn), subs, EpsEqn, 'x'); 
grmap(LHStest(dn,dn), collect, 'x', ord, distributed, factor); 
grmap(LHStest(dn,dn), subs, Traces, 'x');  
grmap(LHStest(dn,dn), collect, 'x', ord, distributed, factor);
grdisplay(LHStest(dn,dn));
 
 
 
 # Assume A Specific Form For The Solution #
 
grdef(`AngM{^a} := [0, 0, 0, A]`);
grcalc(AngM(up)); grdisplay(AngM(up));
 
 
grdef(`h1{(a b)} := 
kdelta{^$t a}*kdelta{^$t b}*(mu/((x^2+y^2+z^2)^(3/2)))* 
   (att*etadn{i l}*AngM{^l}*x{^i}*BB{j k}*x{^j}*x{^k}
  + btt*AngM{^i}*BB{i j}*x{^j}*(x^2+y^2+z^2))           
+kdelta{^$t a}*kdelta{^$x b}*(mu/((x^2+y^2+z^2)^(3/2)))* 
   (cx*epsx{i j}*AngM{^i}*etaup{^j ^n}*EE{n l}*x{^l}*(x^2+y^2+z^2)
  + dx*epsx{i j}*AngM{^i}*x{^j}*EE{l k}*x{^l}*x{^k}
  + fx*epsx{i j}*AngM{^l}*etaup{^i ^n}*EE{n l}*x{^j}*(x^2+y^2+z^2)
  + gx*epsx{i j}*etadn{m l}*AngM{^m}*etaup{^i ^n}*EE{n k}
      *x{^j}*x{^l}*x{^k}
  + px*eps{i j l}*AngM{^i}*etaup{^j ^n}*EE{$x n}*x{^l}*(x^2+y^2+z^2)
  + qx*eps{i j l}*AngM{^i}*etaup{^j ^n}*EE{n k}
      *x{^d}*kdelta{^$x d}*x{^l}*x{^k})
+kdelta{^$t a}*kdelta{^$y b}* (mu/((x^2+y^2+z^2)^(3/2)))*
   (cy*epsy{i j}*AngM{^i}*etaup{^j ^n}*EE{n l}*x{^l}*(x^2+y^2+z^2)
  + dy*epsy{i j}*AngM{^i}*x{^j}*EE{l k}*x{^l}*x{^k}
  + fy*epsy{i j}*AngM{^l}*etaup{^i ^n}*EE{n l}*x{^j}*(x^2+y^2+z^2)
  + gy*epsy{i j}*etadn{m l}*AngM{^m}*etaup{^i ^n}*EE{n k}
      *x{^j}*x{^l}*x{^k}
  + py*eps{i j l}*AngM{^i}*etaup{^j ^n}*EE{$y n}*x{^l}*(x^2+y^2+z^2)
  + qy*eps{i j l}*AngM{^i}*etaup{^j ^n}*EE{n k}
      *x{^d}*kdelta{^$y d}*x{^l}*x{^k})
+kdelta{^$t a}*kdelta{^$z b}* (mu/((x^2+y^2+z^2)^(3/2)))*
   (cz*epsz{i j}*AngM{^i}*etaup{^j ^n}*EE{n l}*x{^l}*(x^2+y^2+z^2)
  + dz*epsz{i j}*AngM{^i}*x{^j}*EE{l k}*x{^l}*x{^k}
  + fz*epsz{i j}*AngM{^l}*etaup{^i ^n}*EE{n l}*x{^j}*(x^2+y^2+z^2)
  + gz*epsz{i j}*etadn{m l}*AngM{^m}*etaup{^i ^n}*EE{n k}
      *x{^j}*x{^l}*x{^k}
  + pz*eps{i j l}*AngM{^i}*etaup{^j ^n}*EE{$z n}*x{^l}*(x^2+y^2+z^2)
  + qz*eps{i j l}*AngM{^i}*etaup{^j ^n}*EE{n k}
      *x{^d}*kdelta{^$z d}*x{^l}*x{^k})
+kdelta{^$x a}*kdelta{^$x b}*(mu/((x^2+y^2+z^2)^(3/2)))*
   (cxx*etadn{i $x}*AngM{^i}*BB{$x k}*x{^k}*(x^2+y^2+z^2)
  + dxx*etadn{i $x}*AngM{^i}*BB{k l}*x{^k}*etadn{$x j}*x{^j}*x{^l}
  + fxx*etadn{k i}*AngM{^i}*BB{$x $x}*x{^k}*(x^2+y^2+z^2)
  + gxx*AngM{^k}*BB{$x k}*etadn{i $x}*x{^i}*(x^2+y^2+z^2)
  + pxx*etadn{i k}*AngM{^i}*BB{$x l}*etadn{j $x}*x{^j}*x{^k}*x{^l}
  + qxx*AngM{^k}*BB{k l}*etadn{i $x}*x{^i}*etadn{j $x}*x{^j}*x{^l}
  + vxx*etadn{$x $x}*etadn{i l}*AngM{^l}*BB{j k}*x{^i}*x{^j}*x{^k}
  + uxx*etadn{$x $x}*AngM{^i}*BB{i j}*x{^j}*(x^2+y^2+z^2))
+kdelta{^$x a}*kdelta{^$y b}*(mu/((x^2+y^2+z^2)^(3/2)))*
  (axy*AngM{^i}*kdelta{i ^$y}*BB{$x k}*x{^k}*(x^2+y^2+z^2)
  + bxy*AngM{^k}*kdelta{k ^$y}*BB{i j}*x*x{^i}*x{^j}
  + cxy*etadn{i $x}*AngM{^i}*BB{$y k}*x{^k}*(x^2+y^2+z^2)
  + dxy*etadn{i $x}*AngM{^i}*BB{k l}*x{^k}*etadn{$y j}*x{^j}*x{^l}
  + fxy*etadn{k i}*AngM{^i}*BB{$x $y}*x{^k}*(x^2+y^2+z^2)
  + gxy*AngM{^k}*BB{$x k}*etadn{i $y}*x{^i}*(x^2+y^2+z^2)
  + pxy*etadn{i k}*AngM{^i}*BB{$x l}*etadn{j $y}*x{^j}*x{^k}*x{^l}
  + qxy*AngM{^k}*BB{k l}*etadn{i $x}*x{^i}*etadn{j $y}*x{^j}*x{^l}
  + sxy*etadn{i k}*AngM{^k}*BB{$y j}*x*x{^i}*x{^j} 
  + txy*AngM{^j}*BB{$y j}*x*(x^2+y^2+z^2))                            
+kdelta{^$x a}*kdelta{^$z b}*(mu/((x^2+y^2+z^2)^(3/2)))*
  (axz*AngM{^i}*kdelta{i ^$z}*BB{$x k}*x{^k}*(x^2+y^2+z^2)
  + bxz*AngM{^k}*kdelta{k ^$z}*BB{i j}*x*x{^i}*x{^j}
  + cxz*etadn{i $x}*AngM{^i}*BB{$z k}*x{^k}*(x^2+y^2+z^2)
  + dxz*etadn{i $x}*AngM{^i}*BB{k l}*x{^k}*etadn{$z j}*x{^j}*x{^l}
  + fxz*etadn{k i}*AngM{^i}*BB{$x $z}*x{^k}*(x^2+y^2+z^2)
  + gxz*AngM{^k}*BB{$x k}*etadn{i $z}*x{^i}*(x^2+y^2+z^2)
  + pxz*etadn{i k}*AngM{^i}*BB{$x l}*etadn{j $z}*x{^j}*x{^k}*x{^l}
  + qxz*AngM{^k}*BB{k l}*etadn{i $x}*x{^i}*etadn{j $z}*x{^j}*x{^l}
  + sxz*etadn{i k}*AngM{^k}*BB{$z j}*x*x{^i}*x{^j} 
  + txz*AngM{^j}*BB{$z j}*x*(x^2+y^2+z^2))                          
+kdelta{^$y a}*kdelta{^$y b}*(mu/((x^2+y^2+z^2)^(3/2)))*
  (cyy*etadn{i $y}*AngM{^i}*BB{$y k}*x{^k}*(x^2+y^2+z^2)
  + dyy*etadn{i $y}*AngM{^i}*BB{k l}*x{^k}*etadn{$y j}*x{^j}*x{^l}
  + fyy*etadn{k i}*AngM{^i}*BB{$y $y}*x{^k}*(x^2+y^2+z^2)
  + gyy*AngM{^k}*BB{$y k}*etadn{i $y}*x{^i}*(x^2+y^2+z^2)
  + pyy*etadn{i k}*AngM{^i}*BB{$y l}*etadn{j $y}*x{^j}*x{^k}*x{^l}
  + qyy*AngM{^k}*BB{k l}*etadn{i $y}*x{^i}*etadn{j $y}*x{^j}*x{^l}
  + vyy*etadn{$y $y}*etadn{i l}*AngM{^l}*BB{j k}*x{^i}*x{^j}*x{^k}
  + uyy*etadn{$y $y}*AngM{^i}*BB{i j}*x{^j}*(x^2+y^2+z^2))
+kdelta{^$y a}*kdelta{^$z b}*(mu/((x^2+y^2+z^2)^(3/2)))*
  (ayz*AngM{^i}*kdelta{i ^$z}*BB{$y k}*x{^k}*(x^2+y^2+z^2)
  + byz*AngM{^k}*kdelta{k ^$z}*BB{i j}*y*x{^i}*x{^j}
  + cyz*etadn{i $y}*AngM{^i}*BB{$z k}*x{^k}*(x^2+y^2+z^2)
  + dyz*etadn{i $y}*AngM{^i}*BB{k l}*x{^k}*etadn{$z j}*x{^j}*x{^l}
  + fyz*etadn{k i}*AngM{^i}*BB{$y $z}*x{^k}*(x^2+y^2+z^2)
  + gyz*AngM{^k}*BB{$y k}*etadn{i $z}*x{^i}*(x^2+y^2+z^2)
  + pyz*etadn{i k}*AngM{^i}*BB{$y l}*etadn{j $z}*x{^j}*x{^k}*x{^l}
  + qyz*AngM{^k}*BB{k l}*etadn{i $y}*x{^i}*etadn{j $z}*x{^j}*x{^l}
  + syz*etadn{i k}*AngM{^k}*BB{$z j}*y*x{^i}*x{^j} 
  + tyz*AngM{^j}*BB{$z j}*y*(x^2+y^2+z^2))                           
+kdelta{^$z a}*kdelta{^$z b}*(mu/((x^2+y^2+z^2)^(3/2)))*
  (czz*etadn{i $z}*AngM{^i}*BB{$z k}*x{^k}*(x^2+y^2+z^2)
  + dzz*etadn{i $z}*AngM{^i}*BB{k l}*x{^k}*etadn{$z j}*x{^j}*x{^l}
  + fzz*etadn{k i}*AngM{^i}*BB{$z $z}*x{^k}*(x^2+y^2+z^2)
  + gzz*AngM{^k}*BB{$z k}*etadn{i $z}*x{^i}*(x^2+y^2+z^2)
  + pzz*etadn{i k}*AngM{^i}*BB{$z l}*etadn{j $z}*x{^j}*x{^k}*x{^l}
  + qzz*AngM{^k}*BB{k l}*etadn{i $z}*x{^i}*etadn{j $z}*x{^j}*x{^l}
  + vzz*etadn{$z $z}*etadn{i l}*AngM{^l}*BB{j k}*x{^i}*x{^j}*x{^k}
  + uzz*etadn{$z $z}*AngM{^i}*BB{i j}*x{^j}*(x^2+y^2+z^2))` );
grcalc(h1(dn,dn));
 
 
grdef(`hh{(a b)} := h0{a b} + h1{a b}`);
grcalc(hh(dn,dn));
 
grdef(`htrace := hh{^c c}`);
grcalc(htrace);  
grmap(htrace, series, 'x', e, 4);  
grmap(htrace, convert, 'x', polynom);
grmap(htrace, expand, 'x');  
grmap(htrace, subs, EpsEqn, 'x');
grmap(htrace, collect, 'x', ord, distributed, factor);
grmap(htrace, subs, Traces, 'x');  
grmap(htrace, collect, 'x', ord, distributed, factor);
grdisplay(htrace);

grdef(`hbar{(a b)} := hh{a b} - (1/2)*g{a b}*htrace`);
grcalc(hbar(dn,dn)) ;  
grmap(hbar(dn,dn), series, 'x', e, 4); 
grmap(hbar(dn,dn), convert, 'x', polynom);  
grmap(hbar(dn,dn), expand,'x');
grmap(hbar(dn,dn), subs, EpsEqn, 'x');  
grmap(hbar(dn,dn), collect, 'x', ord, distributed, factor);
grmap(hbar(dn,dn), subs, Traces, 'x');
grmap(hbar(dn,dn), collect, 'x', ord, distributed, factor);
grdisplay(hbar(dn,dn));

grcalc(hbar(dn,dn,cdn));
grmap(hbar(dn,dn,cdn), series, 'x', e, 4);  
grmap(hbar(dn,dn,cdn), convert, 'x', polynom); 
grmap(hbar(dn,dn,cdn), expand, 'x'); 
grmap(hbar(dn,dn,cdn), subs, EpsEqn, 'x');  
grmap(hbar(dn,dn,cdn), collect, 'x', ord, distributed, factor);
grmap(hbar(dn,dn,cdn), subs, Traces, 'x');
grmap(hbar(dn,dn,cdn), collect, 'x', ord, distributed, factor);
grdisplay(hbar(dn,dn,cdn));

grdef(`Dhbar{a b ^c} := hbar{a b ;^c}`);
grcalc(Dhbar(dn,dn,up)); 
grmap(Dhbar(dn,dn,up), series, 'x', e, 4);
grmap(Dhbar(dn,dn,up), convert, 'x', polynom); 
grmap(Dhbar(dn,dn,up), expand, 'x');  
grmap(Dhbar(dn,dn,up), subs, EpsEqn, 'x');
grmap(Dhbar(dn,dn,up), collect, 'x', ord, distributed, factor);
grmap(Dhbar(dn,dn,up), subs, Traces, 'x'); 
grmap(Dhbar(dn,dn,up), collect, 'x', ord, distributed, factor); 
grdisplay(Dhbar(dn,dn,up));

grdef(`DDhbar{a b} := Dhbar{a b ^c ;c}`);
grcalc(DDhbar(dn,dn)); 
grmap(DDhbar(dn,dn), series, 'x', e, 4); 
grmap(DDhbar(dn,dn), convert, 'x', polynom);  
grmap(DDhbar(dn,dn), expand, 'x');
grmap(DDhbar(dn,dn), subs, EpsEqn, 'x'); 
grmap(DDhbar(dn,dn), collect, 'x', ord, distributed, factor); 
grmap(DDhbar(dn,dn), subs, Traces, 'x');  
grmap(DDhbar(dn,dn), collect, 'x', ord, distributed, factor);
grdisplay(DDhbar(dn,dn));

grdef(`Rhbar{a b} := 2*R{a ^c b ^d}*hbar{c d}`);
grcalc(Rhbar(dn,dn)); 
grmap(Rhbar(dn,dn), series, 'x', e, 4); 
grmap(Rhbar(dn,dn), convert, 'x', polynom);  
grmap(Rhbar(dn,dn), expand, 'x');
grmap(Rhbar(dn,dn), subs, EpsEqn, 'x'); 
grmap(Rhbar(dn,dn), collect, 'x', ord, distributed, factor); 
grmap(Rhbar(dn,dn), subs, Traces, 'x');  
grmap(Rhbar(dn,dn), collect, 'x', ord, distributed, factor);
grdisplay(Rhbar(dn,dn));

grdef(`LHS{(a b)} := DDhbar{a b} + Rhbar{a b}`);
grcalc(LHS(dn,dn)); 
grmap(LHS(dn,dn), series, 'x', e, 4); 
grmap(LHS(dn,dn), convert, 'x', polynom);  
grmap(LHS(dn,dn), expand, 'x');
grmap(LHS(dn,dn), subs, EpsEqn, 'x');  
grmap(LHS(dn,dn), collect, 'x', ord, distributed, factor);
grmap(LHS(dn,dn), subs, Traces, 'x');
grmap(LHS(dn,dn), collect, 'x', ord, distributed, factor);
grdisplay(LHS(dn,dn));
 
grmap(LHS(dn,dn), subs, {e=1}, 'x');  
grmap(LHS(dn,dn), collect, 'x', ord, distributed, factor);
grdisplay(LHS(dn,dn));
 
grmap(LHS(dn,dn), collect, 'x', [t, x, y, z, Exx, Eyy], 
       distributed, factor);



 # Solution Of The Resulting Algebraic Equations Follows #


 
 #Final Verification Of The Solution #

grdef(`h1{(a b)} :=  kdelta{^$t a}*kdelta{^$t b}*(-2)*mu*etadn{i l}*
            AngM{^l}*BB{j k}*x{^i}*x{^j}*x{^k}/((x^2+y^2+z^2)^(3/2))
+ kdelta{^$t a}*kdelta{^$x b}*mu*
   (eps{$x i j}*etadn{l n}*AngM{^n}*etaup{^i ^m}*EE{m k}
      *x{^j}*x{^l}*x{^k}
   - eps{$x i j}*AngM{^i}*EE{l k}*x{^j}*x{^l}*x{^k}
   + 3*eps{i j l}*AngM{^i}*EE{$x n}*etaup{^n ^j}*x{^l}*(x^2+y^2+z^2) )
   /((x^2+y^2+z^2)^(3/2))
+ kdelta{^$t a}*kdelta{^$y b}*mu*
   (eps{$y i j}*etadn{l n}*AngM{^n}*etaup{^i ^m}*EE{m k}
      *x{^j}*x{^l}*x{^k} 
   - eps{$y i j}*AngM{^i}*EE{l k}*x{^j}*x{^l}*x{^k}
   + 3*eps{i j l}*AngM{^i}*EE{$y n}*etaup{^n ^j}*x{^l}*(x^2+y^2+z^2) )
   /((x^2+y^2+z^2)^(3/2))   
+ kdelta{^$t a}*kdelta{^$z b}*mu*
   (eps{$z i j}*etadn{l n}*AngM{^n}*etaup{^i ^m}*EE{m k}
      *x{^j}*x{^l}*x{^k} 
   - eps{$z i j}*AngM{^i}*EE{l k}*x{^j}*x{^l}*x{^k}
   + 3*eps{i j l}*AngM{^i}*EE{$z n}*etaup{^n ^j}*x{^l}*(x^2+y^2+z^2) )
   /((x^2+y^2+z^2)^(3/2))  
+ kdelta{^$x a}*kdelta{^$x b}*mu*
   (2*etadn{i k}*AngM{^k}*BB{$x $x}*x{^i}*(x^2+y^2+z^2)
   + (2/3)*AngM{^i}*BB{i j}*etadn{$x k}*x{^k}*etadn{$x l}*x{^l}*x{^j}
   + 2*AngM{^i}*BB{$x i}*etadn{$x k}*x{^k}*(x^2+y^2+z^2)
   - (2/3)*etadn{i k}*AngM{^k}*BB{$x j}*etadn{$x l}*x{^l}*x{^i}*x{^j}
   - (10/3)*etadn{$x k}*AngM{^k}*BB{$x j}*x{^j}*(x^2+y^2+z^2)
   - (2/3)*etadn{$x k}*AngM{^k}*BB{i j}*etadn{$x l}*x{^l}*x{^i}*x{^j}
   - (2/3)*AngM{^i}*BB{i j}*x{^j}*(x^2+y^2+z^2) 
   + (2/3)*etadn{i l}*AngM{^l}*BB{j k}*x{^i}*x{^j}*x{^k})
   ((x^2+y^2+z^2)^(3/2))   
+ kdelta{^$y a}*kdelta{^$y b}*mu*
   (2*etadn{i k}*AngM{^k}*BB{$y $y}*x{^i}*(x^2+y^2+z^2)
   + (2/3)*AngM{^i}*BB{i j}*etadn{$y k}*x{^k}*etadn{$y l}*x{^l}*x{^j}
   + 2*AngM{^i}*BB{$y i}*etadn{$y k}*x{^k}*(x^2+y^2+z^2)
   - (2/3)*etadn{i k}*AngM{^k}*BB{$y j}*etadn{$y l}*x{^l}*x{^i}*x{^j}
   - (10/3)*etadn{$y k}*AngM{^k}*BB{$y j}*x{^j}*(x^2+y^2+z^2)
   - (2/3)*etadn{$y k}*AngM{^k}*BB{i j}*etadn{$y l}*x{^l}*x{^i}*x{^j}
   - (2/3)*AngM{^i}*BB{i j}*x{^j}*(x^2+y^2+z^2)
   + (2/3)*etadn{i l}*AngM{^l}*BB{j k}*x{^i}*x{^j}*x{^k})
   ((x^2+y^2+z^2)^(3/2))
+ kdelta{^$z a}*kdelta{^$z b}*mu*
   (2*etadn{i k}*AngM{^k}*BB{$z $z}*x{^i}*(x^2+y^2+z^2)
   + (2/3)*AngM{^i}*BB{i j}*etadn{$z k}*x{^k}*etadn{$z l}*x{^l}*x{^j}
   + 2*AngM{^i}*BB{$z i}*etadn{$z k}*x{^k}*(x^2+y^2+z^2)
   - (2/3)*etadn{i k}*AngM{^k}*BB{$z j}*etadn{$z l}*x{^l}*x{^i}*x{^j}
   - (10/3)*etadn{$z k}*AngM{^k}*BB{$z j}*x{^j}*(x^2+y^2+z^2)
   - (2/3)*etadn{$z k}*AngM{^k}*BB{i j}*etadn{$z l}*x{^l}*x{^i}*x{^j}
   - (2/3)*AngM{^i}*BB{i j}*x{^j}*(x^2+y^2+z^2) 
   + (2/3)*etadn{i l}*AngM{^l}*BB{j k}*x{^i}*x{^j}*x{^k})
   /((x^2+y^2+z^2)^(3/2))      
+ kdelta{^$x a}*kdelta{^$y b}*mu*
   (2*etadn{i k}*AngM{^k}*BB{$x $y}*x{^i}*(x^2+y^2+z^2)
   + (2/3)*AngM{^i}*BB{i j}*etadn{$x k}*x{^k}*etadn{$y l}*x{^l}*x{^j}
   + AngM{^i}*BB{i $x}*etadn{$y k}*x{^k}*(x^2+y^2+z^2)
   + AngM{^i}*BB{i $y}*etadn{$x k}*x{^k}*(x^2+y^2+z^2)
   - (1/3)*etadn{i k}*AngM{^k}*BB{j $x}*etadn{$y l}*x{^l}*x{^i}*x{^j}
   - (1/3)*etadn{i k}*AngM{^k}*BB{j $y}*etadn{$x l}*x{^l}*x{^i}*x{^j}
   - (5/3)*AngM{^k}*etadn{k $x}*BB{$y j}*x{^j}*(x^2+y^2+z^2)
   - (5/3)*AngM{^k}*etadn{k $y}*BB{$x j}*x{^j}*(x^2+y^2+z^2)
   - (1/3)*AngM{^k}*etadn{k $x}*etadn{$y l}*BB{i j}*x{^l}*x{^i}*x{^j}
   - (1/3)*AngM{^k}*etadn{k $y}*etadn{$x l}*BB{i j}*x{^l}*x{^i}*x{^j})
   /((x^2+y^2+z^2)^(3/2))   
+ kdelta{^$x a}*kdelta{^$z b}*mu*
   (2*etadn{i k}*AngM{^k}*BB{$x $z}*x{^i}*(x^2+y^2+z^2)
   + (2/3)*AngM{^i}*BB{i j}*etadn{$x k}*x{^k}*etadn{$z l}*x{^l}*x{^j}
   + AngM{^i}*BB{i $x}*etadn{$z k}*x{^k}*(x^2+y^2+z^2)
   + AngM{^i}*BB{i $z}*etadn{$x k}*x{^k}*(x^2+y^2+z^2)
   - (1/3)*etadn{i k}*AngM{^k}*BB{j $x}*etadn{$z l}*x{^l}*x{^i}*x{^j}
   - (1/3)*etadn{i k}*AngM{^k}*BB{j $z}*etadn{$x l}*x{^l}*x{^i}*x{^j}
   - (5/3)*AngM{^k}*etadn{k $x}*BB{$z j}*x{^j}*(x^2+y^2+z^2)
   - (5/3)*AngM{^k}*etadn{k $z}*BB{$x j}*x{^j}*(x^2+y^2+z^2)
   - (1/3)*AngM{^k}*etadn{k $x}*etadn{$z l}*BB{i j}*x{^l}*x{^i}*x{^j}
   - (1/3)*AngM{^k}*etadn{k $z}*etadn{$x l}*BB{i j}*x{^l}*x{^i}*x{^j})
   /((x^2+y^2+z^2)^(3/2))   
+ kdelta{^$y a}*kdelta{^$z b}*mu*
   (2*etadn{i k}*AngM{^k}*BB{$y $z}*x{^i}*(x^2+y^2+z^2) 
   + (2/3)*AngM{^i}*BB{i j}*etadn{$y k}*x{^k}*etadn{$z l}*x{^l}*x{^j} 
   + AngM{^i}*BB{i $y}*etadn{$z k}*x{^k}*(x^2+y^2+z^2) 
   + AngM{^i}*BB{i $z}*etadn{$y k}*x{^k}*(x^2+y^2+z^2)
   - (1/3)*etadn{i k}*AngM{^k}*BB{j $y}*etadn{$z l}*x{^l}*x{^i}*x{^j}
   - (1/3)*etadn{i k}*AngM{^k}*BB{j $z}*etadn{$y l}*x{^l}*x{^i}*x{^j}
   - (5/3)*AngM{^k}*etadn{k $y}*BB{$z j}*x{^j}*(x^2+y^2+z^2)
   - (5/3)*AngM{^k}*etadn{k $z}*BB{$y j}*x{^j}*(x^2+y^2+z^2)
   - (1/3)*AngM{^k}*etadn{k $y}*etadn{$z l}*BB{i j}*x{^l}*x{^i}*x{^j}
   - (1/3)*AngM{^k}*etadn{k $z}*etadn{$y l}*BB{i j}*x{^l}*x{^i}*x{^j})
   /((x^2+y^2+z^2)^(3/2))  ` ):
grcalc(h1(dn,dn)):

grdef(`hh{(a b)} := h0{a b} + h1{a b}`):
grcalc(hh(dn,dn)):

grdef(`htrace := hh{^c c}`):
grcalc(htrace);  grmap(htrace, series, 'x', e, 4): 
grmap(htrace, convert, 'x', polynom):
grmap(htrace, expand, 'x'):  
grmap(htrace, subs, EpsEqn, 'x'):
grmap(htrace, collect, 'x', ord, distributed, factor):
grmap(htrace, subs, Traces, 'x'):  
grmap(htrace, collect, 'x', ord, distributed, factor):
grdisplay(htrace):

grdef(`hbar{(a b)} := hh{a b} - (1/2)*g{a b}*htrace`):
grcalc(hbar(dn,dn)):
grmap(hbar(dn,dn), series, 'x', e, 4):
grmap(hbar(dn,dn), convert, 'x', polynom):  
grmap(hbar(dn,dn), expand, 'x'):
grmap(hbar(dn,dn), subs, EpsEqn, 'x'):  
grmap(hbar(dn,dn), collect, 'x', ord, distributed, factor):
grmap(hbar(dn,dn), subs, Traces, 'x'):
grmap(hbar(dn,dn), collect, 'x', ord, distributed, factor):
grdisplay(hbar(dn,dn)):

grcalc(hbar(dn,dn,cdn)):
grmap(hbar(dn,dn,cdn), series, 'x', e, 4):  
grmap(hbar(dn,dn,cdn), convert, 'x', polynom):
grmap(hbar(dn,dn,cdn), expand, 'x'):
grmap(hbar(dn,dn,cdn), subs, EpsEqn, 'x'):  
grmap(hbar(dn,dn,cdn), collect, 'x', ord, distributed, factor):
grmap(hbar(dn,dn,cdn), subs, Traces, 'x'):
grmap(hbar(dn,dn,cdn), collect, 'x', ord, distributed, factor):
grdisplay(hbar(dn,dn,cdn)):

grdef(`Dhbar{(a b) ^c} := hbar{a b ;^c}`):
grcalc(Dhbar(dn,dn,up)): 
grmap(Dhbar(dn,dn,up), series, 'x', e, 4):
grmap(Dhbar(dn,dn,up), convert, 'x', polynom):  
grmap(Dhbar(dn,dn,up), expand, 'x'):
grmap(Dhbar(dn,dn,up), subs, EpsEqn, 'x'):
grmap(Dhbar(dn,dn,up), collect, 'x', ord, distributed, factor):
grmap(Dhbar(dn,dn,up), subs, Traces, 'x'):  
grmap(Dhbar(dn,dn,up), collect, 'x', ord, distributed, factor):
grdisplay(Dhbar(dn,dn,up)):

grdef(`DDhbar{(a b)} := Dhbar{a b ^c ;c}`):
grcalc(DDhbar(dn,dn)): 
grmap(DDhbar(dn,dn), series, 'x', e, 4):
grmap(DDhbar(dn,dn), convert, 'x', polynom):  
grmap(DDhbar(dn,dn), expand, 'x'):
grmap(DDhbar(dn,dn), subs, EpsEqn, 'x'):
grmap(DDhbar(dn,dn), collect, 'x', ord, distributed, factor):
grmap(DDhbar(dn,dn), subs, Traces, 'x'):  
grmap(DDhbar(dn,dn), collect, 'x', ord, distributed, factor):
grdisplay(DDhbar(dn,dn)):

grdef(`Rhbar{a b} := 2*R{a ^c b ^d}*hbar{c d}`):
grcalc(Rhbar(dn,dn)): 
grmap(Rhbar(dn,dn), series, 'x', e, 4):
grmap(Rhbar(dn,dn), convert, 'x', polynom):  
grmap(Rhbar(dn,dn), expand, 'x'):
grmap(Rhbar(dn,dn), subs, EpsEqn, 'x'):
grmap(Rhbar(dn,dn), collect, 'x', ord, distributed, factor):
grmap(Rhbar(dn,dn), subs, Traces, 'x'):  
grmap(Rhbar(dn,dn), collect, 'x', ord, distributed, factor):
grdisplay(Rhbar(dn,dn));

grdef(`LHS{(a b)} := DDhbar{a b} + Rhbar{a b}`):
grcalc(LHS(dn,dn)): 
grmap(LHS(dn,dn), series, 'x', e, 4):
grmap(LHS(dn,dn), convert, 'x', polynom):  
grmap(LHS(dn,dn), expand, 'x'):
grmap(LHS(dn,dn), subs, EpsEqn, 'x'):  
grmap(LHS(dn,dn), collect, 'x', ord, distributed, factor):
grmap(LHS(dn,dn), subs, Traces, 'x'):
grmap(LHS(dn,dn), collect, 'x', ord, distributed, factor):
grdisplay(LHS(dn,dn));
grmap(LHS(dn,dn), subs, {e=1}, 'x'):  
grmap(LHS(dn,dn), collect, 'x', ordtxyz, distributed, factor):
grdisplay(LHS(dn,dn)):
\end{verbatim}
\end{singlespace}

      \chapter{Decomposition of $\tilde{\rho}^{p} \chi^{-q}$}
\label{A1}

A detailed description of the spherical harmonic decomposition of the product $\tilde{\rho}^{p} \chi^{-q}$ is presented in this appendix, where $p$ is any integer and $q$ can be any real number, even though in every case in which this decomposition will be used $q$ will be either an integer or a half-integer.
The notation used is
\begin{eqnarray}
\tilde{\rho}^2 &=& \frac{r_o \Delta^2}{r_o - 2M} + 2 \: r_o^2 \: \frac{r_o - 2M}{r_o - 3M} \: \chi \: (1-\cos\Theta) \\
\chi &=& 1 - \frac{M}{r_o - 2M} \sin^{2}\Phi \\
\Delta &=& r - r_o. 
\end{eqnarray}
In order to simplify the equations, the following symbols are used:
\begin{eqnarray}
A &=& \frac{r_o \Delta^2}{r_o - 2M} \\ \label{EqnForA}
B &=& 2 \: r_o^2 \: \frac{r_o - 2M}{r_o - 3M} \\
C &=& \frac{M}{r_o - 2M} \\
\gamma^2 &=& \frac{\Delta^2 (r_o - 3M)}{2 r_o (r_o-2M)^2} \\
\delta^2 &=& \frac{\Delta^2 (r_o - 3M)}{2 r_o (r_o-2M)^2 \chi} = \frac{\gamma^2}{\chi}. \label{EqnFordeltaSqr}
\end{eqnarray}
Using Equations (\ref{EqnForA})-(\ref{EqnFordeltaSqr}) one gets
\begin{eqnarray}
\tilde{\rho} &=& B^{\frac{1}{2}} \: \chi^{\frac{1}{2}} \: \big ( \frac{\gamma^2}{\chi} + 1 - \cos\Theta \big )^{\frac{1}{2}} = B^{\frac{1}{2}} \: \chi^{\frac{1}{2}} \: \big ( \delta^2 + 1 - \cos\Theta \big )^{\frac{1}{2}} \\
\tilde{\rho}^{p} \chi^{-q} &=& B^{\frac{p}{2}} \: \chi^{\frac{p}{2} - q} \: \big ( \frac{\gamma^2}{\chi} + 1 - \cos\Theta \big )^{\frac{p}{2}}  = B^{\frac{p}{2}} \: \chi^{\frac{p}{2} - q} \: \big ( \delta^2 + 1 - \cos\Theta \big )^{\frac{p}{2}}. 
\end{eqnarray}

For the decomposition then it is assumed that
\begin{equation}
\begin{split}
\tilde{\rho}^{p} \chi^{-q} &= B^{\frac{p}{2}} \: \chi^{\frac{p}{2} - q} \: \big ( \frac{\gamma^2}{\chi} + 1 - \cos\Theta \big )^{\frac{p}{2}} \\
			&= B^{\frac{p}{2}} \sum_{l = 0}^{\infty} \sum_{m = -l}^{l} E_{l,m}^{p,q}(\gamma^2, C) \: Y_{lm}(\Theta, \Phi) \\
\end{split}
\label{firstDec}
\end{equation}
and the coefficients $E_{l,m}^{p,q}$ are calculated by multiplying both sides of the above equation with the complex-conjugate spherical harmonics $Y_{lm}^{\ast}(\Theta, \Phi)$ and integrating over $\Theta = [0,\pi]$ and $\Phi = [0,2\pi]$.
Using the results given in the discussion of spherical harmonics in \cite{Jackson}, the following expression for the coefficients is obtained
\begin{equation}
E_{l,m}^{p,q}(\gamma^2, C) = (-1)^{m} \sqrt{ \frac{2l+1}{4 \pi} \frac{(l+m)!}{(l-m)!}} \int_{0}^{2\pi} d\Phi e^{-i m \Phi} \chi^{\frac{p}{2}-q} \int_{-1}^{1} du (\delta^2 + 1 - u)^{\frac{p}{2}} P_{l}^{-m}(u)
\label{Elmpq}
\end{equation}
where $u = \cos\Theta$ and $P_{l}^{m}(u)$ are the Associated Legendre polynomials.
In the following it is also assumed that $P_{l}^{m=0}(u) = P_{l}(u)$, which are the Legendre polynomials.
It should be noted that, because only even powers of $\sin\Phi$ show up in the expression for the coefficients $E_{l,m}^{p,q}$, only the even values of $m$ contribute to the sum (\ref{firstDec}) and for all odd $m$ the coefficients $E_{l,m}^{p,q}$ are equal to 0.
In all the equations that follow, $m$ is assumed to be even.

First, the integral
\begin{equation}
I_{lm}^{p} = \int_{-1}^{1} du (\delta^2 + 1 - u)^{\frac{p}{2}} P_{l}^{-m}(u)
\label{Ilmp1}
\end{equation}
is calculated. 
Assuming that $m$ is positive, Equation 8.911 of \cite{GR} gives the $m$-derivative of the Legendre polynomials
\begin{equation}
\frac{d^{m} P_{l}(u)}{du^{m}} = \frac{1}{2^l} \sum_{k=0}^{[\frac{l}{2}]} \frac{(-1)^{k} \Gamma(2l - 2k +1)}{\Gamma(k+1) \Gamma(l-k+1) \Gamma(l-2k-m+1)} u^{l-2k-m}
\end{equation}
where $\Gamma(x)$ is the usual $\Gamma$-function and $[\frac{l}{2}]$ stands for the largest integer smaller or equal to $\frac{l}{2}$.
Then, Equation 8.810 of \cite{GR} gives for the associated Legendre polynomials, for positive $m$
\begin{equation}
P_{l}^{m}(u) = (1-u^2)^{\frac{m}{2}} \frac{1}{2^l} \sum_{k=0}^{[\frac{l}{2}]} \frac{(-1)^{k} \Gamma(2l - 2k +1)}{\Gamma(k+1) \Gamma(l-k+1) \Gamma(l-2k-m+1)} u^{l-2k-m}
\end{equation}
where the right-hand side is a polynomial, since $m$ is even.
The case of negative $m$ can be dealt with if one takes into account the fact that for any $m$:
\begin{equation}
P_{l}^{-m}(u) = \frac{\Gamma(l- |m| + 1)}{\Gamma(l + m +1)} P_{l}^{|m|}(u)
\end{equation}
which can be easily derived from Equation 8.752 of \cite{GR}.
Finally, the expression of the associated Legendre polynomials that is used in the calculation of the integral $I_{lm}^{p}$ is found to be
\begin{equation}
\begin{split}
P_{l}^{-m}(u) = \frac{\Gamma(l- |m| + 1)}{\Gamma(l + m +1)} &(1-u^2)^{\frac{|m|}{2}} \frac{1}{2^l} \times \\
		& \sum_{k=0}^{[\frac{l}{2}]} \frac{(-1)^{k} \Gamma(2l - 2k +1)}{\Gamma(k+1) \Gamma(l-k+1) \Gamma(l-2k-|m|+1)} u^{l-2k-|m|} \\
\end{split}
\label{PolLegM}
\end{equation}
which is valid for all even $m$, positive or negative.
Substituting Equation (\ref{PolLegM}) into Equation (\ref{Ilmp1}) gives
\begin{equation}
\begin{split}
I_{lm}^{p} = \frac{\Gamma(l - |m| +1)}{\Gamma(l+m+1)} \frac{1}{2^l} \sum_{k=0}^{[\frac{l}{2}]} &\frac{(-1)^k \Gamma(2l-2k+1)}{\Gamma(k+1) \Gamma(l-k+1) \Gamma(l-2k-|m|+1)} \times \\ 
	&\qquad \qquad \int_{-1}^{1} du (\delta^2 +1 -u)^{\frac{p}{2}} (1-u^2)^{\frac{|m|}{2}} u^{l-|m|-2k} .\\
\end{split}
\label{Ilmp2}
\end{equation}
The integral of Equation (\ref{Ilmp2}) can be calculated if the term $(1-u^2)^{\frac{|m|}{2}}$ is expanded in powers of $u$, specifically
\begin{equation} 
\begin{split}
I &= \int_{-1}^{1} du (\delta^2 +1 -u)^{\frac{p}{2}} (1-u^2)^{\frac{|m|}{2}} u^{l-|m|-2k} \\
  &= \sum_{\nu = 0}^{\frac{|m|}{2}} \frac{\Gamma(\frac{|m|}{2} + 1)}{\Gamma(\nu + 1) \Gamma(\frac{|m|}{2} - \nu +1)} \int_{-1}^{1} du (\delta^2 +1 -u)^{\frac{p}{2}} u^{l-|m| - 2k + 2\nu}. \\
\end{split}
\label{Iu}
\end{equation}
The task, then, shifts to the calculation of the integral
\begin{equation}
I_{(a,b)} = \int_{-1}^{1} du \: u^{a} (\delta^2 +1 -u)^{\frac{b}{2}}
\end{equation}
where $a$ and $b$ are integers, with $a \geq 0$.
Performing integration by parts once results in 
\begin{equation}
\begin{split}
I_{(a,b)} &= - \int_{-1}^{1} u^a \frac{1}{\frac{b}{2} + 1} d(\delta^2 +1 -u)^{\frac{b}{2} + 1} \\
	&= - \frac{1}{\frac{b}{2}+1} (\delta^2)^{\frac{b}{2}+1} + (-1)^a \frac{1}{\frac{b}{2}+1} (2+\delta^2)^{\frac{b}{2}+1} + \frac{2 a}{b +2} \int_{-1}^{1} u^{a-1} (\delta^2 +1 -u)^{\frac{b+2}{2}} \\
	&= \frac{2}{b+2} \Big [ -(\delta^2)^{\frac{b}{2}+1} + (-1)^a (2+\delta^2)^{\frac{b}{2}+1} \Big ] + \frac{2a}{b+2} I_{(a-1, b+2)} \\
\end{split}
\end{equation}
which is a recursion relation for $I_{(a,b)}$. 
Using the fact that
\begin{equation}
I_{(0,b)} = \frac{2}{b+2} \Big [ -(\delta^2)^{\frac{b}{2}+1} + (2+\delta^2)^{\frac{b}{2}+1} \Big ]
\end{equation}
it can be proven by induction that the general integral $I_{(a,b)}$ is equal to
\begin{equation}
\begin{split}
I_{(a,b)} = \sum_{n=0}^{a} \frac{2^{n+1}}{(b+2)(b+4)\ldots[b+2(n+1)]} \: \frac{\Gamma(a+1)}{\Gamma(a-n+1)} & \: \Big [ (-1)^{a-n} (2+\delta^2)^{\frac{b}{2} + n +1} \\
			& - (\delta^2)^{\frac{b}{2} + n +1} \Big ]. \\
\end{split}
\label{Iab}
\end{equation}
Finally, using Equations (\ref{Iab}) and (\ref{Iu}) into Equation (\ref{Ilmp2}), the integral $I_{lm}^{p}$ ends up being 
\begin{equation}
\begin{split}
I_{lm}^{p}  &= \frac{\Gamma(l - |m| +1)}{\Gamma(l+m+1)} \frac{1}{2^l} \sum_{k=0}^{[\frac{l}{2}]} \frac{(-1)^k \Gamma(2l-2k+1)}{\Gamma(k+1) \Gamma(l-k+1) \Gamma(l-2k-|m|+1)} \times \\
		& \sum_{\nu = 0}^{\frac{|m|}{2}} \frac{\Gamma(\frac{|m|}{2}+1)}{\Gamma(\nu +1) \Gamma(\frac{|m|}{2}-\nu +1)}  \sum_{n=0}^{l-|m| -2k+2\nu} \frac{2^{n+1}}{(p+2)(p+4)\ldots[p+2(n+1)]} \times \\
		& \frac{\Gamma(l-|m|-2k+2\nu+1)}{\Gamma(l-|m|-2k+2\nu-n+1)} \: \Big [ (-1)^{l-n} (2+\delta^2)^{\frac{p}{2} + n +1} - (\delta^2)^{\frac{p}{2} + n +1} \Big ]. \\
\end{split}
\label{Ilmpfinal}
\end{equation}

A short comment about the term $(\delta^2)^{\frac{p}{2}+n+1}$ should be made at this point.
If $p$ is an odd integer, that term implies that the square root of $\delta^2$, or of $\gamma^2$, must be taken.
One thing that may be ambiguous is the sign of the square root, $+$ or $-$, that must be considered.
That point is discussed in Chapter \ref{RP} of this dissertation, where the physical interpretation of the quantities becomes clear and the correct sign to use can be easily recognized.
For the calculations in this appendix it is enough to interpret the term $(\delta^2)^{\frac{1}{2}}$ as the \emph{appropriate} square root of $\delta^2$.
 
Continuing the calculation of the coefficients $E_{l,m}^{p,q}$, substituting Equation (\ref{Ilmpfinal}) into Equation (\ref{Elmpq}) shows that the next two integrals to be done are
\begin{equation}
\begin{split}
I_{1} &= (-1)^{l-n} \int_{0}^{2 \pi} d\Phi e^{-im\Phi} \chi^{\frac{p}{2}-q} (2+\delta^2)^{\frac{p}{2}+n+1} \\
      &= (-1)^{l-n} \int_{0}^{2 \pi} d\Phi e^{-im\Phi} \chi^{\frac{p}{2}-q} (2+\frac{\gamma^2}{\chi})^{\frac{p}{2}+n+1}
\end{split}
\end{equation}
\begin{equation}
\begin{split}
I_{2} &= \int_{0}^{2 \pi} d\Phi e^{-im\Phi} \chi^{\frac{p}{2}-q} (\delta^2)^{\frac{p}{2}+n+1} \\
      &= \int_{0}^{2 \pi} d\Phi e^{-im\Phi} \chi^{\frac{p}{2}-q} (\frac{\gamma^2}{\chi})^{\frac{p}{2}+n+1}
\end{split}
\end{equation}
and their difference must be multiplied by the appropriate factor given by Equation (\ref{Ilmpfinal}).
For both these integrals, the exponential $e^{-im\Phi}$ must be written in terms of the variable $\chi$.
In general
\begin{equation}
e^{-im\Phi} = \cos(m\Phi) - i \sin(m\Phi)
\end{equation}
but, since only even $m$'s are considered and the remaining of the integrand is a function of $\sin^{2}\Phi$, the $\sin(m\Phi)$ part of the exponential gives 0 once the integration is performed.
So, as far as the integrals are concerned, the exponential can be replaced by the $\cos(m\Phi)$.
Then Equation 1.331 of \cite{GR} gives that for any even $m$, positive or negative,
\begin{equation}
\cos(m\Phi) = \sum_{\lambda = 0}^{\frac{|m|}{2}} \frac{\Gamma(|m|+1) (-1)^{\lambda}}{\Gamma(|m|-2\lambda+1) \Gamma(2\lambda+1)} \cos^{|m|-2\lambda}\Phi \sin^{2\lambda}\Phi.
\end{equation}
The differential $d\Phi$ is written in terms of the differential $d\chi$
\begin{equation}
d\Phi = - \frac{d\chi}{2 C \sin\Phi \cos\Phi}.
\end{equation}
Also
\begin{eqnarray}
\sin^{2}\Phi &=& \frac{1-\chi}{C} \\
\cos^{2}\Phi &=& 1 - \frac{1-\chi}{C}
\end{eqnarray}
and caution should be used when the square root of these expressions is taken.
Specifically, for $\Phi$ in $[0,\frac{\pi}{2})$ or in $[\pi,\frac{3\pi}{2})$, $\sin\Phi \cos\Phi \geq 0$, so the equation
\begin{equation}
\sin\Phi \cos\Phi = \sqrt{\Big (\frac{1-\chi}{C}\Big ) \Big  (1 - \frac{1-\chi}{C}\Big )}
\end{equation}
should be used.
On the other hand, for $\Phi$ in $[\frac{\pi}{2},\pi)$ or in $[\frac{3\pi}{2},2\pi)$, $\sin\Phi \cos\Phi \leq 0$, so the equation
\begin{equation}
\sin\Phi \cos\Phi = -\sqrt{ \Big (\frac{1-\chi}{C}\Big ) \Big (1 - \frac{1-\chi}{C}\Big)}
\end{equation}
should be used.

In $I_{1}$, the factor $(2+\frac{\gamma^2}{\chi})^{\frac{p}{2}+n+1}$ is first written as an infinite sum
\begin{equation}
(2+\frac{\gamma^2}{\chi})^{\frac{p}{2}+n+1} = \sum_{s=0}^{\infty} \frac{\Gamma(\frac{p}{2}+n+1)}{\Gamma(s+1) \Gamma(\frac{p}{2}+n-s+1)} 2^{\frac{p}{2}+n-s+1} \frac{\gamma^{2s}}{\chi^s}.
\end{equation}
The integral becomes
\begin{equation}
\begin{split}
I_{1} = (-1)^{l-n} 2 \sum_{\lambda=0}^{\frac{|m|}{2}} & \frac{\Gamma(|m|+1) (-1)^{\lambda}}{\Gamma(|m|-2\lambda+1) \Gamma(2\lambda+1)} C^{-\frac{|m|}{2}} \sum_{s=0}^{\infty} \frac{\Gamma(\frac{p}{2}+n+1)}{\Gamma(s+1) \Gamma(\frac{p}{2}+n-s+1)} \times \\ 
& 2^{\frac{p}{2}+n-s+1} \gamma^{2s} \int_{1-C}^{1} d\chi \: \chi^{\frac{p}{2}-q-s} (1-\chi)^{\lambda-\frac{1}{2}} (C-1+\chi)^{\frac{|m|}{2}-\lambda-\frac{1}{2}}. \\ 
\end{split}
\end{equation} 
Changing the variable of integration from $\chi$ to $y = \frac{1-\chi}{C}$ and using the definition of the hypergeometric functions
\begin{equation}
F(\alpha,\beta;\gamma;z) = {}_{2}F_{1} (\alpha, \beta; \gamma; z) = \frac{\Gamma(\gamma)}{\Gamma(\beta) \Gamma(\gamma-\beta)} \int_{0}^{1} t^{\beta-1} (1-t)^{\gamma-\beta-1} (1-tz)^{-\alpha} dt
\end{equation}
where: $\Re(\gamma) > \Re(\beta) > 0$, result in the final expression for the integral
\begin{equation}
\begin{split}
I_{1} = \sqrt{\pi} (-1)^{l-n} & \sum_{\lambda=0}^{\frac{|m|}{2}} (-1)^{\lambda} \frac{\Gamma(\frac{|m|+1}{2})}{\Gamma(\lambda+1) \Gamma(\frac{|m|}{2} - \lambda +1)} \sum_{s=0}^{\infty} \gamma^{2s} 2^{\frac{p}{2}+n-s+2} \times \\
		& \frac{\Gamma(\frac{p}{2}+n+1)}{\Gamma(s+1) \Gamma(\frac{p}{2}+n-s+1)} F(-\frac{p}{2}+q+s, \lambda+\frac{1}{2}; \frac{|m|}{2}+1; C). \\
\end{split}
\label{Int1}
\end{equation}

To calculate the second integral $I_{2}$ one follows the same procedure that was used for $I_{1}$, the only difference being that the step of expanding the factor $(2+\frac{\gamma^2}{\chi})^{\frac{p}{2}+n+1}$ is now not necessary.
The result is
\begin{equation}
I_{2} = \sum_{\lambda=0}^{\frac{|m|}{2}} 2 (-1)^{\lambda} \sqrt{\pi} \frac{\Gamma(\frac{|m|+1}{2})}{\Gamma(\lambda+1) \Gamma(\frac{|m|}{2}-\lambda+1)} (\gamma^2)^{\frac{p}{2}+n+1} F(q+n+1, \lambda+\frac{1}{2}; \frac{|m|}{2}+1;C).
\label{Int2}
\end{equation}

The final expression for the coefficients $E_{l,m}^{p,q}(\gamma^2,C)$ can be obtained by combining Equations (\ref{Ilmpfinal}), (\ref{Int1}), (\ref{Int2}) and (\ref{Elmpq}).
It is
\begin{equation}
\begin{split}
E_{l,m}^{p,q}(\gamma^2,&C) = \sqrt{\frac{2l+1}{\pi}} \frac{\Gamma(l-|m|+1)}{[\Gamma(l-m+1) \Gamma(l+m+1)]^{\frac{1}{2}}} \times \\
		&\sum_{\lambda=0}^{\frac{|m|}{2}} \frac{(-1)^{\lambda} \Gamma(\frac{|m|+1}{2})}{\Gamma(\lambda+1) \Gamma(\frac{|m|}{2}-\lambda+1)} \times \\
		&\sum_{k=0}^{[\frac{l}{2}]} (-1)^{k} 2^{l-2k} \frac{\Gamma(l-k+\frac{1}{2})}{\Gamma(k+1) \Gamma(l-2k-|m|+1)} \times \\
		&\sum_{\nu=0}^{\frac{|m|}{2}} \frac{\Gamma(\frac{|m|}{2}+1)}{\Gamma(\nu+1) \Gamma(\frac{|m|}{2}-\nu+1)} \times \\ 
		&\sum_{n=0}^{l-|m|-2k+2\nu} \frac{2^{n}}{(p+2) (p+4) \ldots [p+2(n+1)]} \frac{\Gamma(l-|m|-2k+2\nu+1)}{\Gamma(l-|m|-2k+2\nu-n+1)} \times \\
		&\Bigg \{ - 2 (\gamma^2)^{\frac{p}{2}+n+1} F(q+n+1, \lambda+\frac{1}{2}; \frac{|m|}{2}+1; C) \: \: + \\
		&\quad (-1)^{l+n} \sum_{s=0}^{\infty} \gamma^{2s} \frac{\Gamma(\frac{p}{2}+n+1)}{\Gamma(s+1) \Gamma(\frac{p}{2}+n-s+1)} 2^{\frac{p}{2}+n-s+2} \times \\
		&\quad \qquad \qquad \qquad \qquad \qquad \qquad \qquad \: F(-\frac{p}{2}+q+s, \lambda+\frac{1}{2}; \frac{|m|}{2}+1; C) \Bigg \}. \\
\end{split}
\end{equation}
It is reminded that this expression is valid for all even $m$'s, and that for odd $m$'s these coefficients are equal to zero.

If this expression for the coeffients is used, the equations become very long and their dependence on $\gamma^2$, which is of particular interest for the calculation of the regularization parameters, becomes unclear.
In order for the expressions that appear in this dissertation to look simpler, the following notation is used for the three finite sums over $k$, $\nu$ and $\lambda$ which appear in the previous equation and which have no $\gamma^2$-dependence
\begin{equation}
\begin{split}
\Big ( \sum_{k,\nu,\lambda} &\Big ) = \sqrt{\frac{2l+1}{\pi}} \frac{\Gamma(l-|m|+1)}{[\Gamma(l-m+1) \Gamma(l+m+1)]^{\frac{1}{2}}} \times \\
			&\sum_{k=0}^{[\frac{l}{2}]} (-1)^{k} 2^{l-2k} \frac{\Gamma(l-k+\frac{1}{2})}{\Gamma(k+1) \Gamma(l-2k-|m|+1)} \sum_{\nu=0}^{\frac{|m|}{2}} \frac{\Gamma(\frac{|m|}{2}+1)}{\Gamma(\nu+1) \Gamma(\frac{|m|}{2}-\nu+1)} \times \\
			&\sum_{\lambda=0}^{\frac{|m|}{2}} (-1)^{\lambda} \frac{\Gamma(\frac{|m|+1}{2})}{\Gamma(\lambda+1) \Gamma(\frac{|m|}{2}-\lambda+1)}. \\
\end{split}
\end{equation}
It is implied that $(\sum_{\kappa,\nu,\lambda})$ is a multiplicative factor and any quantities appearing to the right of it can depend on $k$, $\nu$ or $\lambda$.
Also, the notation $F_{(a)}^{\lambda, m}$ is used for the hypergeometric function $F(a,\lambda+\frac{1}{2};\frac{|m|}{2}+1;C)$. 

Finally
\begin{equation}
\begin{split}
E_{l,m}^{p,q}(\gamma^2, C) = &\Big ( \sum_{k,\nu,\lambda} \Big ) \sum_{n=0}^{l-|m|-2k+2\nu} \frac{2^{n}}{(p+2) (p+4) \ldots [p+2(n+1)]} \times \\
			   &\frac{\Gamma(l-|m|-2k+2\nu+1)}{\Gamma(l-|m|-2k+2\nu-n+1)} \Bigg \{ - 2 (\gamma^2)^{\frac{p}{2}+n+1} F_{(q+n+1)}^{\lambda,m} \: \: + \\
			   &\sum_{s=0}^{\infty} \gamma^{2s} \frac{(-1)^{l+n} \Gamma(\frac{p}{2}+n+1)}{\Gamma(s+1) \Gamma(\frac{p}{2}+n-s+1)} 2^{\frac{p}{2}+n-s+2} \: F_{(-\frac{p}{2}+q+s)}^{\lambda,m} \Bigg \}. \\
\end{split}
\label{Efinal}
\end{equation}

	\chapter{Numerical Code for the Retarded Field}
\label{A2}

\begin{singlespace}
\begin{verbatim}
#include <stdlib.h>
#include <iostream.h>
#include <iomanip.h>
#include <fstream.h>
#include <assert.h>
#include <string.h>
#include <stdio.h>
#include <complex.h>
#include <math.h>



// Convenient macros for debugging

#define SHOW(a)    " "<<#a<<" = "<<a<<" "
#define PRT(a) cerr << " "<<#a<<" = " << a << endl;
#define PRINT(a) cerr <<__LINE__<< " |" <<__FILE__<<": "<< 
                 #a <<" = "<<a<< endl;
#define WHEREAMI  __LINE__ << " |" << __FILE__ << ": "
#define TRACE cerr << __LINE__ << " |" << __FILE__ << ": " 
#define TRACK cerr << __LINE__ << " |" << __FILE__ << endl; 



typedef complex<double> cmplx ;


int kmax,kount;
double *xp,dxsav;
cmplx **yp;



cmplx mu(int l, int m);

void fkseh(cmplx *f1, cmplx *f2, cmplx *f3, int k);

void fksinf(cmplx *f1, cmplx *f2, cmplx *f3, int k);

cmplx coefkeh(int k);

cmplx coefkinf(int k);

void rkqs(cmplx y[],cmplx dydr[], int n, double *r, double htry, 
        double eps, double yscal[], double *hdid, double *hnext,
        void (*derivs)(double, cmplx [], cmplx [])) ;

void rkck(cmplx y[],cmplx dydr[], int n, double r, double h,
          cmplx yout[], cmplx yerr[], void (*derivs)(double, 
          cmplx [], cmplx [])) ;

void derivs(double r, cmplx y[], cmplx dydr[]) ;

void factor(cmplx& coefr, cmplx& coefq, double r); 

void strline(double x1,double x2,cmplx y1,cmplx y2, 
             double R,cmplx& yR);

void odeint(cmplx ystart[],int nvar,double x1,double x2,double eps,
            double h1, double hmin, int *nok, int *nbad,
            cmplx& ylast1, cmplx& ylast2,
            void (*derivs)(double,cmplx [],cmplx []),
            void (*rkqs)(cmplx [],cmplx [], int,double *,
            double,double,double [], double *, double *,
            void (*)(double,cmplx [],cmplx []))) ;



const cmplx I(0,1);

namespace ScalarWave{
  double M = 1;                      // Mass of the Black Hole
  int l;                             // Spherical Harmonic Index
  int m;                             // Spherical Harmonic Index
  double omega;                      // Frequency of Radiation 
  double w;                          // Angular Frequency of Orbit
  double R = 20*M;                   // Radius of the Orbit
  double dtdtau = 1/sqrt(1-(3*M/R)); // t-Component of 4-velocity
}



#include "rkqs.c"
#include "rkck.c"
#include "nrutil.c"
#include "odeint1.c"
#include "sum.c"
#include "nrutil.h"
#include "sphharm.c"
#include "zero.cpp"



int main(){ 

   using namespace ScalarWave;
   M = 1.0 ;             
   w = sqrt(M/(R*R*R)) ; 
   double pi = 4.0*atan(1.0);


   char* buff;
   buff = new char[513];
   char bh;
   

   int nvar = 2;    // Number of First Order Differential Equations.
   int nok = 0;
   int nbad = 0; 
   int lmax = 31;


   cmplx ystart[3] ;
   double mulm;
   cmplx yLR1, yLR2, yRR1, yRR2;
   cmplx W1, W2, W3, W4 ;


   cout << endl;
   cout << "Radius of the orbit:" << SHOW(R) << endl;
   

   ofstream out("DPsi20.dat");
   out << "l m       mulm          yLN2            yRN2" << endl;


// Initialization of the Self-Force Contributions.

  double Flmt =0;
  double Flt =0;
  double Ft =0;
  double FlmrL =0;
  double FlrL =0;
  double FrL =0;
  double FlmrR =0;
  double FlrR =0;
  double FrR =0;


   for (l=0; l<lmax; l++) {

 
      for(m=l ; m>-1 ; m=m-2) { 
     
         
         if (m!=0) {
	  
          omega = m*w ;
	  

          cout << endl;
          cout << SHOW(l) << SHOW(m) << endl;
          

          cmplx b1 = (l*l+l+1)/(2*M+8*I*omega*M*M);
          cmplx a1 = -I*l*(l+1)/(2*omega) ;
	  

          double Reh = 1/(100*abs(b1)) + 2*M ; 
                             // Starting Radius Close to Event Horizon 
          double Rinf = 100*abs(a1) ;          
                             // Starting Radius Close to Infinity


          // Integrate in the Region (Rs, R] to Get yL(r).

          double h = 1.0e-5 ;     
          double hmin = 1.0e-10 ; 
          double eps = 1.0e-12 ;  
          kmax = 0;
          dxsav = 0.0;
	  

          // Boundary Conditions at the Event Horizon :
   
          cmplx expeh = 1;
          cmplx Dexpeh = 0;
          coefkeh(0); 
          double Xn = 1;
          double last = 1.e+33;


          for(int k=1; k<100; k++) {  

             Xn *= (Reh-2*M);      
             cmplx bk = coefkeh(k);
             cmplx next = bk*Xn;


             if (expeh + next == expeh) {

                cout << "Machine precision in event-horizon loop" 
                     << endl;
                cout << setprecision(13) << SHOW(k) << SHOW(expeh) 
                     << SHOW(next) << endl;
                break;

             }


             if (abs(next) > last) {

                cerr << "No convergence in event-horizon loop" 
                     << endl;
                cerr << setprecision(13) << SHOW(k) << SHOW(expeh) 
                     << SHOW(next) << endl;
                break;

             }


             last = abs(next);
             expeh += next;
             Dexpeh += k*next/(Reh-2*M);


             if (k < 5) {

             cout << setprecision(13) << SHOW(k) << SHOW(bk) 
                  << SHOW(expeh) << endl;

             }

          } 
    

         expeh =expeh* 
           exp(cmplx(0.0, omega*Reh + omega*2.0*M*log(Reh/(2*M)-1) ));
         Dexpeh = exp(cmplx(0.0, omega*Reh + 
                   omega*2.0*M*log(Reh/(2*M)-1) ))*Dexpeh 
                + expeh*cmplx(0,omega)*Reh/(Reh-2*M);


         ystart[1]=expeh/Reh;
         ystart[2]=Dexpeh/Reh - expeh/(Reh*Reh);


         cout << setprecision(13) << SHOW(ystart[1]) 
              << SHOW(ystart[2]) << endl;


         // Wronskian at the event horizon.

         W1 = Reh*(Reh-2*M)*( ystart[1]*conj(ystart[2]) 
                            - ystart[2]*conj(ystart[1]) );


         cout << SHOW(W1) << SHOW(-2*I*omega) << endl ;


         // Integration of the differential equation for Psi.
	
         odeint(ystart,nvar,Reh,R,eps,h,hmin,&nok,&nbad,yLR1,yLR2,
                (*derivs),(*rkqs));


         // Wronskian at the orbit.

         W2 = R*(R-2*M)* (yLR1*conj(yLR2) - yLR2*conj(yLR1)) ;
	 

         //Solution before the matching.

         cout << setprecision(13) << SHOW(Reh) << SHOW(yLR1) 
              << SHOW(yLR2) << endl;
         cout << setprecision(13) << SHOW(W1) << SHOW(W2) 
              << SHOW(-2*I*omega) << endl;
	 
	 
         //Integrate in the region [R,Rinf) to Get yR(r).
	 
         h = -1.0e-5;
         hmin = 1.e-10;
         eps = 1.e-12;
         kmax = 0;
         dxsav = 0.0;
	 

         //Boundary Conditions at Infinity:
	 
         cmplx expinf = 1;
         cmplx Dexpinf = 0;
         coefkinf(0);
         Xn = 1;
         last = 1.e+33 ;
	 

         for(int k=1; k<100; k++) {

         Xn *= Rinf;
         cmplx ak = coefkinf(k) ;
         cmplx next = ak/Xn;
	 

         if (expinf + next ==expinf) {

            cout << "Machine precision in infinity loop" << endl;
            cout << setprecision(13) << SHOW(k) << SHOW(expinf) 
                 << SHOW(next) << endl;
            break; 

         } 
	 
         if (abs(next) > last) {

            cout << "No convergence in infinity loop" << endl;
            cout << setprecision(13) << SHOW(k) << SHOW(expinf) 
                 << SHOW(next) << endl;
            break;

         }
	 

         last = abs(next);
         expinf += next;
         Dexpinf += -k*next/Rinf;

	 
          if (k<5) {

          cout << setprecision(13) << SHOW(k) << SHOW(ak) 
               << SHOW(expinf) << endl;

         }

	 
         }
	 
	  
          double rstar = Rinf + 2*M*log(Rinf/(2*M)-1) ;	    
	  
	  
          expinf = exp(I*omega*rstar)*expinf;
          Dexpinf = I*omega*Rinf*expinf/(Rinf-2*M) 
                    + exp(I*omega*rstar)*Dexpinf;
	  
	  
          ystart[1] = expinf/Rinf;
          ystart[2] = Dexpinf/Rinf - expinf/(Rinf*Rinf);
	  

          cout << setprecision(13) << SHOW(ystart[1]) 
               << SHOW(ystart[2]) << endl;
	 

          //Wronskian at infinity

          W3 = Rinf*(Rinf-2*M)*(ystart[1]*conj(ystart[2])
                                -ystart[2]*conj(ystart[1]));
	  

          //Integration of the differential equation for Psi.
	 
          odeint(ystart,nvar,Rinf,R,eps,h,hmin,&nok,&nbad,yRR1,yRR2,
                 (*derivs),(*rkqs));
	  

          //Wronskian at the orbit.

          W4 = R*(R-2*M)*(yRR1*conj(yRR2)-yRR2*conj(yRR1));
	  

          //Solution before the matching.

          cout << setprecision(13) << SHOW(Rinf) << SHOW(yRR1) 
               << SHOW(yRR2) << endl;
          cout << setprecision(13) << SHOW(W3) << SHOW(W4) 
               << SHOW(-2*I*omega)<< endl ;
	  

          //Matching of the solutions at the orbit.
	  
          cmplx A = -mu(l,m) * yRR1 /(R-2*M)/ (yLR1*yRR2-yRR1*yLR2);
          cmplx B = -mu(l,m) * yLR1 /(R-2*M)/ (yLR1*yRR2-yRR1*yLR2);
	  

          //Normalized solution.

          cmplx yLN1 = yLR1*A ;
          cmplx yLN2 = yLR2*A ;
          cmplx yRN1 = yRR1*B ;
          cmplx yRN2 = yRR2*B ;
	  

          cout << SHOW(yLR1) << SHOW(yLR2) << endl;
          cout << SHOW(yRR1) << SHOW(yRR2) << endl;
          cout << SHOW(A) << SHOW(B) << endl;
          cout << SHOW(yLN1) << SHOW(yLN2) << endl;
          cout << SHOW(yRN1) << SHOW(yRN2) << endl << endl;
	  

          // Contribution to the self-force

          Flmt = 2*real(I*omega*yLN1*sharm(l,m,pi/2.0,0.0));
          FlmrL = 2*real(yLN2*sharm(l,m,pi/2.0,0.0));
          FlmrR = 2*real(yRN2*sharm(l,m,pi/2.0,0.0));


          out << l <<" "<< m <<" "<< mu(l,m) <<" "<< yLN2 
              <<" "<< yRN2 << endl;
	  

       } 
       

       
       if (m==0) {
    
          cmplx PsiH, dPsiH, PsiI, dPsiI;
       

          cout << endl;
          cout << SHOW(l) << SHOW(m) << endl;
       

          getZeroPsiH(&PsiH,&dPsiH);
          getZeroPsiI(&PsiI,&dPsiI);


          cmplx wronsk = R*(R-2.0*M)*(PsiH*dPsiI-PsiI*dPsiH);
       

          cmplx PsiHN = -mu(l,m) * PsiH*R*PsiI/wronsk;
          cmplx PsiIN = -mu(l,m) * PsiI*R*PsiH/wronsk;
          cmplx DPsiH = -mu(l,m)*dPsiH*R*PsiI/wronsk;
          cmplx DPsiI = -mu(l,m)*dPsiI*R*PsiH/wronsk;
       

          Flmt = real(I*omega*PsiHN*sharm(l,m,pi/2.0,0.0));
          FlmrL = real(DPsiH*sharm(l,m,pi/2.0,0.0));
          FlmrR = real(DPsiI*sharm(l,m,pi/2.0,0.0));


          cout << SHOW(DPsiH) << SHOW(DPsiI) << endl << endl;
          out <<l<<" "<<m<<" "<<mu(l,m)<<" "<<DPsiH<<" "
              <<DPsiI<< endl;
       
       } 
    

    Flt += Flmt;
    FlrL += FlmrL;
    FlrR += FlmrR;
    
    }
    
    
 Ft += Flt;
 FrL += FlrL;
 FrR += FlrR;


 }
    

 out.close();
 cout << "File DPsi20.dat closed" << endl;
    
}
	  
    
  

// Factor mu_(l,m).

cmplx mu(int l, int m){

  using namespace ScalarWave;
  
  double mmu = 1.0;
  double pi = 4.0*atan(1.0);

  cout <<SHOW(R)<<SHOW(M)<<SHOW(l)<<SHOW(m)
       <<SHOW(sharm(l,m,pi/2,0))<< endl;

  cmplx mulm = 4.0*pi*mmu * conj(sharm(l,m,pi/2.0,0.0))
               *sqrt(1-(3*M/R))/R;
  
  return mulm ;

}




// Derivatives, For Use in rkqs.c .

void derivs(double r, cmplx y[], cmplx dydr[]) {
 
   cmplx coefr,coefq ;

   factor(coefr, coefq, r); 

   dydr[1] = y[2] ;                       // dy/dr
   dydr[2] = coefr*y[1] - coefq * y[2] ;  // d^2y/dr^2 

   return  ;

}




// coefr(r) and coefq(r) 

void factor(cmplx& coefr, cmplx& coefq, double r) {

   using namespace ScalarWave;

   int  l2 = l*(l+1);
   double r2 = r*r;
   double rm = r-2*M;

   coefr = -( omega*omega*r2/rm/rm - l2/r/(r-2.0*M));
   coefq = 2.0*(r-M)/r/rm ;

}





// Spherical Harmonics.

cmplx sharm(int l, int m, double theta, double phi){

   cmplx I = cmplx(0,1.0);
   double c = cos(theta) ;
   cmplx sh = exp(I*abs(m)*phi) * plgndr(l,abs(m),c);

   float fact =
   sqrt((2*l+1)*factorial(l-abs(m))
   /(4.0*(4.0*atan(1.0)))/factorial(l+abs(m))) ;

   sh = sh*fact;

   if(m<0) sh = pow(-1,abs(m))*conj(sh) ;
   
  return sh;

}




// Terms in the expansion of the intial guess for the solution
// at the event horizon and at infinity

void fksinf(cmplx *f1, cmplx *f2, cmplx *f3, int k){
  using namespace ScalarWave;

  *f1 = cmplx(-k*(k-1)+l*(l+1) , 4*omega*M*(k-1) ) 
        / cmplx(0.0, 2*omega*k) ;
  *f2 = cmplx( 0.0, -M*( (k-1)*(2*k-3) - l*(l+1) ) ) /omega/k ;
  *f3 = cmplx( 0.0, 2*M*M*(k-2)*(k-2)/omega/k );
}



void fkseh(cmplx *f1, cmplx *f2, cmplx *f3, int k){

  using namespace ScalarWave;

  *f1 = cmplx( -(2*k-3)*(k-1)+(l*l+l+1), -12*omega*M*(k-1) ) 
        /2.0/M/ cmplx(k*k, 4*omega*M*k) ;
  *f2 = cmplx( -(k-2)*(k-3)+l*(l+1), -12*omega*M*(k-2) ) 
        /4.0/M/M/ cmplx(k*k, 4*omega*M*k) ;
  *f3 = cmplx( 0.0, -omega*(k-3) ) 
        / 2.0/M/M/ cmplx(k*k, 4*omega*M*k) ;

}




cmplx coefkeh(int KK) {

  using namespace ScalarWave;

  static cmplx bk_1 = 1 ;
  static cmplx bk_2 = 0;
  static cmplx bk_3 = 0;
  static int k = 0;

  if (KK == 0) {
    bk_1 = 1;
    bk_2 = 0;
    bk_3 = 0;
    k = 0;
  }

  if (KK != k) {
    cerr << "sum::coefkeh called out of order, k = " 
         << KK << " expected " << k << endl;
    assert(KK == k);
  }

  cmplx f1 = 0;
  cmplx f2 = 0;
  cmplx f3 = 0;
  cmplx ret;

  if (k == 0) {
    k++;
    return 1;
  } else {
    fkseh(&f1, &f2, &f3, k);
    ret = f1*bk_1 + f2*bk_2 + f3*bk_3;
  }

  bk_3 = bk_2;
  bk_2 = bk_1;
  bk_1 = ret;

  k++;

  return ret;

}




cmplx coefkinf(int KK) {

  using namespace ScalarWave;

  static cmplx ak_1 = 1;
  static cmplx ak_2 = 0;
  static cmplx ak_3 = 0;
  static int k = 0;

  if (KK == 0) {
    ak_1 = 1;
    ak_2 = 0;
    ak_3 = 0;
    k = 0;
  }

  if (KK != k) {
    cerr << "sum::coefkinf called out of order, k = " 
         << KK << " expected " << k << endl;
    assert(KK == k);
  }

  cmplx f1, f2, f3, ret;

  if (k == 0) {
    k++;
    return 1;
  } else {
    fksinf(&f1, &f2, &f3, k);
    ret = f1*ak_1 + f2*ak_2 + f3*ak_3;
  }

  ak_3 = ak_2;
  ak_2 = ak_1;
  ak_1 = ret;

  k++;

  return ret;

}




// Scalar wave, zero frequency 


define D(a) double(a)

typedef complex<double> cmplx ;

cmplx Ylm(int l, int m, double theta, double phi);

void test_zero(double radius);

void getZeroPsiH(cmplx* PsiH, cmplx* dPsiH);
void getZeroPsiI(cmplx* PsiI, cmplx* dPsiI);

const double PI(M_PI);

using namespace ScalarWave ;




double Bang2LDL2(int L) {

  assert(L >= 0);

  double ret = 1;

  for (int i = L; i > 0; i--) {
    ret *= (2.*i)*(2.*i-1)/(i*i);
  }

  return ret;

}




double AnHorizon(double* Anm1, int n, int l) {
  return *Anm1 = -(*Anm1/(n*n))*(l+n)*(l-n+1);
}




double AnInfinity(double* Anm1, int n, int l) {
  *Anm1 = (*Anm1/(n*(2*l+1+n)))*(1.*(l+n)*(l+n));
  return *Anm1;
}



// Wronskian/M for zero frequency scalar modes

double WzeroDM(int l) {
  return (l%2==0?-1:+1)*Bang2LDL2(l)*2*(2*l+1);
}




void getZeroPsiH(cmplx* PsiH, cmplx* dPsiH) {

  *PsiH = 0;
  *dPsiH = 0;

  double b = 1;

  for (int n = 0; n <= l; n++) {
    cerr << SHOW(b) << endl;
    double Rn = pow(R/(2*M),n);
    *PsiH  += b*Rn;
    *dPsiH += n*b*Rn/R;
    cout << SHOW(*PsiH) << SHOW(*dPsiH) << endl;
    AnHorizon(&b,n+1, l);
  }

  assert(b == 0);

  cout << SHOW(*PsiH) << SHOW(*dPsiH) << endl;

}




void getZeroPsiI(cmplx* PsiI, cmplx* dPsiI) {

  *PsiI = 0;
  *dPsiI = 0;

  double a = 1;

  for (int n = 0; n <= 100; n++) {
    cerr << SHOW(a) << endl;
    double Rn = pow(2*M/R,n);
    *PsiI  += a*Rn;
    *dPsiI += -n*a*Rn/R;
    cout << SHOW(*PsiI) << SHOW(*dPsiI) << endl;

    if (n>1) {
      double corr = abs(a*Rn/(*PsiI)) + abs(-n*a*Rn*R/(2*M)/(*dPsiI));
      if (corr < 1.e-17) {
        cerr << SHOW(n) << SHOW(corr) << endl;
        break;
      }

    }

    AnInfinity(&a,n+1,l);

  }

  *PsiI *= pow(R/(2*M),-(l+1));
  *dPsiI = pow(R/(2*M),-(l+1))*(*dPsiI) -(l+1)*(*PsiI)/R;

}




void test_zero() {

  cout << "Entering test_zero" << endl;
  time_t tenter = time(0);


  // Set ScalarWave variables:

  M = 1;     // mass of the hole
  m = 0;     // spherical harmonic index
  R = 20*M;  // radius of the orbit
  w = M/(R*R*R);  // angular frequency of orbit
  omega = m*w;    // frequency of the radiation

  for (int l=0; l < 41; l++) {

    cmplx PsiH;
    cmplx dPsiH;

    getZeroPsiH(&PsiH, &dPsiH);

    cmplx PsiI = 0;
    cmplx dPsiI = 0;

    getZeroPsiI(&PsiI, &dPsiI);

    cout << "Here is a Wronskian test" << endl;
    cmplx W = R*R*(PsiH*dPsiI - PsiI*dPsiH)*(1-2*M/R);
    cout << setprecision(12)<< SHOW(l) << SHOW(W) 
         << SHOW(M*WzeroDM(l))  << endl;
    cout << SHOW(l) << SHOW(abs((W - M*WzeroDM(l))/W)) 
         << SHOW(W - M*WzeroDM(l)) << endl << endl;

    cout << SHOW(PsiI) << SHOW(dPsiI) << endl;
    cout << SHOW(PsiH) << SHOW(dPsiH) << endl;

  }

  time_t texit = time(0);

  cout<< "test_zero execution time is " << difftime(texit, tenter)
       << " seconds" << endl;

  cout << "Leaving test_zero" << endl;

}

\end{verbatim}
\end{singlespace}

   \backmatter		
      \bibliographystyle{unsrt}	
      \bibliography{References}	
\end{document}